\documentclass[useAMS,usenatbib,usegraphicx]{mn2e}   

\newcommand{\hi}{\hbox{H~I}}          
\newcommand{\hii}{\hbox{H~II}}        

\newcommand{\msun}{\hbox{M$_{\odot}$}}

\newcommand{\otwo}{\hbox{[O II] $\lambda3727$}}

\newcommand{\hdelta}{\hbox{H$\delta$}}
\newcommand{\hgamma}{\hbox{H$\gamma$}}
\newcommand{\othreea}{\hbox{[O III] $\lambda4363$}}
\newcommand{\hbeta}{\hbox{H$\beta$}}
\newcommand{\othree}{\hbox{[O III] $\lambda\lambda4959,5007$}}

\newcommand{\othreec}{\hbox{[O III] $\lambda5007$}}

\newcommand{\halpha}{\hbox{H$\alpha$}}
\newcommand{\ntwob}{\hbox{[N II] $\lambda6583$}}

\newcommand{\stwob}{\hbox{[S II] $\lambda6731$}}
\newcommand{\stwo}{\hbox{[S II] $\lambda\lambda6716,6731$}}

\newcommand{\otwored}{\hbox{[O II] $\lambda\lambda7320,7330$}}
\newcommand{\ntwootwo}{\hbox{[N II]/[O II]}}

\newcommand{\phs}{\phantom{$-$}}

\newcommand{\aap}{\hbox{A\&A}}

\newcommand{\aj}{\hbox{AJ}}
\newcommand{\apj}{\hbox{ApJ}}
\newcommand{\apjs}{\hbox{ApJS}}
\newcommand{\araa}{\hbox{ARA\&A}}
\newcommand{\mnras}{\hbox{MNRAS}}
\newcommand{\pasp}{\hbox{PASP}}

\title[O/H for Dwarf Irregulars in the Cen A Group]{
Interstellar Medium Oxygen Abundances of Dwarf Irregular Galaxies
in Centaurus~A and Nearby Groups\thanks{
Based on EFOSC2 observations collected at the European Southern
Observatory, Chile: proposal \#70.B-0180(B). 
} 
} 
\author[H. Lee et al.]{Henry Lee,$^{1,2}$\thanks{
E-mail: {\tt hlee@gemini.edu}
}
D. B. Zucker,$^{3}$
and 
E. K. Grebel$\,^{4}$
\vspace*{2mm}\\
$^1\,$Gemini Observatory, Southern Operations Center,
Casilla 603, La Serena, Chile. \\
$^2\,$Dept. of Astronomy, University of Minnesota, 
116 Church St. S.E., Minneapolis, MN 55455 USA. \\
$^3\,$Institute of Astronomy, 
University of Cambridge, Madingley Road, Cambridge CB3 0HA,
United Kingdom. \\
$^4\,$Astronomical Institute, Dept. of Physics \& Astronomy,
University of Basel, Venusstrasse 7, CH-4102 Binningen, Switzerland.
} 

\begin{document}

\date{\large {\em 
Accepted: 8 January 2007
}}

\pagerange{\pageref{firstpage}--\pageref{lastpage}} \pubyear{2007}

\maketitle

\label{firstpage}

\begin{abstract}		
%
%
We present results of optical spectroscopy of 35 \hii\ regions
from eight dwarf galaxies in the Centaurus~A group.
\othreea\ is detected in ESO272$-$G025 and ESO324$-$G024, and
direct oxygen abundances of 
12$+$log(O/H) = $7.76 \pm 0.09$ and $7.94 \pm 0.11$ 
are derived, respectively.
For the remaining galaxies, abundances are derived using
common bright-line methods.
To compare the influence of group environments on dwarf galaxies, we
have also gathered data for additional dwarf irregular galaxies from
the Cen~A and the Sculptor groups from the literature.
We have examined possible relationships between oxygen abundance, gas
fraction, effective chemical yield, and tidal indices.
Despite large positive tidal indices for a number of Cen~A dwarfs in 
the present sample, there is no clear separation between 
galaxies with positive tidal indices and galaxies with negative tidal
indices in the luminosity-metallicity, metallicity-gas fraction, and
metallicity-tidal index diagrams. 
The \hi\ surface mass density decreases with increasing positive tidal
index, which is expected in strong tidal encounters.
There are no strong trends between oxygen abundances or yields
and projected distances of galaxies within their respective
groups.
We also present spectra for 13 \hii\ regions in three nearby dwarf
irregular galaxies: DDO~47, NGC~3109, and Sextans~B.
For DDO~47, the \othreea\ oxygen abundance ($7.92 \pm 0.06$) for the
\hii\ region SHK91 No.~18 agrees with recently published values.
For Sextans~B, the \othreea\ oxygen abundance ($7.80 \pm 0.13$) 
for \hii\ region SHK91 No.~5 agrees with published work in which O$^+$
abundances were determined entirely from \otwored\ fluxes.
\end{abstract}


\section{Introduction}


The problem of the formation and evolution of dwarf galaxies has
partly been motivated by the long-standing questions about 
possible connections and transformations between different types of
dwarf galaxies. 
The two basic morphological types are:
gas-poor dwarf spheroidals with very little present-day star formation,
and gas-rich dwarf irregulars (dI's) and blue compact dwarf galaxies
which contain recent bursts of star formation.
Intrinsic properties (e.g., total mass) and external forces (e.g.,
environment) all have a role to play, but key processes
may be difficult to identify and disentangle 
(e.g.,
\citealp{bmcm86,grebel97,grebel99,ggh03,lee03virgo,vanzee04a,vanzee04b}).
The study of dwarf galaxies in different environments of varying
galaxy number density (i.e., field, groups, clusters) provides
valuable insights about the key parameters controlling evolution, and
constraints to the various models for galaxy evolution.

\subsection{Centaurus A Group}


The Centaurus A (Cen~A) group is a loose collection
of galaxies, and contains a very rich population of galaxies with 
the largest dispersion in morphological types \citep{devau79}.
The average distance of the Cen~A group is comparable to
that of the M81 group (cf. \citealp{kara02a,kara05}).
The Cen~A group may already be virialized as indicated by the
relatively short crossing timescale 
($\sim$ 2--3 Gyr; \citealp{kara05,tully05}). 

The Cen~A group is separated into two ``subgroups'' 
\citep[their Fig.~1]{kara02b} 
with one collection of galaxies surrounding NGC~5128 and the
other collection surrounding the spiral galaxy M~83.
The M~83 collection is more compact and contains
more late-type galaxies, whereas the NGC~5128 collection is
more dispersed and contains fewer late-type galaxies.
The NGC~5128 subgroup has a heliocentric velocity $v_{\odot} = +312$
km~s$^{-1}$ and an average distance $\langle D \rangle = 3.66$~Mpc.
There are 31 known members, of which ten (32\%) are classified 
as late-type dwarf galaxies 
(RC3 morphological~type $T > 8$; \citealp{rc3}).
The M~83 subgroup has $v_{\odot} = +308$ km~s$^{-1}$ and an
average distance $\langle D \rangle = 4.56$~Mpc.
There are 19 known members, of which 12 (63\%) are classified as
late-type dwarf galaxies ($T > 8$).

The Cen~A group is uniquely dominated by NGC~5128 (Centaurus~A), which
is a large massive radio-loud elliptical galaxy 
(e.g., \citealp{meier89}).
%
%
The remaining massive members in the group all exhibit disturbed
morphologies or abnormal properties, suggestive of a recent infall
episode in which a population of gas-rich dwarf galaxies 
has been accreted into the group 
(e.g., \citealp{graham79,vangorkom90,mirabel99}). 
This is precisely the scenario presented by \cite{peng02}, who
discovered a long, thin, blue arc in the northeast halo of NGC~5128.
This feature is thought to have once been a low-mass dI which fell
into the halo of the elliptical and is undergoing tidal disruption.
This may also help explain why there are relatively few dI's 
in the vicinity of NGC~5128, compared to M~83 \citep{kara04}.

Additional studies of dwarf galaxies in the Cen~A group have been
carried out in \halpha\ (S. C\^ot\'e et al., in prep.) and
in \hi\ (\citealp{cote97,banks99,cote00}) to examine the 
total and spatial distribution of gas and recent star-formation.
\cite{rejkuba06} studied the red giant stellar populations in two
Cen~A dwarf elliptical galaxies, and found the fraction of
intermediate-age stars to all stars is smaller than that found in
Local Group dwarf ellipticals.
\cite{jbf00} identified 13 new dwarf elliptical galaxies,
and confirmed the membership of two dwarf irregular galaxies
in the group.
They also identified AM~1318$-$444 and ESO~381$-$G018 as two new
dwarf irregulars in the Cen~A group.
\cite{grossi07} examined three gas-rich dwarf spheroidals with old
($\ga 2$ Gyr) stellar populations, and showed that the relatively high
gas content could be explained by the low level of past star
formation.


Oxygen abundances are a reliable measure of the present-day
gas-phase metallicity within \hii\ regions in gas-rich star-forming
dwarf galaxies (e.g., \citealp{dinerstein90,skillman98}).
In fact, they provide useful constraints as the most
recent and maximum metallicity to anchor the star-formation
history.
\cite{ws83} and \cite{webster83} obtained oxygen abundances for
southern irregular and spiral galaxies, including galaxies in the Cen~A
group and in the field, with a measured range of abundances from about
ten to sixty per cent of the solar value.
Although these papers have been, until recently, the primary work
regarding abundances in Cen~A late-type dwarf galaxies, the spectra
were obtained with inherently non-linear detectors.
In many cases, the character of the non-linearities were
not understood until after publication (e.g., \citealp{jenkins87});
subsequent corrections for non-linearity were not possible.

For a comparison of gas-phase abundances with published work on
nearby dwarf galaxies (e.g.,
\citealp{scm03b,ls04,lsv05,lsv06,vzsh06}), we undertook a 
program of determining new and confirming previous nebular oxygen
abundances for gas-rich dwarf galaxies in the Local Volume
\citep{grebel00}, which would help answer the question
about whether recent chemical enrichment is sensitive to the (group)
environment.
Here, we present spectra of 35 \hii\ regions in eight Cen~A dI's,
as well as spectra of 13 \hii\ regions in three additional nearby
dwarf galaxies.
The properties of galaxies in the present sample are listed in
Table~\ref{table_gxylist}.


\begin{table*}
\begin{minipage}{160mm}
\caption{
Basic properties of galaxies in the present sample.
}
\label{table_gxylist}
\scriptsize 
\begin{center}
\renewcommand{\tabcolsep}{4.3pt} 
\vspace*{0mm}
\begin{tabular}{lcccccccccc}
\hline \hline
& Other & & $v_{\odot}$ & $B_T$ & $F_{21}$ & $D$ & & & & 12$+$ \\
Galaxy & Name(s) & $T$ & (km s$^{-1}$) & (mag) & (Jy km s$^{-1}$) &
(Mpc) & Method & $\Theta$ & MD & log(O/H) \\
(1) & (2) & (3) & (4) & (5) & (6) & (7) & (8) & (9) & (10) & (11) \\
\hline
\multicolumn{11}{c}{{\sf Centaurus A group dwarfs}} \\
\hline
AM 1318$-$444 & KK 196 & 
10 & 
$+741$ & 16.1 & \ldots & $3.98 \pm 0.32$ & TRGB & $+2.2$ & Cen A & 
$7.87 \pm 0.20$ \\ 
AM 1321$-$304 & KK 200, KDG 15 & 
9 & 
$+490$ & 16.67 & 1.7 & $4.63 \pm 0.46$ & TRGB & $+1.2$ & M 83 & 
\ldots \\
ESO 272$-$G025 & PGC 52591 & 
8 & 
$+624$ & 14.79 & 3.0 & 5.9 & HF & $-1.5$ & Cen A & 
$7.76 \pm 0.09$ \\
ESO 274$-$G001 & UKS 1510$-$466 & 
7 & 
$+522$ & 11.70 & 117$\,^a$ & $3.05 \pm 0.24$ & TRGB & $-1.0$ & Cen A &
$8.38 \pm 0.20$ \\
ESO 321$-$G014 & AM 1211$-$375 & 
10 & 
$+613$ & 15.16 & 2.9 & $3.19 \pm 0.26$ & TRGB & $-0.3$ & Cen A & 
\ldots \\
ESO 324$-$G024 & AM 1324$-$411 & 
10 & 
$+513$ & 12.91 & 52.1$\,^a$ & $3.73 \pm 0.43$ & TRGB & $+2.4$ & Cen A &
$7.94 \pm 0.18$ \\
ESO 325$-$G011 & AM 1342$-$413 & 
10 & 
$+541$ & 13.99 & 25.4$\,^a$ & $3.40 \pm 0.39$ & TRGB & $+1.1$ & Cen A &
$7.94 \pm 0.20$ \\
ESO 381$-$G020 & AM 1243$-$333 & 
10 & 
$+590$ & 14.44 & 
36$\,^b$ & $5.45 \pm 0.44$ & TRGB & $-0.3$ & M 83 & 
$7.90 \pm 0.20$ \\
IC 4247 & ESO 444$-$G034 & 
10 & 
$+274$ & 14.41 & 3.0 & $4.97 \pm 0.40$ & TRGB & $+1.5$ & M 83 &
$8.27 \pm 0.20$ \\
IC 4316 & ESO 445$-$G006 & 
10 & 
$+674$ & 14.97 & 7.8$\,^a$ & $4.41 \pm 0.44$ & TRGB & $+2.4$ & M 83 &
$8.16 \pm 0.20$ \\
\hline
\multicolumn{11}{c}{{\sf Nearby dwarfs}} \\
\hline
DDO 47 & UGC 3974 & 
10 & 
$+272$ & 13.60 & 67.0$\,^c$ & $5.18 \pm 0.57$ & TRGB & $-2.1$ & NGC 2683 &
$7.92 \pm 0.06$ \\
NGC 3109$\,^d$ & DDO 236 & 
9 & 
$+404$ & 10.26 & 1110$\,^e$ & $1.33 \pm 0.08$ & TRGB & $-0.1$ & Antlia &
$7.94 \pm 0.20$ \\
Sextans B$\,^d$ & DDO 70, UGC 5373 &
10 & 
$+301$ & 11.85 & 102.4$\,^c$ & $1.36 \pm 0.07$ & TRGB & $-0.7$ & MWay &
$7.80 \pm 0.13$ \\
\hline
\end{tabular}
\end{center}
%
Galaxies are are listed in alphabetical order by their primary name.
All properties are obtained from NED, unless otherwise noted.
Cols.~(1) and (2): Galaxy name used in the present work, and other
names from NED, respectively. 
Col.~(3): Morphological type \citep{rc3}.
Col.~(4): Heliocentric velocity.
Col.~(5): Total apparent $B$ magnitude.
Col.~(6): Total \hi\ 21-cm flux; additional reference: \citet{kara04}.
Col.~(7): Measured or estimated distances.
References:
Centaurus~A group galaxies --- \citet{kara02b,kara04,kara06};
DDO~47 --- \citet{kara03b};
NGC~3109 --- \citet{kara02c};
Sextans~B --- \citet{sakai97}, \cite{kara02c}.
Col.~(8): Method of determining distances:
TRGB --- tip of the red giant branch;
HF --- Hubble flow.
Cols.~(9) and (10): Tidal index, and ``closest'' or main disturber
(MD) galaxy, respectively \citep{kara04}. 
Col.~(11): Nebular oxygen abundances derived in the present work.
NOTES
$^a$~From \citet{cote97}.
$^b$~From \citet{banks99}.
$^c$~From \citet{hoffman96}.
$^d$~Common group of dwarfs including NGC~3109, Sextans~A,
Sextans~B, and the Antlia dwarf \citep{vdb99,tully02}.
$^e$~From \citet{bdb01}.
\end{minipage}
\end{table*}

The paper is organized as follows.
Observations and reductions of the data are presented in
\S~\ref{sec_obs}, 
the measurements and analyses are given in \S~\ref{sec_analysis},
derivations of chemical abundances are described in
\S~\ref{sec_abund}, 
individual galaxies are presented in
\S~\ref{sec_discuss},
a discussion of environmental effects is given in
\S~\ref{sec_enveffects}, 
and the conclusions are given in \S~\ref{sec_concl}.
For the remainder of this paper, we adopt 12$+$log(O/H) = 8.66 as the
revised solar value for the oxygen abundance, and $Z_{\odot}$ = 0.0126
as the revised solar mass fraction in the form of metals
\citep{asplund04,melendez04}.

\section{Observations and Reductions}
\label{sec_obs}

Long-slit spectroscopic observations were carried out on 
2003 Mar. 6--8 UT with the EFOSC2 imaging spectrograph on the 3.6-m
telescope at ESO La Silla Observatory. 
Details of the instrumentation employed and the log of observations
are listed in Tables~\ref{table_obsprops} and \ref{table_obslog},
respectively.
Observations were obtained just after new moon phase.
Conditions were mostly clear on the first two nights with patchy thin
cloud towards the end of the second night; 
the third night was clear and photometric.

\begin{table*}
\begin{minipage}{160mm}
\caption{
Properties of EFOSC2 spectrograph employed at the ESO La Silla
3.6-m telescope.
}
\label{table_obsprops}
\begin{center}
\renewcommand{\arraystretch}{1.0}
\vspace*{0mm}
\begin{tabular}{lc}
\hline \hline
%
CCD & Loral, No.\ 40 \\
Total area & 2048 $\times$ 2048 pix$^2$ \\
Field of view & 5\farcm2 $\times$ 5\farcm2 \\
Pixel size & 15 $\mu$m \\
Image scale & 0\farcs{157} pix$^{-1}$ \\
Gain & 1.3 $e^-$ ADU$^{-1}$ \\
Read-noise (rms) & 8.5 $e^-$ \\ 
%
%
Grating & No.\ 11 \\
Groove density & 300 lines mm$^{-1}$ \\
Blaze $\lambda$ (1st order) & 4000 \AA \\
Dispersion & 2.0 \AA\ pix$^{-1}$ \\
Effective $\lambda$ range & 3380--7520 \AA \\ 
%
%
Long-slit length & $\simeq 5$\arcmin \\
Slit width & 1\arcsec \\
\hline
\end{tabular}
\end{center}
\end{minipage}
\end{table*}

\begin{table*}
\begin{minipage}{160mm}
\caption{
Log of observations.
}
\label{table_obslog}
\footnotesize 
\begin{center}
\renewcommand{\arraystretch}{1.0}
\begin{tabular}{lcccccc}
\hline \hline
& 2003 Date & & $t_{\rm total}$ & & & RMS \\ 
Galaxy & (UT) & $N_{\rm exp}$ & (s) & $\langle X \rangle$ & 
\othreea & (mag) \\
(1) & (2) & (3) & (4) & (5) & (6) & (7)  \\
\hline
AM 1318$-$444 & 7 Mar & $3 \times 1200 + 1 \times 1800$ & 5400 & 
1.11 & \ldots & 0.021  \\
AM 1321$-$304$\,^a$ & 7 Mar & 
\ldots & \ldots & \ldots & \ldots & \ldots \\
%
DDO 47 & 7 Mar &
	  $4 \times 1200$ & 4800 & 1.44 & yes & 0.021 \\
ESO 272$-$G025 & 6 Mar & $5 \times 1200$ & 6000 & 1.04 & yes & 0.021 \\
ESO 274$-$G001 & 8 Mar & $4 \times 600$ & 2400 & 
1.05 & \ldots & 0.024 \\
ESO 321$-$G014$\,^b$ & 6 Mar & $1 \times 1200$ & 1200 & 
\ldots & \ldots & \ldots \\
ESO 324$-$G024 & 7 Mar & $4 \times 1200$ & 4800 & 1.02 & yes & 0.021 \\
ESO 325$-$G011 & 8 Mar & $3 \times 1200$ & 3600 & 1.04 & \ldots & 0.024 \\
ESO 381$-$G020 & 8 Mar & $2 \times 1200$ & 2400 & 1.22 & \ldots & 0.024 \\
IC 4247 & 8 Mar & $3 \times 1200$ & 3600 & 1.17 & \ldots & 0.024 \\
IC 4316 & 6 Mar & $4 \times 1200$ & 4800 & 1.03 & \ldots & 0.021 \\
%
NGC 3109 slit A & 7 Mar & 
	  $4 \times 1200$ & 4800 & 1.01 & \ldots & 0.021 \\
%
NGC 3109 slit B & 8 Mar &
	  $4 \times 1200$ & 4800 & 1.10 & \ldots & 0.024 \\
%
Sextans B slit A & 6 Mar & %
	  $3 \times 1200$ & 3600 & 1.43 & \ldots & 0.021 \\
%
Sextans B slit B & 6 Mar &
	  $1 \times 548 + 4 \times 1200$ & 5348 & 1.23 & yes & 0.021 \\
\hline
\end{tabular}
\end{center}
Col.~(1): Galaxy name (in alphabetical order).
Col.~(2): Date of observation.
Col.~(3): Number of exposures obtained and the length of each
exposure in seconds. 
Col.~(4): Total exposure time.
Col.~(5): Mean effective airmass.
Col.~(6): \othreea\ detection.
Col.~(7): Relative root-mean-square error in the sensitivity
function obtained from observations of standard stars.
NOTES:
$^a$~No compact \halpha\ emission observed in a three-minute
acquisition frame.
%
$^b$~Faint diffuse \halpha\ emission. 
The long-slit was placed along major axis of the galaxy, but no
emission lines were seen.
%
%
\end{minipage}
\end{table*}

Three- to five-minute \halpha\ acquisition images were obtained in
order to position the slit.
The slit angle was aligned to obtain spectra of multiple \hii\ 
regions.
H~II region targets for each galaxy are shown
in Figs.~\ref{fig_dwarfs1} and \ref{fig_dwarfs2}.
Identifications and locations of various \hii\ regions in each target
are labeled.
The acquisition images were obtained in ``fast'' readout mode, which
is about a factor of two faster in readout time than ``normal'' mode.
However, two different amplifiers are used during ``fast'' readout,
which causes different characteristics for the two halves of the CCD
detector.
The ``split'' appearance was corrected by adding a constructed image
with zeros on one-half of the frame and the average zero-point
difference on the other half of the frame.
All subsequent science exposures (biases, flats, arcs, targets) 
were obtained in ``normal'' readout.

\begin{figure*}
\centering
\caption{
Raw (unreduced) \halpha\ acquisition images in three-minute exposures 
taken with EFOSC2. 
Black objects indicate bright sources.
North and east are to the top and the left, respectively, in each frame.
Thin white stripes in the frames indicates bad rows or columns.
Numbers in each panel indicate the locations of extracted \hii\ region
spectra tabulated in Tables~\ref{table_data1} to \ref{table_data8}.
All images shown have approximately 1\farcm6 by 1\farcm3 fields of
view.
(a) AM1318$-$444.
(b) ESO~272$-$G025.
(c) ESO~274$-$G001.
(d) ESO~324$-$G024.
(e) ESO~325$-$G011.
(f) ESO~381$-$G020.
\hii\ regions 1 and 3 correspond to \hii\ regions A and B,
respectively, from \protect\cite{webster83}; 
their \hii\ region C is the dark blob to the southwest of 
\hii\ region 1.
}
\label{fig_dwarfs1}
\end{figure*}

\begin{figure*}
\centering
\caption{
Additional \halpha\ acquisition images.
(a) IC~4247.
(b) IC~4316.
(c) DDO~47.
(d) NGC~3109 (approx. 5\farcm2 by 5\farcm2 field of view).
(e) Sextans~B (approx. 5\farcm2 by 5\farcm2 field of view).
See Fig.~\ref{fig_dwarfs1} for additional comments.
}
\label{fig_dwarfs2}
\end{figure*}

Data reductions were carried out using standard IRAF\footnote{
IRAF is distributed by the National Optical 
Astronomical Observatories, which is operated by the Associated
Universities for Research in Astronomy, Inc., under contract to the
National Science Foundation.}
routines.
Data obtained on each separate night were reduced independently.
The raw two-dimensional images were bias-subtracted and trimmed.
Dome flat exposures were used to remove pixel-to-pixel variations 
in response, and
twilight flats were acquired at dusk each night to correct
for variations over larger spatial scales.
To correct for the ``slit function'' in the spatial direction, the
variation of illumination along the slit was taken into account
using dome and twilight flats. 
Cosmic rays were removed in the addition of multiple exposures.
Wavelength calibration was obtained using helium-argon (He-Ar) arc
lamp exposures taken throughout each night.
Flux calibration was obtained using exposures of standard stars
CD$-32^{\circ}9927$, Feige~56, LTT~1788, and LTT~3864. 
The flux accuracy is listed in Table~\ref{table_obslog}.
One-dimensional spectra for each \hii\ region were obtained with
unweighted summed extractions.
Some of the spectra are shown in Fig.~\ref{fig_spectra}.
%

The \otwored\ doublet was detected and reported in a number of
spectra below.
However, we did not use a ultraviolet-cutoff filter, and we caution
the reader that the reported \otwored\ fluxes may suffer
from second-order contamination.
In the present work, we used only \otwo\ to derive O$^+$/H ionic
abundances and total oxygen abundances.

\begin{figure*}
\caption{
{\em Left panel}:
[O III] $\lambda$4363 detections : 
DDO~47 SHK91 No.~18, ESO~272$-$G025 No.~1, ESO~324$-$G024 No.~4, and 
Sextans~B SHK91 No.~5.
The SHK identifications for \hii\ regions are found in \citet{strobel91}.
The observed flux per unit wavelength is plotted versus wavelength.
Within each panel, the full spectrum and an expanded view of the
spectrum to highlight faint emission lines are shown.
The \othreea\ line is indicated by an arrow in each panel.
{\em Right panel}:
Other example \hii\ region spectra for 
AM~1318$-$444 No.~1, ESO~274$-$G001 No.~4, ESO~325$-$G01 No.~4,
ESO~381$-$G020 No.~3, IC~4316 No.~2, and NGC~3109 RM92 No.~8.
RM92 refers to \citet{rm92}. 
}
\label{fig_spectra}
\end{figure*}
 
\section{Measurements and Analysis}
\label{sec_analysis}

Emission-line strengths were measured using software developed
by M. L. McCall and L. Mundy; see \cite{lee03south,lee03field,lee03virgo}.
The temperature-sensitive \othreea\ emission line was detected in four
galaxies: 
DDO~47 (\hii\ region SHK91 No.~18), ESO~272-G025 (\hii\ region No.~1),
ESO~324-G024 (\hii\ region No.~4), and 
Sextans~B (\hii\ region SHK91 No.~5).

We have followed the general procedure described in
\cite{ls04} and \cite{lsv05,lsv06}.
To derive reddening values, \halpha\ and \hbeta\ fluxes were
used with the following
\begin{equation}
\log\frac{I(\lambda)}{I(\hbeta)} =
\log\frac{F(\lambda)}{F(\hbeta)} + c(\hbeta) \, f(\lambda),
\label{eqn_corr}
\end{equation}
where 
$F$ and $I$ are the observed flux and corrected intensity ratios,
respectively,
$c(\hbeta)$ is the logarithmic extinction at \hbeta, and 
$f(\lambda)$ is the wavelength-dependent reddening function 
\citep{aller84,osterbrock}.
The logarithmic extinction can be expressed as
\begin{equation}
c(\hbeta) = 1.43 \, E(B-V) = 0.47 \, A_V,
\label{eqn_chbebv}
\end{equation}
where $E(B-V)$ is the reddening and $A_V$ is the extinction
in $V$.
Intrinsic case-B Balmer line ratios determined by \cite{sh95} were
assumed.
In the absence of \hdelta, a best solution to the expected 
$I$(\halpha)/$I$(\hbeta) and $I$(\hgamma)/$I$(\hbeta) intensity
ratios was obtained simultaneously.
Only the $F$(\halpha)/$F$(\hbeta) flux ratio was used to determine a
reddening in the absence of both \hgamma\ and \hdelta. 
An initial temperature $T_e = 10^4$~K was used to derive
the reddening.
As the \stwo\ lines were generally unresolved, $n_e$ = 100~cm$^{-3}$
was adopted for the electron density.

The reddening function normalized to \hbeta\ is derived from
the \cite{cardelli89} reddening law, assuming $R_V$ = 3.07.
As described in \cite{scm03b}, values of $c(\hbeta)$
were derived from the error weighted average of values for
$F(\halpha)/F(\hbeta)$, $F(\hgamma)/F(\hbeta)$, and
$F(\hdelta)/F(\hbeta)$ ratios while simultaneously solving for the
effects of underlying Balmer absorption with equivalent
width, EW$_{\rm abs}$.
We assumed that EW$_{\rm abs}$ was the same for \halpha, \hbeta,
\hgamma, and \hdelta.
Uncertainties in $c(\hbeta)$ and EW$_{\rm abs}$ were determined
from Monte Carlo simulations \citep{os01,scm03b}.
Errors derived from these simulations are larger than errors
quoted in the literature by either assuming a constant value for
the underlying absorption or derived from a $\chi^2$ analysis in
the absence of Monte Carlo simulations for the errors; 
Fig.~\ref{fig_monte} shows an example of these simulations.

\begin{figure*}
\centering
\caption{
Monte Carlo simulations of solutions for the reddening, $c(\hbeta)$,
and the underlying Balmer absorption with equivalent width, 
EW$_{\rm abs}$, from hydrogen Balmer flux ratios.
The dotted line indicates zero reddening.
The results here are shown for the \hii\ region ESO~272$-$G025 No.~1.
Each small point is a solution derived from a different realization
of the same input spectrum.
The large filled circle with error bars shows the mean result with
$1\sigma$ errors derived from the dispersion in the solutions.
}
\label{fig_monte}
\end{figure*}

Observed flux $(F)$ and corrected intensity $(I)$ ratios, observed
\hbeta\ fluxes, logarithmic reddening values, and the equivalent
widths of underlying Balmer absorption are listed in
Tables~\ref{table_data1} to \ref{table_data9} inclusive.
The listed errors for the observed flux ratios at each wavelength
$\lambda$ account for the errors in the fits to the line profiles,
their surrounding continua, and the relative error in the sensitivity
function stated in Table~\ref{table_obslog}.
At the \hbeta\ reference line, errors for both observed and corrected
ratios do not include the error in the flux.
Where negative values of the reddening were derived, the reddening was
set to zero in correcting line ratios and in abundance calculations.
Where \othreea\ is absent, we have assumed a fiducial temperature of
$T_e = 10^4$~K, and, in a number of cases, we have also assumed
2~\AA\ for the equivalent width of the underlying Balmer absorption.


\begin{table*}
\begin{minipage}{160mm}
\caption{
Line ratios and properties for \hii\ regions in Cen~A dwarf irregular
galaxies. 
}
\label{table_data1}
\scriptsize 
\begin{center}
\renewcommand{\arraystretch}{0.95} 
\begin{tabular}{rccccccc}
\hline \hline
& & \multicolumn{2}{c}{AM 1318$-$444 No. 1} & 
\multicolumn{2}{c}{ESO 272$-$G025 No. 1} &
\multicolumn{2}{c}{ESO 272$-$G025 No. 2} \\
\multicolumn{1}{c}{Property} &
\multicolumn{1}{c}{$f(\lambda)$} &
\multicolumn{1}{c}{$F$} & \multicolumn{1}{c}{$I$} &
\multicolumn{1}{c}{$F$} & \multicolumn{1}{c}{$I$} &
\multicolumn{1}{c}{$F$} & \multicolumn{1}{c}{$I$} \\
\hline
$[\rm{O\;II}]\;3727$ & $+0.325$ &
	  $238.2 \pm 4.6$ & $240 \pm 14$ &
	  $199.6 \pm 6.5$ & $208 \pm 13$ &
	  $297.8 \pm 7.8$ & $414 \pm 11$
\\
$[{\rm Ne\;III}]\;3869$ & $+0.294$ &
	  $8.0 \pm 2.1$ & $8.0 \pm 2.1$ &
	  $35.9 \pm 3.5$ & $36.9 \pm 4.0$ &
	  $22.1 \pm 4.3$ & $29.5 \pm 5.7$
\\
${\rm H}8 + {\rm He\;I}\;3889$ & $+0.289$ &
	  $10.0 \pm 2.2$ & $23.3 \pm 2.5$ &
	  $9.8 \pm 2.7$ & $23.7 \pm 3.0$ &
	   \ldots & \ldots
\\
${\rm H}\epsilon + {\rm He\;I}\;3970\,^a$ & $+0.269$ &
	  $7.5 \pm 1.2$ & $20.5 \pm 1.5$ &
	   \ldots & \ldots &
	   \ldots & \ldots
\\
${\rm H}\delta\;4101$ & $+0.232$ &
	  $13.6 \pm 1.8$ & $25.6 \pm 2.0$ &
	  $13.1 \pm 2.3$ & $25.3 \pm 2.5$ &
	   \ldots & \ldots
\\
${\rm H}\gamma\;4340$ & $+0.158$ &
	  $39.6 \pm 1.7$ & $47.8 \pm 2.1$ &
	  $41.7 \pm 2.6$ & $50.6 \pm 2.9$ &
	  $32.1 \pm 3.6$ & $46.7 \pm 4.0$
\\
$[{\rm O\;III}]\;4363$ & $+0.151$ &
	  $< 3.8$ & $< 3.7$ &
	  $8.2 \pm 2.1$ & $7.9 \pm 2.0$ &
	   \ldots & \ldots
\\
${\rm H}\beta\;4861$ & \phs0.000 &
	  $100.0 \pm 3.9$ & $100.0 \pm 3.6$ &
	  $100.0 \pm 4.0$ & $100.0 \pm 3.6$ &
	  $100.0 \pm 4.4$ & $100.0 \pm 4.0$
\\
$[{\rm O\;III}]\;4959$ & $-0.026$ &
	  $70.0 \pm 4.0$ & $64.4 \pm 3.7$ &
	  $142.7 \pm 8.4$ & $128.4 \pm 7.6$ &
	  $55.4 \pm 5.0$ & $49.0 \pm 4.4$
\\
$[{\rm O\;III}]\;5007$ & $-0.038$ &
	  $203.8 \pm 5.1$ & $186.9 \pm 4.8$ &
	  $414 \pm 10$ & $370.3 \pm 9.7$ &
	  $171.5 \pm 6.4$ & $149.3 \pm 5.6$
\\
${\rm He\;I}\;5876$ & $-0.204$ &
	  $5.9 \pm 1.9$ & $5.2 \pm 1.7$ &
	  \ldots & \ldots &
	  \ldots & \ldots
\\
${\rm H}\alpha\;6563$ & $-0.299$ &
	  $333.1 \pm 6.9$ & $289 \pm 16$ &
	  $342.8 \pm 6.9$ & $283 \pm 15$ &
	  $468.3 \pm 8.8$ & $295.7 \pm 5.5$
\\
$[{\rm N\;II}]\;6583$ & $-0.301$ &
	  $10.4 \pm 5.8$ & $8.9 \pm 5.0$ &
	  $12.0 \pm 5.5$ & $9.6 \pm 4.4$ &
	  $55.5 \pm 7.2$ & $34.4 \pm 4.5$
\\
${\rm He\;I}\;6678$ & $-0.314$ &
	  $6.1 \pm 3.1$ & $5.2 \pm 2.7$ &
	  \ldots & \ldots &
	   &
\\
$[{\rm S\;II}]\;6716, 6731$ & $-0.320$ &
	  $47.4 \pm 4.4$ & $40.3 \pm 4.3$ &
	  $96.4 \pm 8.6$ & $76.8 \pm 8.0$ &
	  $106.7 \pm 9.0$ & $64.7 \pm 5.5$
\\
$[{\rm O\;II}]\;7320, 7330$ & $-0.400$ &
	  $12.1 \pm 3.5$ & $10.1 \pm 3.0$ &
	  \ldots & \ldots &
           \ldots & \ldots
\\[1mm]
\multicolumn{2}{c}{$F(\hbeta)$ (ergs s$^{-1}$ cm$^{-2}$)} & 
	  \multicolumn{2}{c}{$(9.19 \pm 0.36) \times 10^{-16}$} &
	  \multicolumn{2}{c}{$(2.71 \pm 0.11) \times 10^{-16}$} &
	  \multicolumn{2}{c}{$(2.56 \pm 0.11) \times 10^{-16}$}
\\
\multicolumn{2}{c}{EW$_{\rm e}$(\hbeta) (\AA)} &
	  \multicolumn{2}{c}{$61.5 \pm 3.6$} &
	  \multicolumn{2}{c}{$35.4 \pm 1.7$} &
	  \multicolumn{2}{c}{$21.4 \pm 1.0$} 
\\
\multicolumn{2}{c}{$c(\hbeta)$} &
	  \multicolumn{2}{c}{$0.115 \pm 0.074$} &
	  \multicolumn{2}{c}{$0.179 \pm 0.073$} &
	  \multicolumn{2}{c}{0.56} 
\\
\multicolumn{2}{c}{EW$_{\rm abs}$ (\AA)} &
	  \multicolumn{2}{c}{$4.9 \pm 1.8$} &
	  \multicolumn{2}{c}{$3.5 \pm 1.3$} &
	  \multicolumn{2}{c}{2} 
\\[1mm]
\hline
& & \multicolumn{2}{c}{ESO 274$-$G001 No. 1} & 
\multicolumn{2}{c}{ESO 274$-$G001 No. 2} &
\multicolumn{2}{c}{ESO 274$-$G001 No. 3} \\
\multicolumn{1}{c}{Property} &
\multicolumn{1}{c}{$f(\lambda)$} &
\multicolumn{1}{c}{$F$} & \multicolumn{1}{c}{$I$} &
\multicolumn{2}{c}{$F$} & 
\multicolumn{2}{c}{$F$} \\
\hline
$[\rm{O\;II}]\;3727$ & $+0.325$ &
	  $253 \pm 20$ & $302 \pm 24$ &
	  \multicolumn{2}{c}{$47 \pm 10$} &
	  \multicolumn{2}{c}{$44.6 \pm 6.6$} 
\\
${\rm H}\beta\;4861$ & \phs0.000 &
	  $100 \pm 11$ & $100 \pm 10$ &
	  \multicolumn{2}{c}{\ldots} &
	  \multicolumn{2}{c}{\ldots}
\\
$[{\rm O\;III}]\;4959$ & $-0.026$ &
	  $5.5 \pm 7.9$ & $5.2 \pm 7.4$ &
	  \multicolumn{2}{c}{\ldots} &
	  \multicolumn{2}{c}{\ldots} 
\\
$[{\rm O\;III}]\;5007$ & $-0.038$ &
	  $39.6 \pm 8.3$ & $37.0 \pm 7.8$ &
	  \multicolumn{2}{c}{\ldots} &
	  \multicolumn{2}{c}{\ldots}
\\
${\rm H}\alpha\;6563$ & $-0.299$ &
	  $360 \pm 17$ & $286 \pm 13$ &
	  \multicolumn{2}{c}{$100 \pm 14$} &
	  \multicolumn{2}{c}{$100.0 \pm 8.1$}
\\
$[{\rm N\;II}]\;6583$ & $-0.301$ &
	  $45 \pm 14$ & $36 \pm 11$ &
	  \multicolumn{2}{c}{$9.5 \pm 11$} &
	  \multicolumn{2}{c}{$18.0 \pm 6.4$}
\\
$[{\rm S\;II}]\;6716$ & $-0.319$ &
	  $90 \pm 11$ & $69.4 \pm 8.4$ &
	  \multicolumn{2}{c}{\ldots} &
	  \multicolumn{2}{c}{\ldots}
\\
$[{\rm S\;II}]\;6731$ & $-0.321$ &
	  $73 \pm 10$ & $56.1 \pm 8.1$ &
	  \multicolumn{2}{c}{\ldots} &
	  \multicolumn{2}{c}{\ldots}
\\[1mm]
\multicolumn{2}{c}{$F(\hbeta)$ (ergs s$^{-1}$ cm$^{-2}$)} & 
	  \multicolumn{2}{c}{$(3.26 \pm 0.34) \times 10^{-16}$} &
	  \multicolumn{2}{c}{$(2.56 \pm 0.37) \times 10^{-16}\,^b$} &
	  \multicolumn{2}{c}{$(6.12 \pm 0.50) \times 10^{-16}\,^b$}
\\
\multicolumn{2}{c}{EW$_{\rm e}$(\hbeta) (\AA)} &
	  \multicolumn{2}{c}{$45.8 \pm 6.1$} &
	  \multicolumn{2}{c}{\ldots} &
	  \multicolumn{2}{c}{\ldots} 
\\
\multicolumn{2}{c}{$c(\hbeta)$} &
	  \multicolumn{2}{c}{0.29} &
	  \multicolumn{2}{c}{\ldots} &
	  \multicolumn{2}{c}{\ldots} 
\\
\multicolumn{2}{c}{EW$_{\rm abs}$ (\AA)} &
	  \multicolumn{2}{c}{2} &
	  \multicolumn{2}{c}{\ldots} &
	  \multicolumn{2}{c}{\ldots} 
\\
\hline
\end{tabular}
\end{center}
%
Emission lines are listed in Angstroms.
$F$ is the observed flux ratio with respect to \hbeta.
$I$ is the corrected intensity ratio, corrected for the adopted
reddening listed, and for underlying Balmer absorption.
The uncertainties in the observed line ratios account for the
uncertainties in the fits to the line profiles, the surrounding
continua, and the relative uncertainty in the sensitivity function
listed in Table~\ref{table_obslog}. 
Flux uncertainties in the \hbeta\ reference line are not included.
Uncertainties in the corrected line ratios account for uncertainties
in the specified line and in the \hbeta\ reference line.
The reddening function, $f(\lambda)$, from
Equation~(\ref{eqn_corr}) is given.
Also listed are the observed \hbeta\ flux,
and the equivalent width of \hbeta\ in emission, EW$_{\rm e}$(\hbeta).
Where \othreea\ is measured, simultaneous solutions for the
logarithmic reddening, $c(\hbeta)$, from Equation~(\ref{eqn_corr}) and
the equivalent width of the underlying Balmer absorption,
EW$_{\rm abs}$ are listed.
Where \othreea\ is not measured, the equivalent width of the
underlying Balmer absorption was set to 2~\AA.
NOTES:
$^a$~Blended with [Ne III] $\lambda$3967.
$^b$~Observed \halpha\ flux in emission.
\end{minipage}
\end{table*}

\begin{table*}
\begin{minipage}{160mm}
\caption{
Line ratios and properties for \hii\ regions in Cen~A dwarf irregular
galaxies (continued).
}
\label{table_data2}
\scriptsize 
\begin{center}
\renewcommand{\arraystretch}{0.95} 
\vspace*{0mm}
\begin{tabular}{rccccccc}
\hline \hline
& & \multicolumn{2}{c}{ESO 274$-$G001 No. 4} & 
\multicolumn{2}{c}{ESO 274$-$G001 No. 5} &
\multicolumn{2}{c}{ESO 274$-$G001 No. 6} \\
\multicolumn{1}{c}{Property} &
\multicolumn{1}{c}{$f(\lambda)$} &
\multicolumn{1}{c}{$F$} & \multicolumn{1}{c}{$I$} &
\multicolumn{1}{c}{$F$} & \multicolumn{1}{c}{$I$} &
\multicolumn{1}{c}{$F$} & \multicolumn{1}{c}{$I$} \\
\hline
$[\rm{O\;II}]\;3727$ & $+0.325$ &
	  $200.4 \pm 3.2$ & $373 \pm 15$ &
	  $129 \pm 35$ & $139 \pm 38$ &
	  $446 \pm 23$ & $605 \pm 31$
\\
$[{\rm Ne\;III}]\;3869$ & $+0.294$ &
	  $21.3 \pm 2.7$ & $36.9 \pm 4.8$ &
	  \ldots & \ldots &
	  \ldots & \ldots
\\
$[{\rm S\;II}]\;4068$ & $+0.241$ &
	  $23.1 \pm 2.7$ & $35.7 \pm 4.3$ &
	  \ldots & \ldots &
	  \ldots & \ldots
\\
${\rm H}\delta\;4101$ & $+0.232$ &
	  $7.3 \pm 2.1$ & $26.0 \pm 3.3$ &
	  \ldots & \ldots &
	  \ldots & \ldots
\\
${\rm H}\gamma\;4340$ & $+0.158$ &
	  $32.4 \pm 2.1$ & $53.6 \pm 2.9$ &
	  \ldots & \ldots &
	  $48.7 \pm 9.4$ & $59 \pm 11$
\\
$[{\rm O\;III}]\;4363$ & $+0.151$ &
	  $< 5.8$ & $< 7.3$ &
	  \ldots & \ldots &
	  \ldots & \ldots
\\
${\rm H}\beta\;4861$ & \phs0.000 &
	  $100.0 \pm 3.4$ & $100.0 \pm 3.1$ &
	  $100 \pm 33$ & $100 \pm 27$ &
	  $100 \pm 10$ & $100.0 \pm 9.8$
\\
$[{\rm O\;III}]\;4959$ & $-0.026$ &
	  $97.8 \pm 5.2$ & $83.5 \pm 4.4$ &
	  $170 \pm 33$ & $134 \pm 26$ &
	  $33.8 \pm 8.9$ & $31.6 \pm 8.3$
\\
$[{\rm O\;III}]\;5007$ & $-0.038$ &
	  $296.0 \pm 6.5$ & $246.1 \pm 5.5$ &
	  $506 \pm 43$ & $397 \pm 34$ &
	  $126 \pm 11$ & $116 \pm 10$
\\
$[{\rm O\;I}]\;6300 + [{\rm S\;III}]\;6312$ & $-0.264$ &
	  $157.9 \pm 6.5$ & $79.2 \pm 4.0$ &
	  \ldots & \ldots &
          \ldots & \ldots
\\
$[{\rm O\;I}]\;6363$ & $-0.272$ &
	  $97.3 \pm 9.7$ & $23.1 \pm 2.5$ &
	  \ldots & \ldots &
	  \ldots & \ldots
\\
${\rm H}\alpha\;6563$ & $-0.299$ &
	  $625 \pm 12$ & $297 \pm 11$ &
	  $434 \pm 48$ & $286 \pm 30$ &
	  $405 \pm 17$ & $286 \pm 12$
\\
$[{\rm N\;II}]\;6583$ & $-0.301$ &
	  $97.3 \pm 9.7$ & $45.0 \pm 4.7$ &
	  \ldots & \ldots &
	  $68 \pm 14$ & $47.6 \pm 9.7$
\\
$[{\rm S\;II}]\;6716$ & $-0.319$ &
	  $120.2 \pm 3.2$ & $53.5 \pm 2.4$ &
	  \ldots & \ldots &
	  $117 \pm 12\,^a$ & $80.0 \pm 8.0$
\\
$[{\rm S\;II}]\;6731$ & $-0.321$ &
	  $110.5 \pm 3.2$ & $49.0 \pm 2.3$ &
	  \ldots & \ldots &
	  $117 \pm 12\,^a$ & $80.0 \pm 8.0$
\\
$[{\rm Ar\;III}]\;7136$ & $-0.375$ &
	  $54.1 \pm 5.6$ & $21.3 \pm 2.4$ &
	  \ldots & \ldots &
	  \ldots & \ldots
\\
$[{\rm O\;II}]\;7320, 7330$ & $-0.400$ &
	  $143.3 \pm 7.5$ & $53.2 \pm 3.7$ &
	  \ldots & \ldots &
          \ldots & \ldots
\\[1mm]
\multicolumn{2}{c}{$F(\hbeta)$ (ergs s$^{-1}$ cm$^{-2}$)} & 
	  \multicolumn{2}{c}{$(1.971 \pm 0.068) \times 10^{-15}$} &
	  \multicolumn{2}{c}{$(8.6 \pm 2.8) \times 10^{-17}$} &
	  \multicolumn{2}{c}{$(2.77 \pm 0.28) \times 10^{-16}$}
\\
\multicolumn{2}{c}{EW$_{\rm e}$(\hbeta) (\AA)} &
	  \multicolumn{2}{c}{$16.69 \pm 0.60$} &
	  \multicolumn{2}{c}{$8.6 \pm 2.8$} &
	  \multicolumn{2}{c}{$48.6 \pm 6.5$}
\\
\multicolumn{2}{c}{$c(\hbeta)$} &
	  \multicolumn{2}{c}{$0.965 \pm 0.048$} &
	  \multicolumn{2}{c}{0.38} &
	  \multicolumn{2}{c}{0.46} 
\\
\multicolumn{2}{c}{EW$_{\rm abs}$ (\AA)} &
	  \multicolumn{2}{c}{$1.75 \pm 0.49$} &
	  \multicolumn{2}{c}{2} &
	  \multicolumn{2}{c}{2} 
\\[1mm]
\hline
& & \multicolumn{2}{c}{ESO 274$-$G001 No. 7} & 
\multicolumn{2}{c}{ESO 274$-$G001 No. 8} &
\multicolumn{2}{c}{ESO 324$-$G024 No. 1} \\
\multicolumn{1}{c}{Property} &
\multicolumn{1}{c}{$f(\lambda)$} &
\multicolumn{1}{c}{$F$} & \multicolumn{1}{c}{$I$} &
\multicolumn{1}{c}{$F$} & \multicolumn{1}{c}{$I$} &
\multicolumn{1}{c}{$F$} & \multicolumn{1}{c}{$I$} \\
\hline
$[\rm{O\;II}]\;3727$ & $+0.325$ &
	  $366 \pm 15$ & $513 \pm 21$ &
	  $268 \pm 38$ & $287 \pm 41$ &
	  $357.8 \pm 9.4$ & $429 \pm 11$
\\
${\rm H}\gamma\;4340$ & $+0.158$ &
	  $46.0 \pm 6.7$ & $56.0 \pm 7.8$ &
	  \ldots & \ldots &
	  $40.2 \pm 3.5$ & $48.9 \pm 3.7$
\\
${\rm H}\beta\;4861$ & \phs0.000 &
	  $100.0 \pm 8.3$ & $100.0 \pm 8.2$ &
	  $100 \pm 19$ & $100 \pm 16$ &
	  $100.0 \pm 4.6$ & $100.0 \pm 4.4$
\\
$[{\rm O\;III}]\;4959$ & $-0.026$ &
	  $43.4 \pm 6.2$ & $41.6 \pm 5.9$ &
	  \ldots & \ldots &
	  $63.1 \pm 5.3$ & $58.9 \pm 4.9$
\\
$[{\rm O\;III}]\;5007$ & $-0.038$ &
	  $96.3 \pm 6.5$ & $91.2 \pm 6.2$ &
	  $55 \pm 15$ & $47 \pm 13$ &
	  $162.5 \pm 6.6$ & $150.3 \pm 6.1$
\\
${\rm H}\alpha\;6563$ & $-0.299$ &
	  $401 \pm 10$ & $286.9 \pm 7.4$ &
	  $383 \pm 27$ & $286 \pm 20$ &
	  $370.7 \pm 9.7$ & $287.3 \pm 7.4$
\\
$[{\rm N\;II}]\;6583$ & $-0.301$ &
	  $62.9 \pm 8.4$ & $44.8 \pm 6.0$ &
	  $36 \pm 21$ & $26 \pm 16$ &
	  $19.4 \pm 8.1$ & $14.9 \pm 6.2$
\\
$[{\rm S\;II}]\;6716$ & $-0.319$ &
	  $109 \pm 12$ & $76.2 \pm 8.4$ &
	  \ldots & \ldots &
	  $46.0 \pm 5.8\,^a$ & $34.8 \pm 4.4$
\\
$[{\rm S\;II}]\;6731$ & $-0.321$ &
	  $48 \pm 11$ & $33.6 \pm 7.4$ &
	  \ldots & \ldots &
	  $46.0 \pm 5.8\,^a$ & $34.8 \pm 4.4$
\\[1mm]
\multicolumn{2}{c}{$F(\hbeta)$ (ergs s$^{-1}$ cm$^{-2}$)} & 
	  \multicolumn{2}{c}{$(3.97 \pm 0.33) \times 10^{-16}$} &
	  \multicolumn{2}{c}{$(1.32 \pm 0.25) \times 10^{-16}$} &
	  \multicolumn{2}{c}{$(4.14 \pm 0.19) \times 10^{-16}$}
\\
\multicolumn{2}{c}{EW$_{\rm e}$(\hbeta) (\AA)} &
	  \multicolumn{2}{c}{$148 \pm 32$} &
	  \multicolumn{2}{c}{$14.1 \pm 2.7$} &
	  \multicolumn{2}{c}{$38.5 \pm 2.1$}
\\
\multicolumn{2}{c}{$c(\hbeta)$} &
	  \multicolumn{2}{c}{0.47} &
	  \multicolumn{2}{c}{0.27} &
	  \multicolumn{2}{c}{0.31} 
\\
\multicolumn{2}{c}{EW$_{\rm abs}$ (\AA)} &
	  \multicolumn{2}{c}{2} &
	  \multicolumn{2}{c}{2} &
	  \multicolumn{2}{c}{2} 
\\[1mm]
\hline
\end{tabular}
\end{center}
See Table~\ref{table_data1} for comments.
NOTE: $^a$~[S II] lines unresolved.
\end{minipage}
\end{table*}

\begin{table*}
\begin{minipage}{160mm}
\caption{
Line ratios and properties for \hii\ regions in Cen~A dwarf irregular
galaxies (continued).
}
\label{table_data3}
\scriptsize 
\begin{center}
\renewcommand{\arraystretch}{0.95} 
\vspace*{0mm}
\begin{tabular}{rccccccc}
\hline \hline
& & \multicolumn{2}{c}{ESO 324$-$G024 No. 2} & 
\multicolumn{2}{c}{ESO 324$-$G024 No. 3} &
\multicolumn{2}{c}{ESO 324$-$G024 No. 4} \\
\multicolumn{1}{c}{Property} &
\multicolumn{1}{c}{$f(\lambda)$} &
\multicolumn{1}{c}{$F$} & \multicolumn{1}{c}{$I$} &
\multicolumn{1}{c}{$F$} & \multicolumn{1}{c}{$I$} &
\multicolumn{1}{c}{$F$} & \multicolumn{1}{c}{$I$} \\
\hline
$[\rm{O\;II}]\;3727$ & $+0.325$ &
	  $334 \pm 10$ & $375 \pm 12$ &
	  $338 \pm 11$ & $377 \pm 12$ &
	  $144.7 \pm 3.1$ & $152.6 \pm 7.6$
\\
${\rm H}11\;3772$ & $+0.316$ &
	  \ldots & \ldots &
	  \ldots & \ldots &
	  $3.4 \pm 2.4$ & $7.4 \pm 2.6$
\\
${\rm H}10\;3799$ & $+0.310$ &
	  \ldots & \ldots &
	  \ldots & \ldots &
	  $5.8 \pm 2.7$ & $10.9 \pm 2.9$
\\
${\rm H}9\;3835$ & $+0.302$ &
	  \ldots & \ldots &
	  \ldots & \ldots &
	  $4.7 \pm 1.9$ & $11.0 \pm 2.0$
\\
$[{\rm Ne\;III}]\;3869$ & $+0.294$ &
	  \ldots & \ldots &
	  \ldots & \ldots &
	  $22.3 \pm 2.0$ & $23.3 \pm 2.3$
\\
${\rm H}8 + {\rm He\;I}\;3889$ & $+0.289$ &
	  \ldots & \ldots &
	  \ldots & \ldots &
	  $21.6 \pm 2.0$ & $29.2 \pm 2.4$
\\
${\rm H}\epsilon + {\rm He\;I}\;3970\,^a$ & $+0.269$ &
	  \ldots & \ldots &
	  \ldots & \ldots &
	  $14.4 \pm 1.8$ & $21.6 \pm 2.0$
\\
${\rm H}\delta\;4101$ & $+0.232$ &
	  \ldots & \ldots &
	  \ldots & \ldots &
	  $19.1 \pm 1.8$ & $25.5 \pm 2.0$
\\
${\rm H}\gamma\;4340$ & $+0.158$ &
	  $40.6 \pm 5.8$ & $46.8 \pm 6.0$ &
	  $43.7 \pm 5.3$ & $48.8 \pm 5.5$ &
	  $44.5 \pm 1.6$ & $49.2 \pm 1.9$
\\
$[{\rm O\;III}]\;4363$ & $+0.151$ &
	  \ldots & \ldots &
	  \ldots & \ldots &
	  $5.3 \pm 1.3$ & $5.3 \pm 1.3$
\\
${\rm H}\beta\;4861$ & \phs0.000 &
	  $100.0 \pm 6.0$ & $100.0 \pm 5.8$ &
	  $100.0 \pm 7.6$ & $100.0 \pm 7.4$ &
	  $100.0 \pm 3.1$ & $100.0 \pm 3.0$
\\
$[{\rm O\;III}]\;4959$ & $-0.026$ &
	  $27.6 \pm 5.5$ & $26.1 \pm 5.2$ &
	  $25.8 \pm 6.8$ & $24.9 \pm 6.6$ &
	  $129.6 \pm 8.1$ & $124.8 \pm 7.8$
\\
$[{\rm O\;III}]\;5007$ & $-0.038$ &
	  $80.5 \pm 7.0$ & $75.8 \pm 6.6$ &
	  $98.4 \pm 9.2$ & $94.5 \pm 8.8$ &
	  $373 \pm 10$ & $358 \pm 10$
\\
${\rm He\;I}\;5876$ & $-0.204$ &
	  \ldots & \ldots &
	  \ldots & \ldots &
	  $8.3 \pm 1.0$ & $7.63 \pm 0.94$
\\
${\rm H}\alpha\;6563$ & $-0.299$ &
	  $336 \pm 10$ & $283.3 \pm 8.6$ &
	  $331 \pm 12$ & $287 \pm 10$ &
	  $316.0 \pm 8.3$ & $285 \pm 14$
\\
$[{\rm N\;II}]\;6583$ & $-0.301$ &
	  $7.5 \pm 8.3$ & $6.2 \pm 6.9$ &
	  $17.4 \pm 9.2$ & $15.0 \pm 7.9$ &
	  $3.1 \pm 1.4$ & $2.8 \pm 1.3$
\\
${\rm He\;I}\;6678$ & $-0.314$ &
	  \ldots & \ldots &
	  \ldots & \ldots &
	  $5.8 \pm 1.9$ & $5.2 \pm 1.7$
\\
$[{\rm S\;II}]\;6716, 6731$ & $-0.320$ &
	  $77 \pm 12$ & $63.5 \pm 9.9$ &
	  $39.4 \pm 4.9$ & $33.7 \pm 4.2$ &
	  $15.5 \pm 2.5$ & $13.8 \pm 2.3$
\\
$[{\rm Ar\;III}]\;7136$ & $-0.375$ &
	  \ldots & \ldots &
	  \ldots & \ldots &
	  $8.5 \pm 1.6$ & $7.5 \pm 1.5$
\\[1mm]
\multicolumn{2}{c}{$F(\hbeta)$ (ergs s$^{-1}$ cm$^{-2}$)} & 
	  \multicolumn{2}{c}{$(2.91 \pm 0.17) \times 10^{-16}$} &
	  \multicolumn{2}{c}{$(2.32 \pm 0.18) \times 10^{-16}$} &
	  \multicolumn{2}{c}{$(1.070 \pm 0.033) \times 10^{-15}$}
\\
\multicolumn{2}{c}{EW$_{\rm e}$(\hbeta) (\AA)} &
	  \multicolumn{2}{c}{$46.9 \pm 3.7$} &
	  \multicolumn{2}{c}{$81 \pm 11$} &
	  \multicolumn{2}{c}{$164 \pm 16$} 
\\
\multicolumn{2}{c}{$c(\hbeta)$} &
	  \multicolumn{2}{c}{0.21} &
	  \multicolumn{2}{c}{0.18} &
	  \multicolumn{2}{c}{$0.112 \pm 0.060$} 
\\
\multicolumn{2}{c}{EW$_{\rm abs}$ (\AA)} &
	  \multicolumn{2}{c}{2} &
	  \multicolumn{2}{c}{2} &
	  \multicolumn{2}{c}{$5.1 \pm 2.8$} 
\\[1mm]
\hline
& & \multicolumn{2}{c}{ESO 324$-$G024 No. 5} & 
\multicolumn{2}{c}{ESO 325$-$G011 No. 1} &
\multicolumn{2}{c}{ESO 325$-$G011 No. 2} \\
\multicolumn{1}{c}{Property} &
\multicolumn{1}{c}{$f(\lambda)$} &
\multicolumn{1}{c}{$F$} & \multicolumn{1}{c}{$I$} &
\multicolumn{1}{c}{$F$} & \multicolumn{1}{c}{$I$} &
\multicolumn{1}{c}{$F$} & \multicolumn{1}{c}{$I$} \\
\hline
$[\rm{O\;II}]\;3727$ & $+0.325$ &
	  $164.9 \pm 3.4$ & $174.1 \pm 8.8$ &
	  $338 \pm 32$ & $358 \pm 34$ &
	  $246.1 \pm 8.7$ & $234.4 \pm 8.3$
\\
${\rm H}11\;3772$ & $+0.316$ &
	  $4.2 \pm 2.6$ & $8.4 \pm 2.8$ &
	  \ldots & \ldots &
	  \ldots & \ldots
\\
${\rm H}10\;3799$ & $+0.310$ &
	  $5.9 \pm 2.9$ & $11.3 \pm 3.1$ &
	  \ldots & \ldots &
	  \ldots & \ldots
\\
${\rm H}9\;3835$ & $+0.302$ &
	  $4.3 \pm 1.8$ & $10.8 \pm 1.9$ &
	  \ldots & \ldots &
	  \ldots & \ldots
\\
$[{\rm Ne\;III}]\;3869$ & $+0.294$ &
	  $23.1 \pm 1.9$ & $24.2 \pm 2.2$ &
	  \ldots & \ldots &
	  \ldots & \ldots
\\
${\rm H}8 + {\rm He\;I}\;3889$ & $+0.289$ &
	  $21.2 \pm 1.9$ & $29.2 \pm 2.3$ &
	  \ldots & \ldots &
	  \ldots & \ldots
\\
${\rm H}\epsilon + {\rm He\;I}\;3970\,^a$ & $+0.269$ &
	  $15.0 \pm 1.9$ & $22.3 \pm 2.2$ &
	  \ldots & \ldots &
	  \ldots & \ldots
\\
${\rm H}\delta\;4101$ & $+0.232$ &
	  $19.3 \pm 1.8$ & $25.7 \pm 2.0$ &
	  \ldots & \ldots &
	  \ldots & \ldots
\\
${\rm H}\gamma\;4340$ & $+0.158$ &
	  $44.2 \pm 1.9$ & $49.0 \pm 2.2$ &
	  \ldots & \ldots &
	  $40.3 \pm 5.4$ & $43.7 \pm 5.1$
\\
$[{\rm O\;III}]\;4363$ & $+0.151$ &
	  $4.4 \pm 1.5$ & $4.4 \pm 1.5$ &
	  \ldots & \ldots &
	  \ldots & \ldots
\\
${\rm H}\beta\;4861$ & \phs0.000 &
	  $100.0 \pm 3.1$ & $100.0 \pm 3.0$ &
	  $100 \pm 16$ & $100 \pm 16$ &
	  $100.0 \pm 6.1$ & $100.0 \pm 5.8$
\\
$[{\rm O\;III}]\;4959$ & $-0.026$ &
	  $128.7 \pm 8.0$ & $123.9 \pm 7.7$ &
	  $19 \pm 12$ & $18 \pm 12$ &
	  $92.9 \pm 6.7$ & $87.8 \pm 6.3$
\\
$[{\rm O\;III}]\;5007$ & $-0.038$ &
	  $373 \pm 10$ & $357.7 \pm 9.9$ &
	  $43 \pm 13$ & $41 \pm 12$  &
	  $261.5 \pm 8.1$ & $247.0 \pm 7.7$
\\
${\rm He\;I}\;5876$ & $-0.204$ &
	  $7.8 \pm 1.3$ & $7.2 \pm 1.2$ &
	  \ldots & \ldots &
	  \ldots & \ldots
\\
${\rm H}\alpha\;6563$ & $-0.299$ &
	  $317.1 \pm 8.0$ & $285 \pm 14$ &
	  $323 \pm 17$ & $286 \pm 15$ &
	  $294.3 \pm 8.9$ & $281.6 \pm 8.4$
\\
$[{\rm N\;II}]\;6583$ & $-0.301$ &
	  $4.3 \pm 1.5$ & $3.9 \pm 1.4$ &
	  $14 \pm 14$ & $12 \pm 12$ &
	  $7.0 \pm 7.0$ & $6.6 \pm 6.6$
\\
${\rm He\;I}\;6678$ & $-0.314$ &
	  $9.7 \pm 2.5$ & $8.7 \pm 2.3$ &
	  \ldots & \ldots &
	  \ldots & \ldots
\\
$[{\rm S\;II}]\;6716, 6731$ & $-0.320$ &
	  $15.5 \pm 2.8$ & $13.8 \pm 2.6$ &
	  \ldots & \ldots &
	  $34.2 \pm 6.8$ & $32.1 \pm 6.4$
\\
$[{\rm Ar\;III}]\;7136$ & $-0.375$ &
	  $9.3 \pm 1.6$ & $8.2 \pm 1.5$ &
	  \ldots & \ldots &
	  \ldots & \ldots
\\[1mm]
\multicolumn{2}{c}{$F(\hbeta)$ (ergs s$^{-1}$ cm$^{-2}$)} & 
	  \multicolumn{2}{c}{$(1.270 \pm 0.039) \times 10^{-15}$} &
	  \multicolumn{2}{c}{$(1.34 \pm 0.22) \times 10^{-16}$} &
	  \multicolumn{2}{c}{$(3.57 \pm 0.22) \times 10^{-16}$}
\\
\multicolumn{2}{c}{EW$_{\rm e}$(\hbeta) (\AA)} &
	  \multicolumn{2}{c}{$173 \pm 18$} &
	  \multicolumn{2}{c}{$51 \pm 11$} &
	  \multicolumn{2}{c}{$34.6 \pm 2.5$} 
\\
\multicolumn{2}{c}{$c(\hbeta)$} &
	  \multicolumn{2}{c}{$0.113 \pm 0.061$} &
	  \multicolumn{2}{c}{0.13} &
	  \multicolumn{2}{c}{0} 
\\
\multicolumn{2}{c}{EW$_{\rm abs}$ (\AA)} &
	  \multicolumn{2}{c}{$5.4 \pm 3.1$} &
	  \multicolumn{2}{c}{2} &
	  \multicolumn{2}{c}{2} 
\\[1mm]
\hline
\end{tabular}
\end{center}
See Table~\ref{table_data1} for comments.
NOTE: $^a$~Blended with [Ne III] $\lambda$3967.
\end{minipage}
\end{table*}

\begin{table*}
\begin{minipage}{160mm}
\caption{
Line ratios and properties for \hii\ regions in Cen~A dwarf irregular
galaxies (continued).
}
\label{table_data4}
\tiny 
\begin{center}
\renewcommand{\arraystretch}{0.95} 
\vspace*{0mm}
\begin{tabular}{rccccccc}
\hline \hline
& & \multicolumn{2}{c}{ESO 325$-$G011 No. 3} & 
\multicolumn{2}{c}{ESO 325$-$G011 No. 4} &
\multicolumn{2}{c}{ESO 325$-$G011 No. 5$\,^a$} \\
\multicolumn{1}{c}{Property} &
\multicolumn{1}{c}{$f(\lambda)$} &
\multicolumn{1}{c}{$F$} & \multicolumn{1}{c}{$I$} &
\multicolumn{1}{c}{$F$} & \multicolumn{1}{c}{$I$} &
\multicolumn{1}{c}{$F$} & \multicolumn{1}{c}{$I$} \\
\hline
$[\rm{O\;II}]\;3727$ & $+0.325$ &
	  $307 \pm 29$ & $311 \pm 30$ &
	  $200.7 \pm 5.5$ & $217.6 \pm 6.0$ &
	  $166 \pm 13$ & $309 \pm 24$
\\
${\rm H}\delta\;4101$ & $+0.232$ &
	  \ldots & \ldots &
	  $21.0 \pm 3.4$ & $24.5 \pm 3.6$ &
	  \ldots & \ldots
\\
${\rm H}\gamma\;4340$ & $+0.158$ &
	  \ldots & \ldots & 
	  $38.7 \pm 2.9$ & $41.9 \pm 3.0$ &
	  \ldots & \ldots
\\
$[{\rm O\;III}]\;4363$ & $+0.151$ &
	  \ldots & \ldots &
	  $< 8.4$ & $< 8.6$ &
	  \ldots & \ldots
\\
${\rm H}\beta\;4861$ & \phs0.000 &
	  $100 \pm 13$ & $100 \pm 12$ &
	  $100.0 \pm 3.4$ & $100.0 \pm 3.3$ &
	  $100.0 \pm 7.5$ & $100.0 \pm 4.2$
\\
$[{\rm O\;III}]\;4959$ & $-0.026$ &
	  $39 \pm 11$ & $39 \pm 11$ &
	  $80.7 \pm 5.9$ & $78.8 \pm 5.8$ &
	  $36.6 \pm 7.5$ & $18.8 \pm 3.9$
\\
$[{\rm O\;III}]\;5007$ & $-0.038$ &
	  $118 \pm 13$ & $116 \pm 13$ &
	  $244.7 \pm 7.4$ & $238.0 \pm 7.2$ &
	  $66.7 \pm 9.0$ & $32.8 \pm 4.4$
\\
${\rm He\;I}\;5876$ & $-0.204$ &
	  \ldots & \ldots &
	  $12.2 \pm 2.7$ & $11.3 \pm 2.5$ &
	  \ldots & \ldots
\\
${\rm H}\alpha\;6563$ & $-0.299$ &
	  $296 \pm 23$ & $286 \pm 22$ &
	  $318.2 \pm 7.4$ & $287.0 \pm 6.7$ &
	  $1440 \pm 25$ & $286.8 \pm 4.8$
\\
$[{\rm N\;II}]\;6583$ & $-0.301$ &
	  \ldots & \ldots &
	  $2.8 \pm 5.9$ & $2.5 \pm 5.3$ &
	  $556 \pm 23$ & $104.0 \pm 4.2$
\\
$[{\rm S\;II}]\;6716$ & $-0.319$ &
	  \ldots & \ldots &
	  $16.9 \pm 3.5\,^b$ & $15.1 \pm 3.1$ &
	  $119 \pm 13$ & $20.9 \pm 2.3$
\\
$[{\rm S\;II}]\;6731$ & $-0.321$ &
	  \ldots & \ldots &
	  $16.9 \pm 3.5\,^b$ & $15.1 \pm 3.1$ &
	  $133 \pm 13$ & $23.1 \pm 2.3$
\\[1mm]
\multicolumn{2}{c}{$F(\hbeta)$ (ergs s$^{-1}$ cm$^{-2}$)} & 
	  \multicolumn{2}{c}{$(1.67 \pm 0.21) \times 10^{-16}$} &
	  \multicolumn{2}{c}{$(8.75 \pm 0.30) \times 10^{-16}$} &
	  \multicolumn{2}{c}{$(4.54 \pm 0.34) \times 10^{-16}$}
\\
\multicolumn{2}{c}{EW$_{\rm e}$(\hbeta) (\AA)} &
	  \multicolumn{2}{c}{$121 \pm 39$} &
	  \multicolumn{2}{c}{$121 \pm 10$} &
	  \multicolumn{2}{c}{$2.64 \pm 0.20$} 
\\
\multicolumn{2}{c}{$c(\hbeta)$} &
	  \multicolumn{2}{c}{0.04} &
	  \multicolumn{2}{c}{0.13} &
	  \multicolumn{2}{c}{1.6} 
\\
\multicolumn{2}{c}{EW$_{\rm abs}$ (\AA)} &
	  \multicolumn{2}{c}{2} &
	  \multicolumn{2}{c}{2} &
	  \multicolumn{2}{c}{2} 
\\[1mm]
\hline
& & \multicolumn{2}{c}{ESO 381$-$G020 No. 1} & 
\multicolumn{2}{c}{ESO 381$-$G020 No. 2} &
\multicolumn{2}{c}{ESO 381$-$G020 No. 3} \\
\multicolumn{1}{c}{Property} &
\multicolumn{1}{c}{$f(\lambda)$} &
\multicolumn{1}{c}{$F$} & \multicolumn{1}{c}{$I$} &
\multicolumn{1}{c}{$F$} & \multicolumn{1}{c}{$I$} &
\multicolumn{1}{c}{$F$} & \multicolumn{1}{c}{$I$} \\
\hline
$[\rm{O\;II}]\;3727$ & $+0.325$ &
	  $176.1 \pm 8.3$ & $187.8 \pm 8.9$ &
	  $328 \pm 64$ & $475 \pm 92$ &
	  $202.3 \pm 5.5$ & $202.3 \pm 5.5$
\\
$[{\rm Ne\;III}]\;3869$ & $+0.294$ &
	  $28.4 \pm 5.6$ & $29.9 \pm 5.9$ &
	  \ldots & \ldots &
	  $23.7 \pm 5.3$ & $23.7 \pm 5.3$
\\
${\rm H}\epsilon + {\rm He\;I}\;3970\,^c$ & $+0.269$ &
	  \ldots & \ldots &
	  \ldots & \ldots  &
	  $18.9 \pm 3.3$ & $18.9 \pm 3.3$
\\
${\rm H}\delta\;4101$ & $+0.232$ &
	  \ldots & \ldots &
	  \ldots & \ldots &
	  $15.3 \pm 3.2$ & $15.3 \pm 3.2$
\\
${\rm H}\gamma\;4340$ & $+0.158$ &
	  $44.6 \pm 4.5$ & $52.0 \pm 4.5$ &
	  \ldots & \ldots &
	  $38.4 \pm 3.4$ & $38.4 \pm 3.4$
\\
$[{\rm O\;III}]\;4363$ & $+0.151$ &
	  $< 16.4$ & $< 16.3$ &
	  \ldots & \ldots &
	  $< 8.0$ & $< 8.0$
\\
${\rm H}\beta\;4861$ & \phs0.000 &
	  $100.0 \pm 4.1$ & $100.0 \pm 3.8$ &
	  $100 \pm 38$ & $100 \pm 31$ &
	  $100.0 \pm 4.5$ & $100.0 \pm 4.5$
\\
$[{\rm O\;III}]\;4959$ & $-0.026$ &
	  $131.2 \pm 8.7$ & $122.0 \pm 8.1$ &
	  $152 \pm 38$ & $116 \pm 27$ &
	  $90.7 \pm 6.5$ & $90.7 \pm 6.5$
\\
$[{\rm O\;III}]\;5007$ & $-0.038$ &
	  $384 \pm 11$ & $356 \pm 10$ &
	  $335 \pm 52$ & $251 \pm 39$ &
	  $258.7 \pm 8.1$ & $258.7 \pm 8.1$
\\
${\rm He\;I}\;5876$ & $-0.204$ &
	  \ldots & \ldots &
	  \ldots & \ldots &
	  $11.2 \pm 2.2$ & $11.2 \pm 2.2$
\\
${\rm H}\alpha\;6563$ & $-0.299$ &
	  $340.7 \pm 8.7$ & $287.3 \pm 7.3$ &
	  $604 \pm 44$ & $286 \pm 20$ &
	  $271.4 \pm 6.7$ & $271.4 \pm 6.7$
\\
$[{\rm N\;II}]\;6583$ & $-0.301$ &
	  $3.6 \pm 6.9$ & $3.0 \pm 5.8$ &
	  $38 \pm 35$ & $18 \pm 16$ &
	  $0.4 \pm 5.4$ & $0.4 \pm 5.4$
\\
$[{\rm S\;II}]\;6716, 6731$ & $-0.320$ &
	  $48.6 \pm 8.4$ & $40.3 \pm 7.0$ &
	  \ldots & \ldots &
	  $24.3 \pm 3.3$ & $24.3 \pm 3.3$
\\[1mm]
\multicolumn{2}{c}{$F(\hbeta)$ (ergs s$^{-1}$ cm$^{-2}$)} & 
	  \multicolumn{2}{c}{$(8.28 \pm 0.34) \times 10^{-16}$} &
	  \multicolumn{2}{c}{$(1.00 \pm 0.38) \times 10^{-16}$} &
	  \multicolumn{2}{c}{$(1.567 \pm 0.071) \times 10^{-15}$}
\\
\multicolumn{2}{c}{EW$_{\rm e}$(\hbeta) (\AA)} &
	  \multicolumn{2}{c}{$30.8 \pm 1.5$} &
	  \multicolumn{2}{c}{$8.1 \pm 3.1$} &
	  \multicolumn{2}{c}{$53.4 \pm 3.3$} 
\\
\multicolumn{2}{c}{$c(\hbeta)$} &
	  \multicolumn{2}{c}{0.17} &
	  \multicolumn{2}{c}{0.79} &
	  \multicolumn{2}{c}{0} 
\\
\multicolumn{2}{c}{EW$_{\rm abs}$ (\AA)} &
	  \multicolumn{2}{c}{2} &
	  \multicolumn{2}{c}{2} &
	  \multicolumn{2}{c}{0} 
\\[1mm]
\hline
\end{tabular}
\end{center}
See Table~\ref{table_data1} for comments.
NOTES:
$^a$~Background galaxy with $v_{\odot} \approx 26200$ km~s$^{-1}$.
$^b$~[S II] unresolved.
$^c$~Blended with [Ne III] $\lambda$3967.
\end{minipage}
\end{table*}

\begin{table*}
\begin{minipage}{160mm}
\caption{
Line ratios and properties for \hii\ regions in Cen~A dwarf irregular
galaxies (continued).
}
\label{table_data5}
\scriptsize 
\begin{center}
\renewcommand{\arraystretch}{0.95} 
\vspace*{0mm}
\begin{tabular}{rccccccc}
\hline \hline
& & \multicolumn{2}{c}{ESO 381$-$G020 No. 4} & 
\multicolumn{2}{c}{ESO 381$-$G020 No. 5} &
\multicolumn{2}{c}{ESO 381$-$G020 No. 6} \\
\multicolumn{1}{c}{Property} &
\multicolumn{1}{c}{$f(\lambda)$} &
\multicolumn{1}{c}{$F$} & \multicolumn{1}{c}{$I$} &
\multicolumn{1}{c}{$F$} & \multicolumn{1}{c}{$I$} &
\multicolumn{1}{c}{$F$} & \multicolumn{1}{c}{$I$} \\
\hline
$[\rm{O\;II}]\;3727$ & $+0.325$ &
	  $320 \pm 30$ & $422 \pm 39$ &
	  $74.9 \pm 7.9$ & $74.9 \pm 7.9$ &
	  $108 \pm 18$ & $108 \pm 18$
\\
${\rm H}\delta\;4101$ & $+0.232$ &
	  \ldots & \ldots &
	  $12.7 \pm 3.7$ & $12.7 \pm 3.7$ &
	  \ldots & \ldots
\\
${\rm H}\gamma\;4340$ & $+0.158$ &
	  \ldots & \ldots &
	  $43.8 \pm 4.3$ & $43.8 \pm 4.3$ &
	  $33.4 \pm 8.0$ & $33.4 \pm 8.0$
\\
${\rm H}\beta\;4861$ & \phs0.000 &
	  $100 \pm 14$ & $100 \pm 13$ &
	  $100.0 \pm 3.9$ & $100.0 \pm 3.9$ &
	  $100 \pm 11$ & $100 \pm 11$
\\
$[{\rm O\;III}]\;4959$ & $-0.026$ &
	  $30 \pm 11$ & $26.8 \pm 9.8$ &
	  $70.4 \pm 4.8$ & $70.4 \pm 4.8$ &
	  $51.0 \pm 9.6$ & $51.0 \pm 9.6$
\\
$[{\rm O\;III}]\;5007$ & $-0.038$ &
	  $52 \pm 11$ & $45.5 \pm 9.6$ &
	  $199.5 \pm 6.1$ & $199.5 \pm 6.1$ &
	  $137 \pm 12$ & $137 \pm 12$
\\
${\rm H}\alpha\;6563$ & $-0.299$ &
	  $438 \pm 18$ & $285 \pm 11$ &
	  $270.6 \pm 5.9$ & $270.6 \pm 5.9$ &
	  $277 \pm 13$ & $277 \pm 13$
\\
$[{\rm N\;II}]\;6583$ & $-0.301$ &
	  \ldots & \ldots &
	  $5.8 \pm 4.8$ & $5.8 \pm 4.8$ &
	  \ldots & \ldots
\\[1mm]
\multicolumn{2}{c}{$F(\hbeta)$ (ergs s$^{-1}$ cm$^{-2}$)} & 
	  \multicolumn{2}{c}{$(2.27 \pm 0.32) \times 10^{-16}$} &
	  \multicolumn{2}{c}{$(8.62 \pm 0.34) \times 10^{-16}$} &
	  \multicolumn{2}{c}{$(2.96 \pm 0.32) \times 10^{-16}$}
\\
\multicolumn{2}{c}{EW$_{\rm e}$(\hbeta) (\AA)} &
	  \multicolumn{2}{c}{$19.8 \pm 2.9$} &
	  \multicolumn{2}{c}{$330 \pm 76$} &
	  \multicolumn{2}{c}{$289 \pm 165$}
\\
\multicolumn{2}{c}{$c(\hbeta)$} &
	  \multicolumn{2}{c}{0.5} &
	  \multicolumn{2}{c}{0} &
	  \multicolumn{2}{c}{0} 
\\
\multicolumn{2}{c}{EW$_{\rm abs}$ (\AA)} &
	  \multicolumn{2}{c}{2} &
	  \multicolumn{2}{c}{0} &
	  \multicolumn{2}{c}{0} 
\\[1mm]
\hline
& & \multicolumn{2}{c}{IC 4247 No. 1} & 
\multicolumn{2}{c}{IC 4247 No. 2} \\
\multicolumn{1}{c}{Property} &
\multicolumn{1}{c}{$f(\lambda)$} &
\multicolumn{1}{c}{$F$} & \multicolumn{1}{c}{$I$} &
\multicolumn{1}{c}{$F$} & \multicolumn{1}{c}{$I$} \\
\hline
$[\rm{O\;II}]\;3727$ & $+0.325$ &
	  $404.1 \pm 8.2$ & $410.2 \pm 8.3$ &
	  $577 \pm 12$ & $523 \pm 11$ 
\\
$[{\rm Ne\;III}]\;3869$ & $+0.294$ &
	  $50.9 \pm 7.2$ & $51.6 \pm 7.3$ &
	  \ldots & \ldots 
\\
${\rm H}\gamma\;4340$ & $+0.158$ &
	  $33.4 \pm 3.3$ & $33.6 \pm 3.3$ &
	  \ldots & \ldots 
\\
${\rm H}\beta\;4861$ & \phs0.000 &
	  $100.0 \pm 5.1$ & $100.0 \pm 5.1$ &
	  $100.0 \pm 7.0$ & $100.0 \pm 6.2$ 
\\
$[{\rm O\;III}]\;4959$ & $-0.026$ &
	  $95.3 \pm 7.1$ & $95.2 \pm 7.1$ &
	  $26.2 \pm 4.9$ & $23.2 \pm 4.3$ 
\\
$[{\rm O\;III}]\;5007$ & $-0.038$ &
	  $292.3 \pm 9.0$ & $291.8 \pm 9.0$ &
	  $83.4 \pm 6.2$ & $73.6 \pm 5.5$ 
\\
${\rm H}\alpha\;6563$ & $-0.299$ &
	  $291.2 \pm 7.4$ & $287.2 \pm 7.3$ &
	  $322 \pm 12$ & $286 \pm 10$ 
\\
$[{\rm N\;II}]\;6583$ & $-0.301$ &
	  $9.1 \pm 5.9$ & $9.0 \pm 5.8$ &
	  $11.9 \pm 9.3$ & $10.3 \pm 8.1$ 
\\
$[{\rm S\;II}]\;6716, 6731$ & $-0.320$ &
	  $41.7 \pm 6.1$ & $41.1 \pm 6.0$ &
	  $28.3 \pm 4.4$ & $24.5 \pm 3.8$ 
\\[1mm]
\multicolumn{2}{c}{$F(\hbeta)$ (ergs s$^{-1}$ cm$^{-2}$)} & 
	  \multicolumn{2}{c}{$(7.33 \pm 0.38) \times 10^{-16}$} &
	  \multicolumn{2}{c}{$(4.85 \pm 0.34) \times 10^{-16}$} 
\\
\multicolumn{2}{c}{EW$_{\rm e}$(\hbeta) (\AA)} &
	  \multicolumn{2}{c}{$15.40 \pm 0.83$} &
	  \multicolumn{2}{c}{$10.75 \pm 0.77$} 
\\
\multicolumn{2}{c}{$c(\hbeta)$} &
	  \multicolumn{2}{c}{0.02} &
	  \multicolumn{2}{c}{0.03} 
\\
\multicolumn{2}{c}{EW$_{\rm abs}$ (\AA)} &
	  \multicolumn{2}{c}{0} &
	  \multicolumn{2}{c}{1.4} 
\\
\hline
\end{tabular}
\end{center}
%
%
See Table~\ref{table_data1} for comments.
\end{minipage}
\end{table*}

\begin{table*}
\begin{minipage}{160mm}
\caption{
Line ratios and properties for \hii\ regions in the Cen~A dwarf 
galaxy IC 4316.
}
\label{table_data6}
\scriptsize 
\begin{center}
\renewcommand{\arraystretch}{0.95} 
\vspace*{0mm}
\begin{tabular}{rccccccc}
\hline \hline
& & \multicolumn{2}{c}{IC 4316 No. 1} &
\multicolumn{2}{c}{IC 4316 No. 2} & 
\multicolumn{2}{c}{IC 4316 No. 3} \\
\multicolumn{1}{c}{Property} &
\multicolumn{1}{c}{$f(\lambda)$} &
\multicolumn{1}{c}{$F$} & \multicolumn{1}{c}{$I$} &
\multicolumn{1}{c}{$F$} & \multicolumn{1}{c}{$I$} \\
\hline
$[\rm{O\;II}]\;3727$ & $+0.325$ &
	  $415.4 \pm 9.8$ & $541 \pm 13$ &
	  $305.1 \pm 3.8$ & $305.1 \pm 3.8$ &
	  $484 \pm 13$ & $510 \pm 13$
\\
${\rm H}9\;3835$ & $+0.302$ &
          \ldots & \ldots &
	  $2.2 \pm 1.1$ & $2.2 \pm 1.1$ &
	  \ldots & \ldots
\\
$[{\rm Ne\;III}]\;3869$ & $+0.294$ &
          \ldots & \ldots &
	  $21.3 \pm 1.2$ & $21.3 \pm 1.2$ &
	  $48 \pm 10$ & $49 \pm 11$
\\
${\rm H}8 + {\rm He\;I}\;3889$ & $+0.289$ &
          \ldots & \ldots &
	  $18.0 \pm 1.2$ & $18.0 \pm 1.2$ &
	  \ldots & \ldots
\\
${\rm H}\epsilon + {\rm He\;I}\;3970\,^a$ & $+0.269$ &
          \ldots & \ldots &
	  $19.2 \pm 1.2$ & $19.2 \pm 1.2$ &
	  \ldots & \ldots
\\
${\rm H}\delta\;4101$ & $+0.232$ &
          \ldots & \ldots &
	  $22.3 \pm 1.2$ & $22.3 \pm 1.2$ &
	  \ldots & \ldots
\\
${\rm H}\gamma\;4340$ & $+0.158$ &
	  $37.1 \pm 3.3$ & $50.6 \pm 3.6$ &
	  $44.11 \pm 0.97$ & $44.11 \pm 0.97$ &
	  \ldots & \ldots
\\
$[{\rm O\;III}]\;4363$ & $+0.151$ &
          \ldots & \ldots &
	  $< 2.3$ & $< 2.3$ &
	  \ldots & \ldots
\\
${\rm He\;I}\;4471$ & $+0.116$ &
          \ldots & \ldots &
	  $3.06 \pm 0.62$ & $3.06 \pm 0.62$ &
	  \ldots & \ldots
\\
${\rm H}\beta\;4861$ & \phs0.000 &
	  $100.0 \pm 5.2$ & $100.0 \pm 4.8$ &
	  $100.0 \pm 3.9$ & $100.0 \pm 3.9$ &
	  $100.0 \pm 6.8$ & $100.0 \pm 6.0$
\\
$[{\rm O\;III}]\;4959$ & $-0.026$ &
	  $30.1 \pm 4.0$ & $27.0 \pm 3.6$ &
	  $90.2 \pm 6.5$ & $90.2 \pm 6.5$ &
	  $92.3 \pm 6.7$ & $80.8 \pm 5.9$
\\
$[{\rm O\;III}]\;5007$ & $-0.038$ &
          $59.6 \pm 4.8$ & $52.9 \pm 4.3$ &
	  $245.6 \pm 7.8$ & $245.6 \pm 7.8$ &
	  $294.2 \pm 8.5$ & $255.9 \pm 7.4$
\\
${\rm He\;I}\;5876$ & $-0.204$ &
          \ldots & \ldots &
	  $9.46 \pm 0.98$ & $9.46 \pm 0.98$ &
	  \ldots & \ldots
\\
$[{\rm O\;I}]\;6300 + [{\rm S\;III}]\;6312$ & $-0.265$ &
          \ldots & \ldots &
	  $7.0 \pm 1.3$ & $7.0 \pm 1.3$ &
          \ldots & \ldots
\\
${\rm H}\alpha\;6563$ & $-0.299$ &
          $418 \pm 12$ & $285.2 \pm 8.3$ &
	  $258.4 \pm 6.3$ & $258.4 \pm 6.3$ &
	  $368.8 \pm 9.8$ & $286.1 \pm 7.4$
\\
$[{\rm N\;II}]\;6583$ & $-0.301$ &
	  $45 \pm 10$ & $30.4 \pm 6.7$ &
	  $26.9 \pm 4.9$ & $26.9 \pm 4.9$ &
	  $46.7 \pm 7.8$ & $35.3 \pm 5.9$
\\
${\rm He\;I}\;6678$ & $-0.314$ &
          \ldots & \ldots &
	  $4.0 \pm 1.0$ & $4.0 \pm 1.0$ &
	  \ldots & \ldots
\\
$[{\rm S\;II}]\;6716$ & $-0.319$ &
	  $175 \pm 12\,^b$ & $114.9 \pm 7.7$ &
	  $26.2 \pm 1.1$ & $26.2 \pm 1.1$ &
	  $148 \pm 13\,^b$ & $110.7 \pm 9.7$
\\
$[{\rm S\;II}]\;6731$ & $-0.321$ &
	  $175 \pm 12\,^b$ & $114.9 \pm 7.7$ &
	  $22.0 \pm 1.1$ & $22.0 \pm 1.1$ &
	  $148 \pm 13\,^b$ & $110.7 \pm 9.7$
\\
$[{\rm Ar\;III}]\;7136$ & $-0.375$ &
          \ldots & \ldots &
	  $10.08 \pm 0.97$ & $10.08 \pm 0.97$ &
	  \ldots & \ldots
\\[1mm]
\multicolumn{2}{c}{$F(\hbeta)$ (ergs s$^{-1}$ cm$^{-2}$)} & 
	  \multicolumn{2}{c}{$(2.44 \pm 0.13) \times 10^{-16}$} &
	  \multicolumn{2}{c}{$(1.330 \pm 0.052) \times 10^{-15}$} &
	  \multicolumn{2}{c}{$(1.291 \pm 0.088) \times 10^{-16}$}
\\
\multicolumn{2}{c}{EW$_{\rm e}$(\hbeta) (\AA)} &
	  \multicolumn{2}{c}{$24.1 \pm 1.4$} &
	  \multicolumn{2}{c}{$83.1 \pm 6.6$} &
	  \multicolumn{2}{c}{$15.8 \pm 1.1$} 
\\
\multicolumn{2}{c}{$c(\hbeta)$} &
	  \multicolumn{2}{c}{0.46} &
	  \multicolumn{2}{c}{0} &
	  \multicolumn{2}{c}{0.23} 
\\
\multicolumn{2}{c}{EW$_{\rm abs}$ (\AA)} &
	  \multicolumn{2}{c}{2} &
	  \multicolumn{2}{c}{0} &
	  \multicolumn{2}{c}{2} 
\\[1mm]
\hline
& & \multicolumn{2}{c}{IC 4316 No. 4} & 
\multicolumn{2}{c}{IC 4316 No. 5} &
\multicolumn{2}{c}{IC 4316 No. 6} \\
\multicolumn{1}{c}{Property} &
\multicolumn{1}{c}{$f(\lambda)$} &
\multicolumn{1}{c}{$F$} & \multicolumn{1}{c}{$I$} &
\multicolumn{1}{c}{$F$} & \multicolumn{1}{c}{$I$} &
\multicolumn{1}{c}{$F$} & \multicolumn{1}{c}{$I$} \\
\hline
$[\rm{O\;II}]\;3727$ & $+0.325$ &
	  $338.3 \pm 8.3$ & $375.8 \pm 9.2$ &
	  $352.6 \pm 5.9$ & $395.1 \pm 6.6$ &
          $328.6 \pm 5.2$ & $366.2 \pm 5.8$
\\
$[{\rm Ne\;III}]\;3869$ & $+0.294$ &
	  \ldots & \ldots &
	  $12.5 \pm 2.6$ & $13.8 \pm 2.9$ &
	  $9.6 \pm 1.9$ & $10.6 \pm 2.1$
\\
${\rm H}8 + {\rm He\;I}\;3889$ & $+0.289$ &
          \ldots & \ldots &
	  \ldots & \ldots &
	  $7.8 \pm 2.4$ & $14.7 \pm 2.6$
\\
${\rm H}\epsilon + {\rm He\;I}\;3970\,^a$ & $+0.269$ &
          \ldots & \ldots &
          \ldots & \ldots &
	  $8.9 \pm 2.1$ & $15.4 \pm 2.3$
\\
${\rm H}\delta\;4101$ & $+0.232$ &
	  \ldots & \ldots &
	  $11.3 \pm 1.6$ & $20.1 \pm 1.7$ &
	  $13.9 \pm 1.8$ & $20.4 \pm 1.9$
\\
${\rm H}\gamma\;4340$ & $+0.158$ &
	  $31.8 \pm 3.1$ & $40.5 \pm 3.2$ &
	  $37.8 \pm 2.1$ & $45.3 \pm 2.1$ &
	  $39.5 \pm 2.2$ & $45.3 \pm 2.3$
\\
${\rm H}\beta\;4861$ & \phs0.000 &
	  $100.0 \pm 4.3$ & $100.0 \pm 4.0$ &
	  $100.0 \pm 3.1$ & $100.0 \pm 2.9$ &
	  $100.0 \pm 3.8$ & $100.0 \pm 3.7$
\\
$[{\rm O\;III}]\;4959$ & $-0.026$ &
	  $46.8 \pm 4.4$ & $43.2 \pm 4.1$ &
	  $54.6 \pm 3.7$ & $51.2 \pm 3.5$ &
	  $56.8 \pm 3.8$ & $54.3 \pm 3.6$
\\
$[{\rm O\;III}]\;5007$ & $-0.038$ &
	  $138.6 \pm 5.5$ & $127.0 \pm 5.0$ &
	  $150.2 \pm 4.6$ & $140.0 \pm 4.3$ &
	  $161.1 \pm 4.7$ & $153.2 \pm 4.5$
\\
${\rm He\;I}\;5876$ & $-0.204$ &
	  \ldots & \ldots &
	  $7.9 \pm 1.7$ & $6.8 \pm 1.5$ &
	  $7.7 \pm 2.2$ & $6.8 \pm 1.9$
\\
${\rm H}\alpha\;6563$ & $-0.299$ &
	  $353.5 \pm 8.6$ & $286.1 \pm 6.7$ &
	  $344.3 \pm 7.0$ & $284.1 \pm 5.7$ &
	  $334.4 \pm 6.2$ & $285.6 \pm 5.3$
\\
$[{\rm N\;II}]\;6583$ & $-0.301$ &
	  $28.3 \pm 6.9$ & $22.5 \pm 5.5$ &
	  $30.7 \pm 5.6$ & $25.0 \pm 4.6$ &
	  $29.9 \pm 4.9$ & $25.3 \pm 4.2$
\\
$[{\rm S\;II}]\;6716, 6731$ & $-0.320$ &
	  $86.6 \pm 5.6$ & $68.3 \pm 4.4$ &
	  $80.9 \pm 3.6$ & $65.3 \pm 2.9$ &
	  $70.5 \pm 4.9$ & $59.3 \pm 4.1$
\\[1mm]
\multicolumn{2}{c}{$F(\hbeta)$ (ergs s$^{-1}$ cm$^{-2}$)} & 
	  \multicolumn{2}{c}{$(3.09 \pm 0.13) \times 10^{-16}$} &
	  \multicolumn{2}{c}{$(9.26 \pm 0.29) \times 10^{-16}$} &
	  \multicolumn{2}{c}{$(4.82 \pm 0.18) \times 10^{-16}$}
\\
\multicolumn{2}{c}{EW$_{\rm e}$(\hbeta) (\AA)} &
	  \multicolumn{2}{c}{$28.8 \pm 1.4$} &
	  \multicolumn{2}{c}{$38.2 \pm 1.5$} &
	  \multicolumn{2}{c}{$58.1 \pm 3.4$}
\\
\multicolumn{2}{c}{$c(\hbeta)$} &
	  \multicolumn{2}{c}{0.23} &
	  \multicolumn{2}{c}{0.22} & 
	  \multicolumn{2}{c}{0.19}
\\
\multicolumn{2}{c}{EW$_{\rm abs}$ (\AA)} &
	  \multicolumn{2}{c}{2} &
	  \multicolumn{2}{c}{2} &
	  \multicolumn{2}{c}{2} 
\\[1mm]
\hline
\end{tabular}
\end{center}
See Table~\ref{table_data1} for comments.
NOTES:
$^a$~Blended with [Ne III] $\lambda$3967.
$^b$~[S II] unresolved.
\end{minipage}
\end{table*}

\begin{table*}
\begin{minipage}{160mm}
\caption{
Line ratios and properties for \hii\ regions in DDO~47.
}
\label{table_data7} 
\scriptsize 
\begin{center}
\renewcommand{\arraystretch}{0.95} 
\vspace*{0mm}
\begin{tabular}{rccccccc}
\hline \hline
& & \multicolumn{2}{c}{SHK91 No. 16$\,^a$} & 
\multicolumn{2}{c}{SHK91 No. 17} &     			  
\multicolumn{2}{c}{SHK91 No. 18$\,^b$} \\    
\multicolumn{1}{c}{Property} &
\multicolumn{1}{c}{$f(\lambda)$} &
\multicolumn{1}{c}{$F$} & \multicolumn{1}{c}{$I$} &
\multicolumn{1}{c}{$F$} & \multicolumn{1}{c}{$I$} &
\multicolumn{1}{c}{$F$} & \multicolumn{1}{c}{$I$} \\
\hline
$[\rm{O\;II}]\;3727$ & $+0.325$ &
	  $246.7 \pm 6.5$ & $244.6 \pm 6.4$ &
	  $245 \pm 11$ & $245 \pm 11$ &
	  $67.5 \pm 4.3$ & $64.6 \pm 4.1$ 
\\
$[{\rm Ne\;III}]\;3869$ & $+0.294$ &
          \ldots & \ldots &
	  $34.9 \pm 7.0$ & $34.9 \pm 7.0$ &
	  $36.7 \pm 3.6$ & $35.1 \pm 3.4$
\\
${\rm H}\epsilon + {\rm He\;I}\;3970\,^c$ & $+0.269$ &
	  \ldots & \ldots &
          \ldots & \ldots &
	  $23.0 \pm 2.6$ & $28.3 \pm 2.5$
\\
${\rm H}\delta\;4101$ & $+0.232$ &
	  \ldots & \ldots &
          \ldots & \ldots &
	  $19.5 \pm 1.9$ & $24.8 \pm 1.8$
\\
${\rm H}\gamma\;4340$ & $+0.158$ &
          $34.9 \pm 3.0$ & $37.4 \pm 3.0$ &
	  $42.4 \pm 5.4$ & $42.4 \pm 5.4$ &
	  $41.1 \pm 1.9$ & $44.5 \pm 1.8$
\\
$[{\rm O\;III}]\;4363$ & $+0.151$ &
	  \ldots & \ldots &
          \ldots & \ldots &
	  $9.7 \pm 1.5$ & $9.3 \pm 1.4$
\\
$[{\rm He\;II}]\;4686$ & $+0.050$ &
	  \ldots & \ldots &
          \ldots & \ldots &
	  $11.1 \pm 2.0$ & $10.6 \pm 1.9$
\\
${\rm H}\beta\;4861$ & \phs0.000 &
	  $100.0 \pm 3.8$ & $100.0 \pm 3.7$ &
          $100.0 \pm 7.0$ & $100.0 \pm 7.0$ &
	  $100.0 \pm 3.4$ & $100.0 \pm 3.3$
\\
$[{\rm O\;III}]\;4959$ & $-0.026$ &
	  $55.9 \pm 4.0$ & $54.5 \pm 3.9$ &
          $133 \pm 10$ & $133 \pm 10$ &
	  $205 \pm 12$ & $196 \pm 12$
\\
$[{\rm O\;III}]\;5007$ & $-0.038$ &
	  $175.9 \pm 5.0$ & $171.5 \pm 4.9$ &
          $404 \pm 13$ & $404 \pm 13$ &
	  $594 \pm 15$ & $568 \pm 14$
\\
${\rm He\;I}\;5876$ & $-0.204$ &
	  \ldots & \ldots &
          $9.6 \pm 4.5$ & $9.6 \pm 4.5$ &
	  $7.4 \pm 1.5$ & $7.1 \pm 1.4$
\\
${\rm H}\alpha\;6563$ & $-0.299$ &
	  $295.9 \pm 5.3$ & $286.5 \pm 5.1$ &
          $268.7 \pm 7.7$ & $268.7 \pm 7.7$ &
	  $289.9 \pm 6.3$ & $278.8 \pm 6.0$
\\
$[{\rm N\;II}]\;6583$ & $-0.301$ &
	  $10.4 \pm 4.2$ & $10.0 \pm 4.0$ &
          $11.8 \pm 6.2$ & $11.8 \pm 6.2$ &
	  $2.6 \pm 4.9$ & $2.5 \pm 4.7$
\\
$[{\rm S\;II}]\;6716, 6731$ & $-0.320$ &
	  $65.8 \pm 6.2$ & $63.3 \pm 6.0$ &
          $45 \pm 10$ & $45 \pm 10$ &
	  $11.9 \pm 1.6$ & $11.4 \pm 1.5$
\\[1mm]
\multicolumn{2}{c}{$F(\hbeta)$ (ergs s$^{-1}$ cm$^{-2}$)} & 
	  \multicolumn{2}{c}{$(5.91 \pm 0.23) \times 10^{-16}$} &
	  \multicolumn{2}{c}{$(3.22 \pm 0.23) \times 10^{-16}$} &
	  \multicolumn{2}{c}{$(8.96 \pm 0.31) \times 10^{-16}$}
\\
\multicolumn{2}{c}{EW$_{\rm e}$(\hbeta) (\AA)} &
	  \multicolumn{2}{c}{$83.9 \pm 5.9$} &
	  \multicolumn{2}{c}{$118 \pm 19$} &
	  \multicolumn{2}{c}{$105.6 \pm 7.8$} 
\\
\multicolumn{2}{c}{$c(\hbeta)$} &
	  \multicolumn{2}{c}{0.02} &
	  \multicolumn{2}{c}{0} &
	  \multicolumn{2}{c}{0} 
\\
\multicolumn{2}{c}{EW$_{\rm abs}$ (\AA)} &
	  \multicolumn{2}{c}{2} &
	  \multicolumn{2}{c}{0} &
	  \multicolumn{2}{c}{4.7} 
\\[1mm]
\hline
\end{tabular}
\end{center}
SHK91: \citet{strobel91}.
See Table~\ref{table_data1} for additional comments,
and Fig.~\ref{fig_dwarfs2} for location of \hii\ regions.
NOTES:
$^a$~Labeled as \hii\ region No.~2 in \citet{skh89}.
$^b$~Labeled as \hii\ region No.~1 in \citet{skh89}.
$^c$~Blended with [Ne III] $\lambda$3967.
\end{minipage}
\end{table*}

\begin{table*}
\begin{minipage}{160mm}
\caption{
Line ratios and properties for \hii\ regions in NGC~3109.
}
\label{table_data8} 
\tiny 
\begin{center}
\renewcommand{\arraystretch}{0.95} 
\renewcommand{\tabcolsep}{5.5pt}
\vspace*{0mm}
\begin{tabular}{rccccccccc}
\hline \hline
%
& & \multicolumn{2}{c}{(RM92 No. 8)$\,^a$} &
    \multicolumn{2}{c}{(RM92 No. 5)$\,^b$} \\
& & \multicolumn{2}{c}{A1} & 
\multicolumn{2}{c}{A3} &     
\multicolumn{2}{c}{B1} &     
\multicolumn{2}{c}{B2} \\    
\multicolumn{1}{c}{Property} &
\multicolumn{1}{c}{$f(\lambda)$} &
\multicolumn{1}{c}{$F$} & \multicolumn{1}{c}{$I$} &
\multicolumn{1}{c}{$F$} & \multicolumn{1}{c}{$I$} &
\multicolumn{1}{c}{$F$} & \multicolumn{1}{c}{$I$} &
\multicolumn{1}{c}{$F$} & \multicolumn{1}{c}{$I$} \\
\hline
$[\rm{O\;II}]\;3727$ & $+0.325$ &
	  $193.5 \pm 2.9$ & $184.0 \pm 2.8$ &
	  $178.7 \pm 3.3$ & $168.7 \pm 3.1$ &
	  $351 \pm 28$ & $289 \pm 23$ &
	  $307 \pm 53$ & $258 \pm 45$
\\
${\rm H}9\;3835$ & $+0.302$ &
	  $3.1 \pm 1.3$ & $8.1 \pm 1.2$ &
	  \ldots & \ldots &
	  \ldots & \ldots &
	  \ldots & \ldots 
\\
$[{\rm Ne\;III}]\;3869$ & $+0.294$ &
	  $23.3 \pm 1.4$ & $22.2 \pm 1.3$ &
	  $30.1 \pm 3.5$ & $28.4 \pm 3.3$ &
	  \ldots & \ldots &
	  \ldots & \ldots
\\
${\rm H}8 + {\rm He\;I}\;3889$ & $+0.289$ &
	  $18.1 \pm 1.4$ & $23.3 \pm 1.3$ &
	  $7.9 \pm 2.8$ & $19.0 \pm 2.6$ &
	  \ldots & \ldots &
	  \ldots & \ldots
\\
${\rm H}\epsilon + {\rm He\;I}\;3970\,^c$ & $+0.269$ &
	  $14.6 \pm 1.3$ & $19.7 \pm 1.2$ &
	  \ldots & \ldots &
	  \ldots & \ldots &
	  \ldots & \ldots
\\
${\rm H}\delta\;4101$ & $+0.232$ &
	  $19.9 \pm 1.0$ & $24.66 \pm 0.95$ &
	  $9.6 \pm 1.3$ & $18.6 \pm 1.2$ &
	  \ldots & \ldots &
	  \ldots & \ldots
\\
${\rm H}\gamma\;4340$ & $+0.158$ &
	  $41.0 \pm 1.4$ & $44.1 \pm 1.3$ &
	  $33.9 \pm 1.6$ & $40.0 \pm 1.5$ &
	  \ldots & \ldots &
	  \ldots & \ldots
\\
$[{\rm O\;III}]\;4363$ & $+0.151$ &
	  $< 2.9$ & $< 2.8$ &
	  $< 4.3$ & $< 4.1$ &
	  \ldots & \ldots &
	  \ldots & \ldots
\\
${\rm He\;I}\;4471$ & $+0.116$ &
	  $3.16 \pm 0.58$ & $3.01 \pm 0.55$ &
	  \ldots & \ldots &
	  \ldots & \ldots &
	  \ldots & \ldots
\\
$[{\rm He\;II}]\;4686$ & $+0.050$ &
	  $13.9 \pm 1.2$ & $13.2 \pm 1.1$ &
	  \ldots & \ldots &
	  \ldots & \ldots &
	  \ldots & \ldots
\\
${\rm H}\beta\;4861$ & \phs0.000 &
	  $100.0 \pm 3.2$ & $100.0 \pm 3.0$ &
	  $100.0 \pm 3.9$ & $100.0 \pm 3.7$ &
	  $100 \pm 14$ & $100 \pm 12$ &
	  $100 \pm 24$ & $100 \pm 18$
\\
$[{\rm O\;III}]\;4959$ & $-0.026$ &
	  $104.8 \pm 6.8$ & $99.7 \pm 6.5$ &
	  $117.2 \pm 6.8$ & $110.6 \pm 6.4$ &
	  $86 \pm 13$ & $71 \pm 11$ &
	  $243 \pm 25$ & $183 \pm 19$
\\
$[{\rm O\;III}]\;5007$ & $-0.038$ &
	  $308.4 \pm 8.5$ & $293.3 \pm 8.1$ &
	  $350.4 \pm 8.5$ & $330.8 \pm 8.0$ &
	  $247 \pm 18$ & $204 \pm 15$ &
	  $649 \pm 32$ & $486 \pm 24$
\\
${\rm He\;I}\;5876$ & $-0.204$ &
	  $11.89 \pm 0.87$ & $11.31 \pm 0.83$ &
	  $8.7 \pm 1.4$ & $8.2 \pm 1.3$ &
	  \ldots & \ldots &
	  \ldots & \ldots
\\
${\rm H}\alpha\;6563$ & $-0.299$ &
	  $294.8 \pm 6.8$ & $283.5 \pm 6.5$ &
	  $297.2 \pm 5.4$ & $283.2 \pm 5.1$ &
	  $336 \pm 15$ & $287 \pm 12$ &
	  $393 \pm 30$ & $285 \pm 21$
\\
$[{\rm N\;II}]\;6583$ & $-0.301$ &
	  $13.2 \pm 3.5$ & $12.6 \pm 3.3$ &
	  $2.2 \pm 4.3$ & $2.1 \pm 4.1$ &
	  \ldots & \ldots &
	  \ldots & \ldots
\\
${\rm He\;I}\;6678$ & $-0.314$ &
	  $4.18 \pm 0.94$ & $3.98 \pm 0.89$ &
	  \ldots & \ldots &
	  \ldots & \ldots &
	  \ldots & \ldots
\\
$[{\rm S\;II}]\;6716$ & $-0.319$ &
	  $20.3 \pm 1.7$ & $19.3 \pm 1.6$ &
	  $12.7 \pm 1.5\,^d$ & $12.0 \pm 1.4$ &
	  \ldots & \ldots &
	  \ldots & \ldots
\\
$[{\rm S\;II}]\;6731$ & $-0.321$ &
	  $12.6 \pm 1.5$ & $12.0 \pm 1.4$ &
	  $12.7 \pm 1.5\,^d$ & $12.0 \pm 1.4$ &
	  \ldots & \ldots &
	  \ldots & \ldots
\\
$[{\rm Ar\;III}]\;7136$ & $-0.375$ &
	  $8.6 \pm 1.1$ & $8.2 \pm 1.1$ &
	  $9.6 \pm 1.8$ & $9.1 \pm 1.7$ &
	  \ldots & \ldots &
	  \ldots & \ldots
\\[1mm]
\multicolumn{2}{c}{$F(\hbeta)$ (ergs s$^{-1}$ cm$^{-2}$)} & 
	  \multicolumn{2}{c}{$(2.239 \pm 0.073) \times 10^{-15}$} &
	  \multicolumn{2}{c}{$(1.073 \pm 0.042) \times 10^{-15}$} &
	  \multicolumn{2}{c}{$(1.09 \pm 0.16) \times 10^{-16}$} &
	  \multicolumn{2}{c}{$(6.1 \pm 1.5) \times 10^{-17}$}
\\
\multicolumn{2}{c}{EW$_{\rm e}$(\hbeta) (\AA)} &
	  \multicolumn{2}{c}{$38.8 \pm 1.6$} &
	  \multicolumn{2}{c}{$33.7 \pm 1.6$} &
	  \multicolumn{2}{c}{$9.4 \pm 1.4$} &
	  \multicolumn{2}{c}{$6.3 \pm 1.5$}
\\
\multicolumn{2}{c}{$c(\hbeta)$} &
	  \multicolumn{2}{c}{0} &
	  \multicolumn{2}{c}{0} &
	  \multicolumn{2}{c}{0} &
	  \multicolumn{2}{c}{0.14} 
\\
\multicolumn{2}{c}{EW$_{\rm abs}$ (\AA)} &
	  \multicolumn{2}{c}{2} &
	  \multicolumn{2}{c}{2} &
	  \multicolumn{2}{c}{2} &
	  \multicolumn{2}{c}{2} 
\\[1mm]
\hline
%
%
& & \multicolumn{4}{c}{(RM92 No. 6)$\,^e$} &
    \multicolumn{2}{c}{(RM92 No. 1)$\,^f$} \\
& & \multicolumn{2}{c}{B4} & 
\multicolumn{2}{c}{B5} &     
\multicolumn{2}{c}{B6} \\    
\multicolumn{1}{c}{Property} &
\multicolumn{1}{c}{$f(\lambda)$} &
\multicolumn{1}{c}{$F$} & \multicolumn{1}{c}{$I$} &
\multicolumn{1}{c}{$F$} & \multicolumn{1}{c}{$I$} &
\multicolumn{1}{c}{$F$} & \multicolumn{1}{c}{$I$} \\
\hline
$[\rm{O\;II}]\;3727$ & $+0.325$ &
	  $201.9 \pm 4.4$ & $192.2 \pm 4.2$ &
	  $274.5 \pm 3.6$ & $275.6 \pm 3.6$ &
	  $265.2 \pm 3.4$ & $268.9 \pm 3.5$ 
\\
${\rm H}9\;3835$ & $+0.302$ &
	  \ldots & \ldots &
	  \ldots & \ldots &
	  $2.4 \pm 1.1$ & $5.0 \pm 1.1$
\\
$[{\rm Ne\;III}]\;3869$ & $+0.294$ &
	  $19.8 \pm 2.1$ & $18.9 \pm 2.0$ &
	  $12.1 \pm 1.6$ & $12.1 \pm 1.6$ &
	  $18.1 \pm 1.3$ & $18.3 \pm 1.3$
\\
${\rm H}8 + {\rm He\;I}\;3889$ & $+0.289$ &
	  $20.6 \pm 2.1$ & $22.5 \pm 2.0$ &
	  $11.6 \pm 1.6$ & $18.4 \pm 1.6$ &
	  $14.7 \pm 1.2$ & $18.0 \pm 1.2$
\\
${\rm H}\epsilon + {\rm He\;I}\;3970\,^c$ & $+0.269$ &
	  $19.0 \pm 2.5$ & $21.4 \pm 2.4$ &
	  $8.8 \pm 1.3$ & $15.3 \pm 1.3$ &
	  $14.5 \pm 1.2$ & $17.6 \pm 1.2$
\\
${\rm H}\delta\;4101$ & $+0.232$ &
	  $23.7 \pm 2.6$ & $26.3 \pm 2.5$ &
	  $18.7 \pm 1.7$ & $24.5 \pm 1.7$ &
	  $22.1 \pm 1.2$ & $24.7 \pm 1.2$
\\
${\rm H}\gamma\;4340$ & $+0.158$ &
	  $50.1 \pm 2.3$ & $51.4 \pm 2.2$ & 
	  $41.0 \pm 1.8$ & $45.2 \pm 1.8$ &
	  $40.2 \pm 1.4$ & $42.4 \pm 1.4$
\\
$[{\rm O\;III}]\;4363$ & $+0.151$ &
	  $< 7.1$ & $< 6.8$ &
	  $< 3.9$ & $< 3.8$ &
	  $< 3.4$ & $< 3.4$
\\
${\rm H}\beta\;4861$ & \phs0.000 &
	  $100.0 \pm 3.9$ & $100.0 \pm 3.7$ &
	  $100.0 \pm 3.7$ & $100.0 \pm 3.5$ &
	  $100.0 \pm 3.1$ & $100.0 \pm 3.1$
\\
$[{\rm O\;III}]\;4959$ & $-0.026$ &
	  $111.7 \pm 7.2$ & $106.3 \pm 6.9$ &
	  $58.3 \pm 3.6$ & $55.4 \pm 3.4$ &
	  $107.9 \pm 6.0$ & $105.9 \pm 5.9$
\\
$[{\rm O\;III}]\;5007$ & $-0.038$ &
	  $327.4 \pm 9.1$ & $311.6 \pm 8.7$ &
	  $172.9 \pm 4.6$ & $163.9 \pm 4.4$ &
	  $319.6 \pm 7.5$ & $313.4 \pm 7.4$
\\
${\rm He\;I}\;5876$ & $-0.204$ &
	  $10.3 \pm 2.5$ & $9.8 \pm 2.4$ &
	  $8.0 \pm 1.2$ & $7.4 \pm 1.1$ &
	  $10.20 \pm 0.90$ & $9.85 \pm 0.87$
\\
${\rm H}\alpha\;6563$ & $-0.299$ &
	  $296.9 \pm 6.3$ & $286.6 \pm 6.0$ &
	  $312.2 \pm 6.4$ & $286.1 \pm 5.8$ &
	  $297.3 \pm 4.9$ & $285.1 \pm 4.7$
\\
$[{\rm N\;II}]\;6583$ & $-0.301$ &
	  $5.5 \pm 5.1$ & $5.2 \pm 4.9$ &
	  $8.8 \pm 5.1$ & $8.0 \pm 4.6$ &
	  $6.7 \pm 3.9$ & $6.4 \pm 3.7$
\\
$[{\rm S\;II}]\;6716$ & $-0.319$ &
	  $18.2 \pm 2.0\,^d$ & $17.3 \pm 1.9$ &
	  $38.7 \pm 3.3\,^d$ & $35.0 \pm 3.0$ &
	  $16.7 \pm 1.7$ & $16.0 \pm 1.6$
\\
$[{\rm S\;II}]\;6731$ & $-0.321$ &
	  $18.2 \pm 2.0\,^d$ & $17.3 \pm 1.9$ &
	  $38.7 \pm 3.3\,^d$ & $35.0 \pm 3.0$ &
	  $10.0 \pm 1.5$ & $9.6 \pm 1.4$
\\
$[{\rm Ar\;III}]\;7136$ & $-0.375$ &
	  \ldots & \ldots &
	  $6.2 \pm 1.1$ & $5.57 \pm 0.99$ &
	  $6.9 \pm 1.2$ & $6.6 \pm 1.1$
\\
$[{\rm O\;II}]\;7320, 7330$ & $-0.400$ &
	  \ldots & \ldots &
	  \ldots & \ldots &
          $6.8 \pm 1.4$ & $6.5 \pm 1.3$
\\[1mm]
\multicolumn{2}{c}{$F(\hbeta)$ (ergs s$^{-1}$ cm$^{-2}$)} & 
	  \multicolumn{2}{c}{$(9.14 \pm 0.35) \times 10^{-16}$} &
	  \multicolumn{2}{c}{$(2.005 \pm 0.074) \times 10^{-15}$} &
	  \multicolumn{2}{c}{$(2.369 \pm 0.073) \times 10^{-15}$}
\\
\multicolumn{2}{c}{EW$_{\rm e}$(\hbeta) (\AA)} &
	  \multicolumn{2}{c}{$41.5 \pm 2.0$} &
	  \multicolumn{2}{c}{$41.1 \pm 1.9$} &
	  \multicolumn{2}{c}{$122.5 \pm 9.1$} 
\\
\multicolumn{2}{c}{$c(\hbeta)$} &
	  \multicolumn{2}{c}{0} &
	  \multicolumn{2}{c}{0.07} &
	  \multicolumn{2}{c}{0.04} 
\\
\multicolumn{2}{c}{EW$_{\rm abs}$ (\AA)} &
	  \multicolumn{2}{c}{2.1} &
	  \multicolumn{2}{c}{2} &
	  \multicolumn{2}{c}{2} 
\\[1mm]
\hline
\end{tabular}
\end{center}
{\scriptsize
RM92: \citet{rm92}. See Table~\ref{table_data1} for additional comments,
and Fig.~\ref{fig_dwarfs2} for slit orientations.
NOTES:
$^a$~Labeled as complex D2H5 in \citet{bresolin93},
and \hii\ \#26 in \citet{pena06}.
$^b$~Labeled as complex F1H3 in \citet{bresolin93},
and \hii\ \#13 in \citet{pena06}.
$^c$~Blended with [Ne III] $\lambda$3967.
$^d$~[S II] unresolved.
$^e$~Labeled as complex D2H5 in \citet{bresolin93},
and \hii\ \#18 in \citet{pena06}.
$^f$~Labeled as complex F1H4 in \citet{bresolin93},
and \hii\ \#8 in \citet{pena06}.
}
\end{minipage}
\end{table*}

\begin{table*}
\begin{minipage}{160mm}
\caption{
Line ratios and properties for \hii\ regions in Sextans~B.
}
\label{table_data9} 
\scriptsize 
\begin{center}
\renewcommand{\arraystretch}{0.95} 
\renewcommand{\tabcolsep}{2mm}
\vspace*{0mm}
\begin{tabular}{rccccccc}
\hline \hline
& & \multicolumn{6}{c}{Long-slit orientation A$\,^a$}
\\[1mm]
%
%
& & \multicolumn{2}{c}{SHK91 No. 1$\,^b$} & 
\multicolumn{2}{c}{SHK91 No. 2$\,^c$} &     
\multicolumn{2}{c}{SHK91 No. 5$\,^d$} \\    
\multicolumn{1}{c}{Property} &
\multicolumn{1}{c}{$f(\lambda)$} &
\multicolumn{1}{c}{$F$} & \multicolumn{1}{c}{$I$} &
\multicolumn{1}{c}{$F$} & \multicolumn{1}{c}{$I$} &
\multicolumn{1}{c}{$F$} & \multicolumn{1}{c}{$I$} \\
\hline
$[\rm{O\;II}]\;3727$ & $+0.325$ &
	  $291.5 \pm 5.3$ & $282.4 \pm 5.1$ &
	  $310 \pm 11$ & $322 \pm 11$ &
	  $263.8 \pm 5.2$ & $255.5 \pm 5.0$
\\
$[{\rm Ne\;III}]\;3869$ & $+0.294$ &
	  $15.0 \pm 2.2$ & $14.5 \pm 2.1$ &
	  \ldots & \ldots &
	  $18.3 \pm 2.7$ & $17.7 \pm 2.6$
\\
${\rm H}8 + {\rm He\;I}\;3889$ & $+0.289$ &
	  $12.8 \pm 2.1$ & $19.1 \pm 2.0$ &
	  \ldots & \ldots &
	  $23.1 \pm 2.8$ & $27.0 \pm 2.7$
\\
${\rm H}\epsilon + {\rm He\;I}\;3970\,^e$ & $+0.269$ &
	  $8.3 \pm 2.2$ & $14.7 \pm 2.1$ &
	  \ldots & \ldots &
	  $13.7 \pm 2.5$ & $18.0 \pm 2.4$
\\
${\rm H}\delta\;4101$ & $+0.232$ &
	  $14.2 \pm 1.7$ & $19.8 \pm 1.6$ &
	  \ldots & \ldots &
	  $22.0 \pm 1.9$ & $25.8 \pm 1.8$
\\
${\rm H}\gamma\;4340$ & $+0.158$ &
	  $36.2 \pm 1.7$ & $39.9 \pm 1.6$ &
	  $27.0 \pm 4.8$ & $37.9 \pm 4.7$ &
	  $42.8 \pm 2.0$ & $45.3 \pm 1.9$
\\
$[{\rm O\;III}]\;4363$ & $+0.151$ &
	  $< 4.7$ & $< 4.5$ &
	  \ldots & \ldots &
	  $< 6.1$ & $< 5.9$
\\
${\rm H}\beta\;4861$ & \phs0.000 &
	  $100.0 \pm 3.3$ & $100.0 \pm 3.2$ &
	  $100.0 \pm 5.5$ & $100.0 \pm 5.0$ &
	  $100.0 \pm 3.5$ & $100.0 \pm 3.4$
\\
$[{\rm O\;III}]\;4959$ & $-0.026$ &
	  $60.3 \pm 3.5$ & $57.9 \pm 3.4$ &
	  $48.0 \pm 4.6$ & $43.5 \pm 4.2$ &
	  $82.2 \pm 5.2$ & $79.6 \pm 5.0$
\\
$[{\rm O\;III}]\;5007$ & $-0.038$ &
	  $171.9 \pm 4.5$ & $165.1 \pm 4.3$ &
	  $153.3 \pm 5.6$ & $138.1 \pm 5.0$ &
	  $241.4 \pm 6.6$ & $233.8 \pm 6.4$
\\
${\rm H}\alpha\;6563$ & $-0.299$ &
	  $296.0 \pm 6.4$ & $284.6 \pm 6.1$ &
	  $347 \pm 12$ & $286.3 \pm 9.8$ &
	  $292.7 \pm 5.4$ & $285.4 \pm 5.2$
\\
$[{\rm N\;II}]\;6583$ & $-0.301$ &
	  $2.2 \pm 5.0$ & $2.1 \pm 4.8$ &
	  $26.1 \pm 9.3$ & $21.2 \pm 7.6$ &
	  $3.9 \pm 4.3$ & $3.8 \pm 4.2$
\\
$[{\rm S\;II}]\;6716, 6731$ & $-0.320$ &
	  $25.1 \pm 3.9$ & $24.0 \pm 3.7$ &
	  \ldots & \ldots & 
	  $28.6 \pm 4.0$ & $27.7 \pm 3.9$
\\[1mm]
\multicolumn{2}{c}{$F(\hbeta)$ (ergs s$^{-1}$ cm$^{-2}$)} & 
	  \multicolumn{2}{c}{$(1.722 \pm 0.056) \times 10^{-15}$} &
	  \multicolumn{2}{c}{$(4.15 \pm 0.23) \times 10^{-16}$} &
	  \multicolumn{2}{c}{$(1.482 \pm 0.052) \times 10^{-15}$}
\\
\multicolumn{2}{c}{EW$_{\rm e}$(\hbeta) (\AA)} &
	  \multicolumn{2}{c}{$49.9 \pm 2.2$} &
	  \multicolumn{2}{c}{$21.4 \pm 1.3$} &
	  \multicolumn{2}{c}{$61.9 \pm 3.3$} 
\\
\multicolumn{2}{c}{$c(\hbeta)$} &
	  \multicolumn{2}{c}{0.01} &
	  \multicolumn{2}{c}{0.17} &
	  \multicolumn{2}{c}{0} 
\\
\multicolumn{2}{c}{EW$_{\rm abs}$ (\AA)} &
	  \multicolumn{2}{c}{2} &
	  \multicolumn{2}{c}{2} &
	  \multicolumn{2}{c}{2} 
\\[1mm]
\hline
& & \multicolumn{6}{c}{Long-slit orientation B}
\\[1mm]
%
%
& & \multicolumn{2}{c}{SHK91 No. 5 narrow$\,^f$} & 
\multicolumn{2}{c}{SHK91 No. 5 wide$\,^g$} &
\multicolumn{2}{c}{SHK91 No. 10$\,^h$} \\ 
\multicolumn{1}{c}{Property} &
\multicolumn{1}{c}{$f(\lambda)$} &
\multicolumn{1}{c}{$F$} & \multicolumn{1}{c}{$I$} &
\multicolumn{1}{c}{$F$} & \multicolumn{1}{c}{$I$} &
\multicolumn{1}{c}{$F$} & \multicolumn{1}{c}{$I$} \\
\hline
$[\rm{O\;II}]\;3727$ & $+0.325$ &
	  $237.1 \pm 5.7$ & $237.1 \pm 5.7$ & 
	  $260.4 \pm 4.4$ & $260.4 \pm 4.4$ &
	  $370.6 \pm 6.7$ & $370.6 \pm 6.7$ 
\\
${\rm H}9\;3835$ & $+0.302$ &
	  $7.5 \pm 2.5$ & $7.5 \pm 2.5$ &
	  $4.2 \pm 1.8$ & $4.2 \pm 1.8$ &
	  $5.9 \pm 1.6$ & $5.9 \pm 1.6$
\\
$[{\rm Ne\;III}]\;3869$ & $+0.294$ &
	  $19.3 \pm 2.8$ & $19.3 \pm 2.8$ &
	  $21.5 \pm 2.0$ & $21.5 \pm 2.0$ &
	  $4.6 \pm 1.6$ & $4.6 \pm 1.6$
\\
${\rm H}8 + {\rm He\;I}\;3889$ & $+0.289$ &
	  $19.4 \pm 2.8$ & $19.4 \pm 2.8$ &
	  $18.5 \pm 2.0$ & $18.5 \pm 2.0$ &
	  $14.1 \pm 1.9$ & $14.1 \pm 1.9$
\\
${\rm H}\epsilon + {\rm He\;I}\;3970\,^e$ & $+0.269$ &
	  $17.9 \pm 2.7$ & $17.9 \pm 2.7$ &
	  $16.4 \pm 1.9$ & $16.4 \pm 1.9$ &
	  $9.8 \pm 1.4$ & $9.8 \pm 1.4$
\\
${\rm H}\delta\;4101$ & $+0.232$ &
	  $21.9 \pm 2.9$ & $21.9 \pm 2.9$ &
	  $25.9 \pm 2.2$ & $25.9 \pm 2.2$ &
	  $21.0 \pm 1.7$ & $21.0 \pm 1.7$
\\
${\rm H}\gamma\;4340$ & $+0.158$ &
	  $43.1 \pm 2.5$ & $43.1 \pm 2.5$ &
	  $43.8 \pm 1.7$ & $43.8 \pm 1.7$ &
	  $43.2 \pm 1.5$ & $43.2 \pm 1.5$
\\
$[{\rm O\;III}]\;4363$ & $+0.151$ &
	  $< 7.1$ & $< 7.1$ &
	  $4.2 \pm 1.4$ & $4.2 \pm 1.4$ & 
	  $< 3.1$ & $< 3.1$
\\
${\rm H}\beta\;4861$ & \phs0.000 &
	  $100.0 \pm 3.3$ & $100.0 \pm 3.3$ &
	  $100.0 \pm 3.3$ & $100.0 \pm 3.3$ &
	  $100.0 \pm 3.6$ & $100.0 \pm 3.6$
\\
$[{\rm O\;III}]\;4959$ & $-0.026$ &
	  $86.0 \pm 5.2$ & $86.0 \pm 5.2$ &
	  $82.4 \pm 4.7$ & $82.4 \pm 4.7$ &
	  $23.3 \pm 2.0$ & $23.3 \pm 2.0$
\\
$[{\rm O\;III}]\;5007$ & $-0.038$ &
	  $264.3 \pm 6.7$ & $264.3 \pm 6.7$ &
	  $237.7 \pm 5.9$ & $237.7 \pm 5.9$ &
	  $72.5 \pm 2.5$ & $72.5 \pm 2.5$
\\
${\rm He\;I}\;5876$ & $-0.204$ &
	  \ldots & \ldots &
	  $8.0 \pm 1.5$ & $8.0 \pm 1.5$ &
	  \ldots & \ldots
\\
${\rm H}\alpha\;6563$ & $-0.299$ &
	  $265.3 \pm 5.5$ & $265.3 \pm 5.5$ &
	  $265.1 \pm 4.8$ & $265.1 \pm 4.8$ &
	  $283.6 \pm 5.7$ & $283.6 \pm 5.7$
\\
$[{\rm N\;II}]\;6583$ & $-0.301$ &
	  $1.0 \pm 4.4$ & $1.0 \pm 4.4$ &
	  $5.9 \pm 3.8$ & $5.9 \pm 3.8$ &
	  $8.8 \pm 4.5$ & $8.8 \pm 4.5$
\\
$[{\rm S\;II}]\;6716, 6731$ & $-0.320$ &
	  $21.2 \pm 3.4$ & $21.2 \pm 3.4$ &
	  $31.3 \pm 3.2$ & $31.3 \pm 3.2$ &
	  $35.3 \pm 2.8$ & $35.3 \pm 2.8$
\\[1mm]
\multicolumn{2}{c}{$F(\hbeta)$ (ergs s$^{-1}$ cm$^{-2}$)} & 
	  \multicolumn{2}{c}{$(3.24 \pm 0.11) \times 10^{-16}$} &
	  \multicolumn{2}{c}{$(1.043 \pm 0.035) \times 10^{-15}$} &
	  \multicolumn{2}{c}{$(1.294 \pm 0.047) \times 10^{-15}$}
\\
\multicolumn{2}{c}{EW$_{\rm e}$(\hbeta) (\AA)} &
	  \multicolumn{2}{c}{$96.4 \pm 6.6$} &
	  \multicolumn{2}{c}{$96.0 \pm 6.4$} &
	  \multicolumn{2}{c}{$57.2 \pm 3.0$} 
\\
\multicolumn{2}{c}{$c(\hbeta)$} &
	  \multicolumn{2}{c}{0} &
	  \multicolumn{2}{c}{0} &
	  \multicolumn{2}{c}{0} 
\\
\multicolumn{2}{c}{EW$_{\rm abs}$ (\AA)} &
	  \multicolumn{2}{c}{0} &
	  \multicolumn{2}{c}{0} &
	  \multicolumn{2}{c}{0} 
\\[1mm]
\hline
\end{tabular}
\end{center}
{\scriptsize
SHK91: \citet{strobel91}.
See Table~\ref{table_data1} for additional comments,
and Fig.~\ref{fig_dwarfs2} for slit orientations.
NOTES:
$^a$~The long-slit also went through \hii\ region SHK91 No.~4, 
but only weak \otwo\ and \halpha\ emission lines were detected.
$^b$~Labeled as \hii\ region No.~1 in \citet{skh89}.
$^c$~Labeled as \hii\ region No.~4 in \citet{skh89}.
$^d$~Labeled as \hii\ region No.~2 in \citet{skh89}, and
the ``primary'' \hii\ region in \citet{scv86} and \citet{mam90}.
$^e$~Blended with [Ne III] $\lambda$3967.
$^f$~``Narrow'' extraction aperture : 10 pixels (1\farcs{6}). 
$^g$~``Wide'' extraction aperture : 44 pixels (7\farcs{0}).
$^h$~Labeled as \hii\ region No.~3 in \citet{skh89}.
}
\end{minipage}
\end{table*}

\begin{figure*}
\centering
\includegraphics[width=87mm]{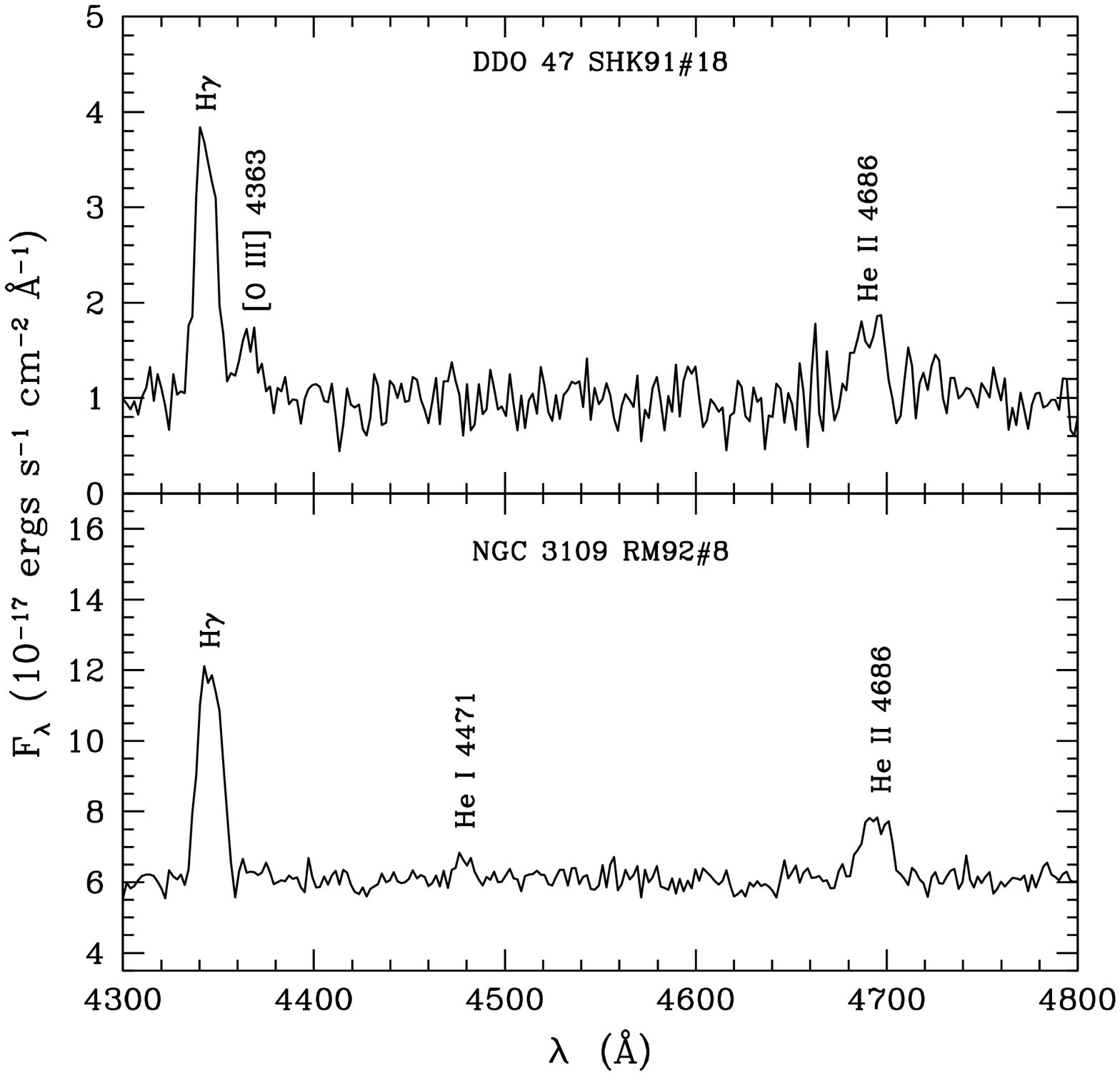}
\caption{
Spectra of \hii\ regions --- DDO~47 SHK91 No.~18, and
NGC~3109 RM92 No.~8 --- between 4300 and 4800~\AA.
The emission lines \hgamma, [O III] $\lambda$4363,
He~I $\lambda$4471, and He~II $\lambda$4686 are labeled.
}
\label{fig_hetwo}
\end{figure*}

He~II $\lambda$4686 emission was detected in \hii\ regions
DDO~47 SHK91 No.~18 and NGC~3109 RM92 No.~8 (A1).
The equivalent widths of the He~II $\lambda4686$
emission line are $9.2 \pm 1.7$~\AA\ and $5.05 \pm 0.45$~\AA,
respectively.
The corresponding fluxes are between 1 and 3 $\times 10^{-16}$ 
ergs s$^{-1}$ cm$^{-2}$ (Fig.~\ref{fig_hetwo};
Tables~\ref{table_data7} and \ref{table_data8}),
and $I$(He~II)/$I$(\hbeta) values are about 10\%.
These values are unusually high compared to predictions
from models of young stellar populations in recent starbursts 
(e.g., \citealp{sv98}).
In particular, the $I$(He~II)/$I$(\hbeta) values do not agree with a
relatively metal-poor ($Z = 0.004$) burst with a Salpeter stellar
initial mass function (upper limit 120~\msun), and an instantaneous
burst of star formation with ages $\la$ 10~Myr.
Shocks from supernovae are an unlikely contributor to the 
He~II emission, as [O~I] emission (e.g., \citealp{skillman85}) is not
present in the spectra.
%

\begin{table*}
\begin{minipage}{160mm}
\caption{
Ionic and total nebular abundances for Cen~A dwarf irregular galaxies.
}
\label{table_abund1}
\tiny 
\begin{center}
\renewcommand{\arraystretch}{1.}
\renewcommand{\tabcolsep}{1.8mm}
\vspace*{0mm}
\begin{tabular}{lcccccc}
\hline \hline
& AM 1318$-$444 & ESO 272$-$G025 & ESO 272$-$G025 & ESO 274$-$G001 &
ESO 274$-$G001 & ESO 274$-$G001 
\\
Property & No. 1 & No. 1 & No. 2 & No. 1 & No. 4$\,^a$ & No. 5
\\
\hline
$T_e$(O$^{+2}$) (K) &
$< 15050$ & $15700 \pm 210$ & \ldots & \ldots & $< 18700$ & \ldots
\\
$T_e$(O$^+$) (K) &
$< 13540$ & $13990 \pm 190$ & \ldots & \ldots & $< 16100$ & \ldots
\\
O$^+$/H $(\times 10^5)$ & 
$> 2.8$ & $2.19 \pm 0.92$ & \ldots & \ldots & $> 2.58$ & \ldots
\\
O$^{+2}$/H $(\times 10^5)$ &
$> 2.0$ & $3.6 \pm 1.0$ & \ldots & \ldots & $> 1.57$ & \ldots
\\
O/H $(\times 10^5)$ & 
$> 4.8$ & $5.7 \pm 1.4$ & \ldots & \ldots & $> 4.15$ & \ldots
\\
12$+$log(O/H) &
$> 7.68$ & $7.76 \pm 0.09\,(^{+0.11}_{-0.15})$ & 
\ldots & \ldots & 
$> 7.62$ & \ldots
\\
12$+$log(O/H) M91$\,^b$ &
\underline{7.87}/8.70 & \underline{8.01}/8.57 & 
\underline{8.14}/8.56 & 7.96/\underline{8.79} &
8.16/\underline{8.51} & 7.90/8.63
\\
12$+$log(O/H) P00,P01a$\,^c$ &
7.86 & 7.84 & \ldots & \ldots &
8.15/\underline{8.24} & 7.72/8.50
\\
\hline
& ESO 274$-$G001 & ESO 274$-$G001 & ESO 274$-$G001 & ESO 324$-$G024 &
ESO 324$-$G024 & ESO 324$-$G024 
\\
Property & No. 6 & No. 7 & No. 8 & No. 1 & No. 2 & No. 3
\\
\hline
$T_e$(O$^{+2}$) (K) &
\ldots & \ldots & \ldots & \ldots & \ldots & \ldots
\\
$T_e$(O$^+$) (K) &
\ldots & \ldots & \ldots & \ldots & \ldots & \ldots
\\
O$^+$/H $(\times 10^5)$ & 
\ldots & \ldots & \ldots & \ldots & \ldots & \ldots
\\
O$^{+2}$/H $(\times 10^5)$ &
\ldots & \ldots & \ldots & \ldots & \ldots & \ldots
\\
O/H $(\times 10^5)$ & 
\ldots & \ldots & \ldots & \ldots & \ldots & \ldots
\\
12$+$log(O/H) &
\ldots & \ldots & \ldots & \ldots & \ldots & \ldots
\\
12$+$log(O/H) M91$\,^b$ &
\underline{8.39}/8.39 & \underline{8.27}/8.49 & 
\underline{7.92}/8.81 & \underline{8.17}/8.54 &
\underline{8.06}/8.66 & \underline{8.06}/8.65
\\
12$+$log(O/H) P00$\,^c$ &
\ldots & \ldots & \ldots & \ldots & \ldots & \ldots
\\
\hline
& ESO 324$-$G024 & ESO 324$-$G024 & ESO 325$-$G011 & ESO 325$-$G011 &
ESO 325$-$G011 & ESO 325$-$G011 \\
Property & No. 4 & No. 5 & No. 1 & No. 2 No. 3 & No. 4 \\
\hline
$T_e$(O$^{+2}$) (K) &
$13400 \pm 1400$ & $12500 \pm 1800$ & \ldots & \ldots &
\ldots & $< 21100$
\\
$T_e$(O$^+$) (K) &
$12400 \pm 1300$ & $11700 \pm 1700$ & \ldots & \ldots &
\ldots & $< 17800$
\\
O$^+$/H $(\times 10^5)$ & 
$2.45 \pm 0.92$ & $3.4 \pm 1.8$ & \ldots & \ldots &
\ldots & $> 1.13$ 
\\
O$^{+2}$/H $(\times 10^5)$ &
$5.2 \pm 1.4$ & $6.3 \pm 2.4$ & \ldots & \ldots &
\ldots & $> 1.16$ 
\\
O/H $(\times 10^5)$ & 
$7.6 \pm 1.7$ & $9.7 \pm 3.0$ & \ldots & \ldots &
\ldots & $> 2.29$ 
\\
12$+$log(O/H) &
$7.88 \pm 0.09\,(^{+0.10}_{-0.14})$ & 
$7.99 \pm 0.11\,(^{+0.14}_{-0.20})$ & 
\ldots & \ldots & \ldots & $> 7.36$
\\
12$+$log(O/H) M91$\,^b$ &
\underline{7.89}/8.64 & \underline{7.93}/8.62 & 
\underline{8.05}/8.72 & \underline{7.92}/8.65 &
7.95/8.70 & \underline{7.86}/8.68
\\
12$+$log(O/H) P00$\,^c$ &
7.72 & 7.77 & \ldots & 7.85 & 8.15 & 7.80
\\
\hline
& ESO 325$-$G011 & ESO 381$-$G020 & ESO 381$-$G020 & ESO 381$-$G020 &
ESO 381$-$G020 & ESO 381$-$G020 
\\
Property & No. 5$\,^d$ & No. 1 & No. 2 & No. 3 & No. 4 & No. 5
\\
\hline
$T_e$(O$^{+2}$) (K) &
\ldots & $< 24700$ & \ldots & $< 19000$ & \ldots & \ldots
\\
$T_e$(O$^+$) (K) &
\ldots & $< 20300$ & \ldots & $< 16300$ & \ldots & \ldots
\\
O$^+$/H $(\times 10^5)$ & 
\ldots & $> 0.69$ & \ldots & $> 1.34$ & \ldots & \ldots
\\
O$^{+2}$/H $(\times 10^5)$ &
\ldots & $> 1.29$ & \ldots & $> 1.61$ & \ldots & \ldots
\\
O/H $(\times 10^5)$ & 
\ldots & $> 1.98$ & \ldots & $> 2.95$ & \ldots & \ldots
\\
12$+$log(O/H) &
\ldots & $> 7.30$ & \ldots & $> 7.47$ & \ldots & \ldots
\\
12$+$log(O/H) M91$\,^b$ &
7.96/\underline{8.78} & \underline{7.96}/8.61 & 
\underline{8.32}/8.39 & \underline{7.87}/8.67 &
8.16/8.63 & \underline{7.43}/8.86 
\\
12$+$log(O/H) P00$\,^c$ &
\ldots & 7.79 & \ldots & 7.77 & \ldots & 7.31
\\
\hline
& ESO 381$-$G020 & IC 4247 & IC 4247 & IC 4316 & IC 4316 & IC 4316 
\\
Property & No. 6 & No. 1 & No. 2 & No. 1 & No. 2 & No. 3  
\\
\hline
$T_e$(O$^+$) (K) &
\ldots & \ldots & \ldots & \ldots & $< 10800$ & \ldots \\
$T_e$(O$^+$) (K) &
\ldots & \ldots & \ldots & \ldots & $< 10800$ & \ldots \\
O$^+$/H $(\times 10^5)$ & 
\ldots & \ldots & \ldots & \ldots & $> 8.11$ & \ldots \\
O$^{+2}$/H $(\times 10^5)$ &
\ldots & \ldots & \ldots & \ldots & $> 6.04$ & \ldots \\
12$+$log(O/H) &
\ldots & \ldots & \ldots & \ldots & $> 8.15$ & \ldots \\
12$+$log(O/H) M91$\,^b$ &
7.46/8.89 & \underline{8.24}/8.44 & \underline{8.29}/8.50 & 
\underline{8.33}/8.49 & \underline{8.05}/8.58 & \underline{8.34}/8.38 
\\
12$+$log(O/H) P00$\,^c$ &
7.38 & \ldots & \ldots & \ldots & 8.01 & \ldots \\
\hline
& IC 4316 & IC 4316 & IC 4316
\\
Property & No. 4 & No. 5 & No. 6
\\
\hline
$T_e$(O$^{+2}$) (K) & \ldots & \ldots & \ldots \\
$T_e$(O$^+$) (K) & \ldots & \ldots & \ldots \\
O$^+$/H $(\times 10^5)$ & \ldots & \ldots & \ldots \\
O$^{+2}$/H $(\times 10^5)$ & \ldots & \ldots & \ldots \\
12$+$log(O/H) & \ldots & \ldots & \ldots \\
12$+$log(O/H) M91$\,^b$ &
\underline{8.07}/8.62 & \underline{8.11}/8.58 & \underline{8.07}/8.60
\\
12$+$log(O/H) P00$\,^c$ & \ldots & \ldots & 8.22 
\\
\hline
\end{tabular}
\end{center}
{\scriptsize
Direct (\othreea) oxygen abundances are shown with two uncertainties.
The first uncertainty is the formal uncertainty in the derivation.
In parentheses is the range of possible values, expressed by the
maximum and minimum values of the oxygen abundance.
For the McGaugh bright-line calibration, two abundances are shown
to reflect the lower and upper branches, respectively. 
The underlined value reflects the choice as judged from the
$I$(\ntwob)/$I$(\otwo) ratio
(e.g., \citealp{mrs85,mcgaugh94,vanzee98,lee03field}).
In the absence of \ntwob, no choice is assigned, and 
abundances from both branches are given.
NOTES:
$^a$~May be a candidate for a supernova remnant; see text in
Sec.~\ref{sec_discuss}.
$^b$~\citet{mcgaugh91} bright-line calibration; see also \citet{chip99}.
$^c$~Bright-line calibration:
lower branch - \citet{pilyugin00};
upper branch - \citet{pilyugin01a}.
$^d$~Background galaxy with $v_{\odot} \approx 26200$ km~s$^{-1}$.
}
\end{minipage}
\end{table*}

\begin{table*}
\begin{minipage}{160mm}
\caption{
Ionic and total abundances for observed \hii\ regions in 
DDO~47, NGC~3109, and Sextans~B.
}
\label{table_abund2} 
\scriptsize 
\begin{center}
\renewcommand{\arraystretch}{1.}
\renewcommand{\tabcolsep}{2.3pt} 
\vspace*{0mm}
\begin{tabular}{lcccccc}
\hline \hline
& & & & NGC 3109 & NGC 3109 \\
& DDO 47 & DDO 47 & DDO 47 & (RM92 No. 8) & (RM92 No. 5) & NGC 3109 \\
Property & SHK91 No. 16 & SHK91 No. 17 & SHK91 No. 18 &
A1 & A3 & B1
\\
\hline
$T_e$(O$^{+2}$) (K) &
\ldots & \ldots & $13910 \pm 940$ &
$< 11300$ & $< 12400$ & \ldots 
\\
$T_e$(O$^+$) (K) &
\ldots & \ldots & $12740 \pm 860$ &
$< 10900$ & $< 11700$ & \ldots 
\\
O$^+$/H $(\times 10^5)$ & 
\ldots & \ldots & $0.94 \pm 0.22$ &
$> 4.8$ & $> 3.3$ & \ldots  
\\
O$^{+2}$/H $(\times 10^5)$ &
\ldots & \ldots & $7.4 \pm 1.2$ &
$> 6.9$ & $> 5.8$ & \ldots 
\\
O/H $(\times 10^5)$ & 
\ldots & \ldots & $8.3 \pm 1.3$ &
$> 11.7$ & $> 9.1$ & \ldots  
\\
12$+$log(O/H) &
\ldots & \ldots & $7.92 \pm 0.06\,(^{+0.07}_{-0.08})$ &
$> 8.07$ & $> 7.96$ & \ldots  
\\
12$+$log(O/H) M91$\,^a$ &
\underline{7.86}/8.72 & \underline{8.11}/8.51 & \underline{7.93}/8.60 &
\underline{7.87}/8.66 & \underline{7.88}/8.65 & \underline{7.98}/8.64 
\\
12$+$log(O/H) P00$\,^b$ & 
7.88 & 7.93 & 7.74 & 7.74 & 7.73 & 7.98  
\\
\hline
& & NGC 3109 & NGC 3109 & NGC 3109 
\\
& NGC 3109 & (RM92 No. 6) & (RM92 No. 6) & (RM92 No. 1) &
Sextans B & Sextans B
\\
Property & B2 & B4 & B5 & B6 & SHK91 No. 1 & SHK91 No. 2
\\
\hline
$T_e$(O$^{+2}$) (K) &
\ldots & 
$< 15800$ & $< 16400$ & $< 11900$ & $< 17800$ & \ldots
\\
$T_e$(O$^+$) (K) &
\ldots & 
$< 14100$ & $< 14500$ & $< 11300$ & $< 15400$ & \ldots 
\\
O$^+$/H $(\times 10^5)$ & 
\ldots & 
$> 2.0$ & $> 2.6$ & $> 6.0$ & $> 2.2$ & \ldots  
\\
O$^{+2}$/H $(\times 10^5)$ &
\ldots & 
$> 2.9$ & $> 1.4$ & $> 6.3$ & $> 1.2$ & \ldots
\\
O/H $(\times 10^5)$ & 
\ldots & 
$> 4.9$ & $> 4.0$ & $> 12.3$ & $> 3.4$ & \ldots
\\
12$+$log(O/H) &
\ldots & 
$> 7.69$ & $> 7.60$ & $> 8.09$ & $> 7.53$ & \ldots 
\\
12$+$log(O/H) M91$\,^a$ &
8.24/8.42 & 
\underline{7.91}/8.63 & \underline{7.92}/8.69 &
\underline{8.06}/8.56 & \underline{7.93}/8.68 & \underline{7.98}/8.67
\\
12$+$log(O/H) P00,P01a$\,^b$ &
8.00/8.25 & 7.77 & 7.97 & 7.94 & 7.99 & 8.14 
\\
\hline
%
%
%
& Sextans B & Sextans B & Sextans B & Sextans B \\
Property & SHK91 No. 5$\,^c$ & 
SHK91 No. 5$\,^{d,e}$ & SHK91 No. 5$\,^{d,f}$ & SHK91 No. 10 
\\
\hline
$T_e$(O$^{+2}$) (K) &
$< 17100$ &
$< 17700$ & $14380 \pm 240$ & $< 23700$
\\
$T_e$(O$^+$) (K) &
$< 15000$ &
$< 15400$ & $13060 \pm 210$ & $< 19600$
\\
O$^+$/H $(\times 10^5)$ & 
$> 2.2$ &
$> 1.9$ & $3.5 \pm 1.9$ & $> 1.5$
\\
O$^{+2}$/H $(\times 10^5)$ &
$> 1.8$ &
$> 1.9$ & $2.8 \pm 1.1$ & $> 0.28$
\\
O/H $(\times 10^5)$ & 
$> 4.0$ &
$> 3.7$ & $6.3 \pm 2.2$ & $> 1.76$
\\
12$+$log(O/H) &
$> 7.60$ &
$> 7.57$ & $7.80 \pm 0.13\,(^{+0.14}_{-0.21})$ & $> 7.25$
\\
12$+$log(O/H) M91$\,^c$ &
\underline{7.95}/8.64 &
\underline{7.94}/8.64 & \underline{7.96}/8.63 & \underline{8.05}/8.68
\\
12$+$log(O/H) P00$\,^d$ &
7.90 & 7.86 & 7.91 & \ldots
\\
\hline
\end{tabular}
\end{center}
{\scriptsize
See Table~\ref{table_abund1} for general comments.
DDO~47 - SHK91 refers to the identifications in \citet{strobel91};
see also Fig.~\ref{fig_dwarfs2} and Table~\ref{table_data7}.
NGC~3109 - Fig.~\ref{fig_dwarfs2} and Table~\ref{table_data8}.
Sextans~B - SHK refers to the identifications in \citet{strobel91};
see also Fig.~\ref{fig_dwarfs2} and Table~\ref{table_data9}.
NOTES:
$^a$~\citet{mcgaugh91} bright-line calibration.
$^b$~Bright-line calibration: lower branch - \citet{pilyugin00};
upper branch - \citet{pilyugin01a}.
$^c$~Slit orientation A.
$^d$~Slit orientation B.
$^e$~``Narrow'' extraction aperture of total 10 pixels.
$^f$~``Wide'' extraction aperture of total 44 pixels.
}
\end{minipage}
\end{table*}


\section{Nebular Abundances}
\label{sec_abund}

Oxygen abundances in \hii\ regions were derived using three methods:
(1) the direct method (e.g., \citealp{dinerstein90,skillman98});
and two bright-line methods discussed by
(2) \cite{mcgaugh91}, which is based on photoionization models;  
and (3) \cite{pilyugin00}, which is purely empirical.
We briefly summarize these methods here;
further details of these methods are found in
\cite{ls04} and \cite{lsv05,lsv06}.

\subsection{Direct (\othreea) Abundances}

The ``direct'' conversion of emission line intensities into ionic
abundances requires a reliable estimate of the electron temperature in
the ionized gas.
We adopt a two-zone model for \hii\ regions, with a low- and a
high-ionization zone characterized by temperatures $T_e($O$^+)$ and
$T_e($O$^{+2})$, respectively. 
The temperature in the O$^{+2}$ zone is measured with
the emission line ratio $I$(\othreec)/$I$(\othreea) 
\citep{osterbrock}.
The temperature in the O$^+$ zone is given by 
\begin{equation}
t_e({\rm O}^+) = 0.7 \, t_e({\rm O}^{+2}) + 0.3,
\label{eqn_toplus}
\end{equation}
where $t_e = T_e/10^4$~K \citep{ctm86,garnett92}.


The total oxygen abundance by number is given by
O/H = O$^+$/H$^+$ $+$ O$^{+2}$/H$^+$.
For conditions found in typical \hii\ regions and those presented
here, very little oxygen in the form of neutral oxygen is expected,
and is not included here.
Ionic abundances for singly- and doubly-ionized oxygen were computed
using O$^+$ and O$^{+2}$ temperatures, respectively, as described
above. 
The O$^{+3}$ contribution is assumed negligible where He~II
emission is absent.
Although He~II emission is reported in two \hii\ regions
(see above), we have not included the small contribution
by O$^{+3}$ to the total oxygen abundance.

In four of the 48 \hii\ region spectra, \othreea\ line fluxes
were measured, and subsequent electron temperatures were derived.
We derived direct oxygen abundances in \hii\ regions with \othreea\
detections using the method described by \cite{scm03b}.
Ionic and total abundances are computed using the emissivities from
the five-level atom program by \cite{sd95}.
As described above, we use the same two-temperature zone model and 
temperatures for the remaining ions.
The error in $T_e$(O$^{+2}$) is derived from the uncertainties in
the corrected emission-line ratios, and does not include any
uncertainties in the atomic data, or the possibility of temperature
variations within the O$^{+2}$ zone.
The fractional error in $T_e$(O$^{+2}$) is applied similarly to
$T_e$(O$^+$) to compute the uncertainty in the latter.
Uncertainties in the resulting ionic abundances are combined in
quadrature for the final uncertainty in the total linear (summed)
abundance.
The appropriate temperatures, \othreea\ abundances, and their
uncertainties are listed in Tables~\ref{table_abund1} and
\ref{table_abund2}.

\subsection{Bright-Line Abundances}

In \hii\ regions without \othreea\ measurements, the bright-line
method has been used to derive oxygen abundances, as the latter
are usually given in terms of bright [O~II] and [O~III] emission
lines. 
\cite{pagel79} devised the $R_{23}$ indicator, defined by
$R_{23}$ = [$I$(\otwo) +  $I$(\othree)]/$I$(\hbeta);
\cite{skillman89} discussed the method further for low-metallicity
galaxies.
\cite{mcgaugh91} developed a grid of photoionization models and
suggested using $R_{23}$ and an ionization proxy, represented by 
$O_{32}$ = $I$(\othree)/$I$(\otwo), 
to estimate oxygen abundances\footnote{
Analytical expressions for the McGaugh calibration 
can be found in \cite{chip99}.
}.
To break the degeneracy in the bright-line method, we have used
the \ntwootwo\ ratio
(e.g., \citealp{mrs85,mcgaugh94,vanzee98,lee03field})
to choose either the ``upper branch'' (high
oxygen abundance) or the ``lower branch'' (low oxygen abundance).
In some instances, oxygen abundances with the McGaugh method could not
be computed, because the $R_{23}$ values were outside of the effective
range for the models.
\cite{pilyugin00} proposed an empirical calibration at low metallicity
with fits of oxygen abundance against bright oxygen lines.
\cite{scm03b} have shown that while the Pilyugin method covers a
large range in $R_{23}$, the calibration applies mostly to \hii\
regions with higher ionizations; see also \cite{vzh06}.

Oxygen abundances derived using the bright-line calibrations are
listed in Tables~\ref{table_abund1} and \ref{table_abund2}.
In the absence of \othreea, oxygen abundances derived using
bright-line methods are in agreement with direct abundances
to within $\approx$ 0.2~dex.

\section{Discussion of Individual Galaxies}		
\label{sec_discuss}

\subsection{Cen A Dwarf Irregulars}

Previously, \cite{lee03south} reported spectra for six Cen~A late-type
dwarf galaxies: 
DDO~161 (UGCA~320), ESO~324$-$G024, ESO~381$-$G020, ESO~383$-$G087,
ESO~444$-$G084, and NGC~5264 (DDO~242).
Only \hbeta\ and \halpha\ were detected in ESO~324$-$G024, and
an oxygen abundance was not derived.
\othreea\ was detected in ESO~383$-$G087, and bright-line 
abundances were derived for the remaining four galaxies.

We briefly comment on a number of Cen~A dI's.
Our measured spectrum of AM~1318$-$444 is indicative of 
a metal-poor \hii\ region, and the resulting (bright-line) oxygen
abundance is about one-sixth of the solar value.
While the bright-line abundance for ESO~274$-$G001 \hii\ region No.~4
implies an unusually high oxygen abundance (compared to the derived
lower limit), our measured intensity ratios 
(Table~\ref{table_data2}) suggest that the nebula is a supernova
remnant; see also the discussion in \cite{lsv06}.
If the object in question is a supernova remnant, the
measured $I$(\ntwob)/$I$(\halpha), $I$(\stwob)/$I$(\halpha), and
the model grid from \citet[their Fig.~8]{dopita84} yield an estimate of
12$+$log(O/H) $\simeq 7.8 \pm 0.1$ for the oxygen abundance.
As an edge-on dwarf galaxy, ESO~274$-$G001 is reminescent of NGC~55 in
the Sculptor group and NGC~1560 in the IC~342 group.
These galaxies may provide further opportunities to investigate the
possibility of abundance gradients in low-luminosity late-type galaxies.
In fact, \cite{lee03south} reported an unusually high (bright-line)
abundance for NGC~5264 in the Cen~A group.
Additional high signal-to-noise spectra would be valuable in
confirming this result.
For ESO~324$-$G024, we obtained a higher-quality spectrum compared to
the observation described in \cite{lee03south}, but we did not detect
\othreea.
The bright-line oxygen abundance derived for ESO~381$-$G020 is
0.12~dex higher than the bright-line value determined in
\cite{lee03south}, but these values are well within the acceptable
uncertainty ($\sim$ 0.2~dex) associated with bright-line abundances.
In all, the derived oxygen abundances for the present sample of Cen~A
dI's (cf. Table~\ref{table_gxylist}) are in the range between
about 10\% to 50\% of the solar value, which are in general agreement
with the results obtained by \cite{ws83} and \cite{webster83}.

\subsection{Nearby Dwarf Irregulars}

\subsubsection{DDO 47}

\cite{skh89} obtained spectra for the \hii\ region SHK91 No.~18 in
DDO~47 using the IIDS spectrometer on the KPNO 2.1-m telescope.
With their \othreea\ measurement, they derived an oxygen abundance of
12$+$log(O/H) = $7.89 \pm 0.20$. 
In our spectrum, we measured \othreea\ and derived an oxygen abundance
of 12$+$log(O/H) = $7.92 \pm 0.06$, in agreement with \cite{skh89}.
We have also measured He~II $\lambda$4686\AA.
As He~II emission is indicative of the presence of O$^{+3}$, the
latter is generally a small contributor to the total
oxygen abundance.
For example, \cite{ks93} measured He~II emission in I~Zw~18 and
found that the resulting O$^{+3}$ contribution was of order one to
four percent.
In \hii\ region SHK91 No.~18, the He~II 4686~\AA\ to \hbeta\ flux
ratio is about 10 percent, which implies that the resulting
contribution by O$^{+3}$ to the total oxygen abundance could
be of order 10\%.
For comparison with other dI's, we have not included any O$^{+3}$
contribution in the total oxygen abundance for DDO~47.

\subsubsection{NGC 3109}

The optical appearance (e.g., \citealp{carignan85,jc90}) is 
suggestive of a low-luminosity spiral galaxy.
From the data reported previously by \citet{lee03south},
the authors suggested the presence of an abundance gradient in 
NGC~3109, which had been previously hinted by 
\citeauthor{grebel01a} (\citeyear{grebel01a}, \citeyear{grebel01b}).
We did not measure \othreea\ in any of the spectra shown here.
The subsequent mean of six bright-line abundances is $7.94 \pm 0.2$,
which is larger than the \cite{lee03field} value by about 0.2~dex, but
is in better agreement with other dwarfs at comparable optical
luminosity.
Since this galaxy is known to contain multiple \hii\ regions and
planetary nebulae (i.e., \citealp{bresolin93,pena06}),
deep spectra of these nebulae would be valuable to confirm the
presence or absence of a radial gradient in oxygen abundance.

\subsubsection{Sextans B}

Published nebular abundances for Sextans~B have differed by as much
as 0.5~dex.
\cite{scv86} published a limit of 12$+$log(O/H) $>$ 7.38 from data
obtained with the IIDS detector on the ESO 3.6-m telescope.
\cite{skh89} reported a higher lower limit to the oxygen abundance:
12$+$log(O/H) $>$ 7.56.
From data obtained with the IDS on the 2.5-m INT telescope,
\cite{mam90} measured \othreea\ and subsequently derived
an oxygen abundance 12$+$log(O/H) = 8.11.
If the higher value is adopted, the present-day metallicity
appears $\sim$ 0.3~dex too high for its galaxy optical luminosity,
compared to other dwarf irregulars at comparable luminosity.

\cite{kniazev05} reported NTT spectra for
one planetary nebula and six \hii\ regions in Sextans~B, from
which \othreea\ was detected in the planetary nebula and in
three \hii\ regions.
They reported \othreea\ in SHK91 No.~5, but their spectrum is cut off
below about 3800~\AA, and there is no measured \otwo. 
Alternatively, they used their \otwored\ measurement to derive the
O$^+$/H abundance with the method developed to derive oxygen
abundances for metal-poor galaxies observed in the Sloan Digital Sky
Survey \citep{kniazev04}. 
Their resulting oxygen abundance for \hii\ region No.~5 is
12$+$log(O/H) = $7.84 \pm 0.05$.
\cite{kniazev05} claimed that the recent chemical enrichment 
was spatially inhomogeneous over length scales as large as
1~kpc, based on a dispersion of 0.31~dex for direct abundances in
three \hii\ regions.
With FORS2 on VLT, \cite{magrini05} measured spectra of \hii\ regions
and planetary nebulae, and detected \othreea\ in three \hii\ regions,
including No.~5.
Their direct oxygen abundance is 12$+$log(O/H) = $7.69 \pm 0.14$,
which agrees with the \cite{kniazev05} result.

We measured \othreea\ only in \hii\ region No.~5, but our spectrum
includes the full range of optical emission lines from \otwo\ to
\stwo\ (Fig.~\ref{fig_spectra}, Table~\ref{table_data9}). 
Our resulting direct abundance is 
12$+$log(O/H) = $7.80 \pm 0.13$, which also agrees with the values
reported by \cite{kniazev05} and \cite{magrini05}.
The data presented here shows that the derived bright-line
abundances for \hii\ regions SHK91 Nos.~1, 2, 5, and 10 all
agree to within $\approx$ 0.1~dex (Table~\ref{table_abund2}).
Although the direct abundance for SHK91 No.~5 and the bright-line
abundance for SHK91 No.~10 differ by 0.25~dex, the difference is
comparable to our adopted uncertainty of 0.2~dex for a bright-line
abundance.

\section{Exploration of Environmental Effects}
\label{sec_enveffects}

If dwarf galaxies are more robust to the effects of internal
processes, such as supernova feedback, external processes found within
environments of galaxy groups may be more damaging to the ``health''
of gas-rich dwarf galaxies.
Galaxy-galaxy encounters create tidal interactions, which can
remove stellar and/or gaseous material and eject the stripped material
into the intragroup medium.
Disturbed galaxy morphologies are also indicative of these interactions. 
Groups may contain hot ($T \sim 10^7$ K) dense X-ray emitting gas,
which could provide an agent for ram-pressure stripping, if group
galaxies traverse the intragroup medium at high speeds.
However, only the highest-mass groups have significant X-ray
luminosities (see, e.g., \citealp{mulchaey00}).
In most nearby groups (at least within the Local Volume), it is
difficult to measure strong X-ray emission. 
\cite{bouchard07} have suggested that ram-pressure stripping may have
been responsible to explain the lower \hi\ content in Cen~A dwarfs
relative to dwarfs in the Local Group and the Scl Group.
At present, we assume that the presence of X-ray gas is negligible,
although the possibility of warm gas ($T \sim 10^6$ K) is not entirely
ruled out \citep{mulchaey00}.
Higher star-formation rates can also be induced by tidal interactions,
as molecular gas is subsequently compressed and shock-heated.
A subsequent result would be a larger fraction of galaxies with strong
emission-lines and/or blue colors in these environments.
Unfortunately, extensive galaxy surface photometry has not yet been
obtained for Cen~A or Sculptor (Scl) group dwarf galaxies, although
narrow-band \halpha\ images have been obtained for some of the dI's in
the two groups (\citealp{scm03a}; S. C\^ot\'e, personal communication).
Although we cannot yet compare the star formation properties, we
can compare the end results to their present-day chemical evolution.
In the discussion which follows, we have also included additional
spectra for Cen~A dwarfs from \cite{lee03south} and Scl group dwarfs
from \cite{scm03b}.
We have adopted a distance of 3.0~Mpc for the Scl group dwarfs
AM0106$-$382 and ESO348$-$G009.

\subsection{Luminosity-Metallicity Relation}

The optical luminosity-metallicity relation has long been used as 
a diagnostic for the evolution of nearby star-forming dI's;
e.g., \cite{skh89,rm95,lee03field,vzh06}.
To augment the local sample described by \cite{lee03field},
additional galaxies with \othreea\ measurements and/or
distances from stellar indicators are taken from the literature:
DDO~167 \citep{skh89,kara03b}, 
ESO~489$-$G056 \citep{rb95,kara02c},
NGC~1705 \citep{ls04},
NGC~3738 \citep{martin97,kara03a},
NGC~4449 \citep{martin97,kara03a},
NGC~6822 \citep{lsv06},
and
WLM \citep{lsv05}.
We have also included direct and bright-line abundances for
additional Cen~A and Scl group dwarf irregulars from 
\cite{lee03south} and \cite{scm03b}.

\begin{figure*}
\centering
\includegraphics[width=81mm]{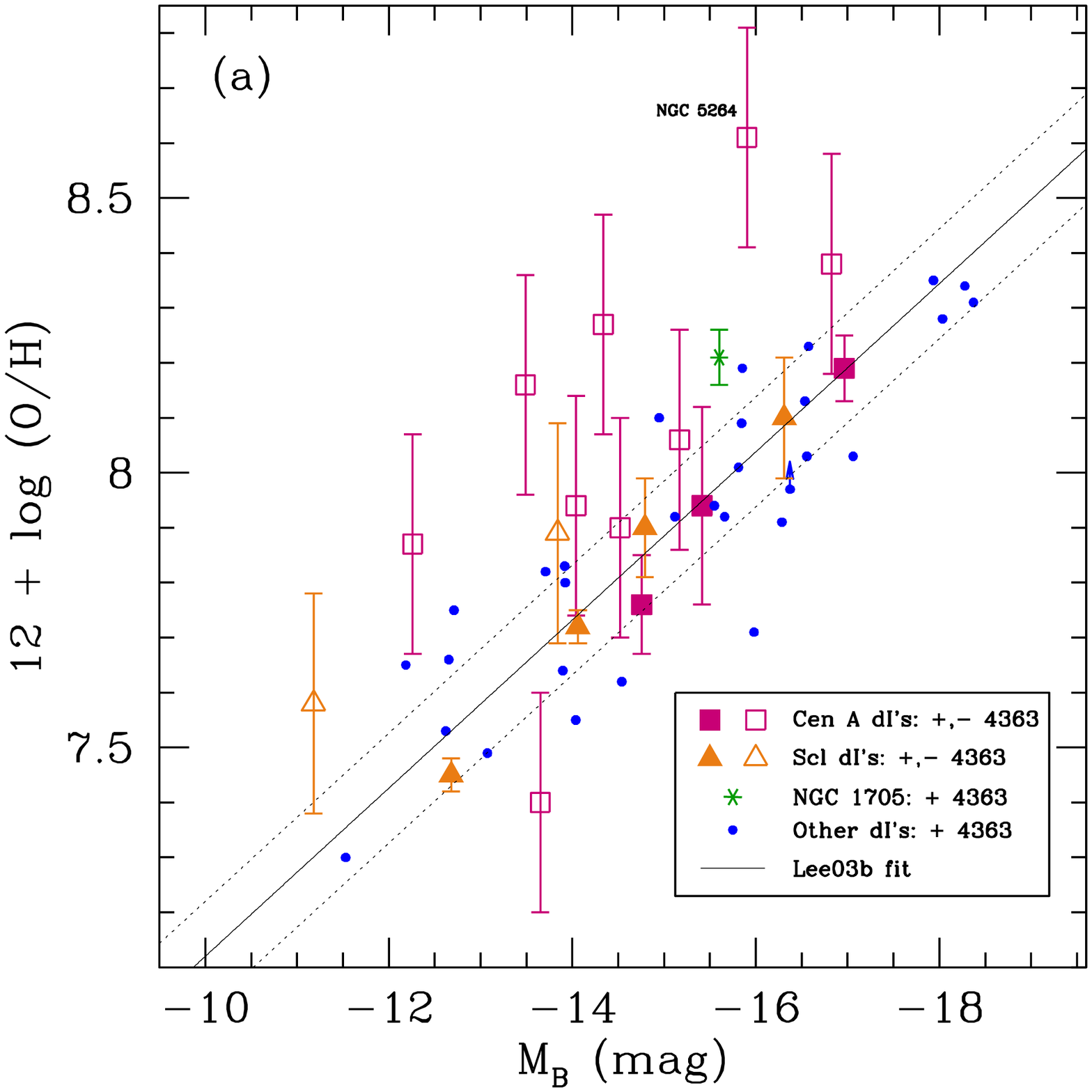} 
\includegraphics[width=81mm]{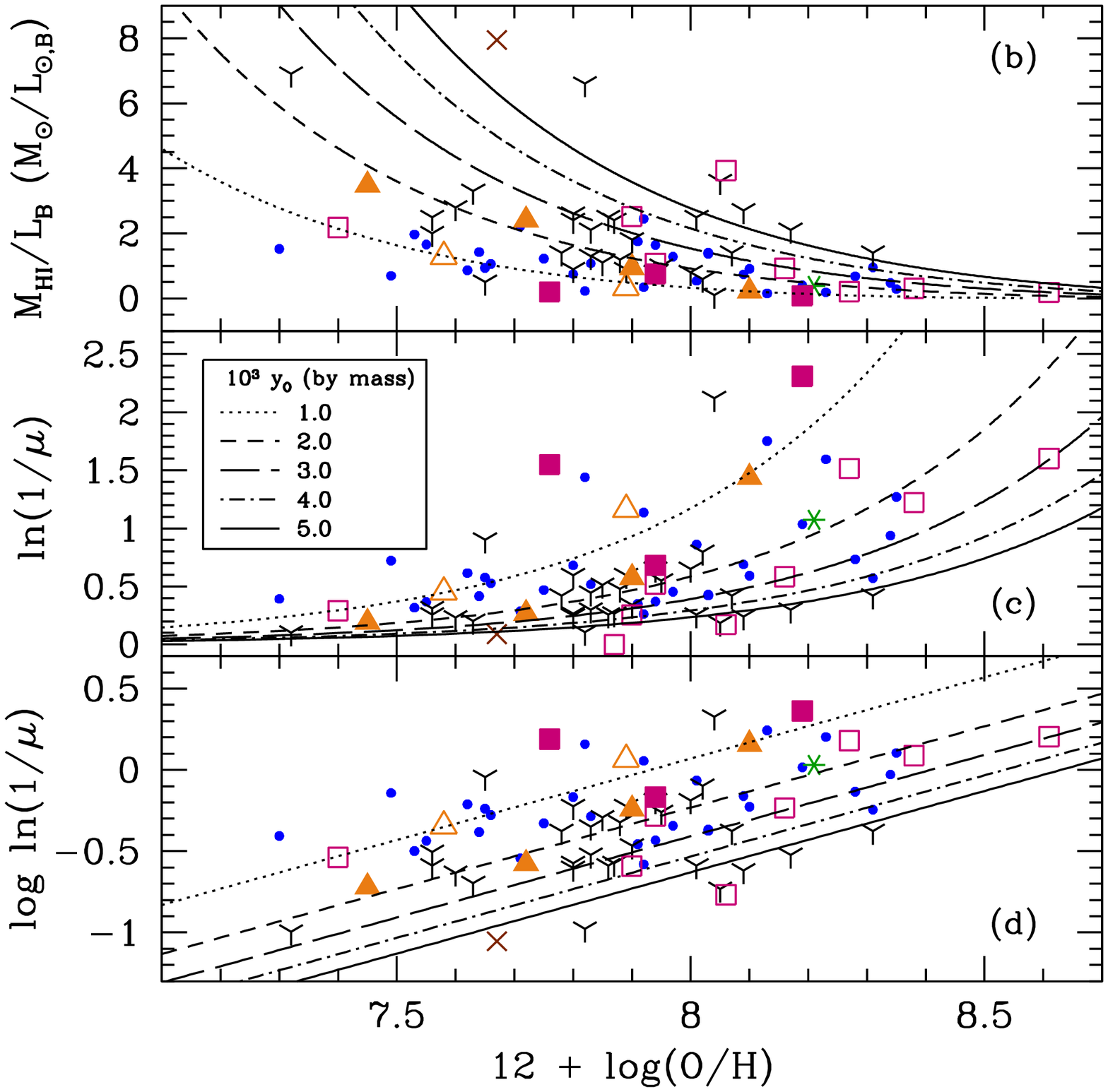}
\caption{
{\em Panel a}:
$B$-band luminosity-metallicity relation for dwarf irregular galaxies.
Squares mark locations of Cen~A dwarf galaxies from the
present data and \citet{lee03south}.
Triangles mark locations of Sculptor group (Scl) dwarf galaxies from
\citet{scm03b} and \citet{lee03south}.
Small circles indicate nearby dwarf irregular galaxies 
in the Local Volume \citep{lee03south,lee03field} with updates
to WLM and NGC~6822 \citep{lsv05,lsv06}; DDO~47 and Sextans~B from
the present spectra have also been included.
Additional objects have been added to the sample of ``nearby dwarfs'';
see text for discussion.
\othreea\ detections are shown as filled symbols.
The star marks the location of the nearby dwarf starburst galaxy
NGC~1705 \citep{ls04}.
The solid line is a fit to a sample of nearby dwarf
irregulars described in \citet{lee03field}.
We note that Cen~A and Scl group dwarfs with \othreea\ abundances are
within 0.1~dex (dotted lines) of the fit.
NGC~5264 exhibits an unusually high oxygen abundance for its
luminosity, although the spectra are of low signal-to-noise.
{\em Panels b--d}:
Gas-to-$B$ light fraction and gas fraction parameters as 
a function of oxygen abundance.
Symbols are the same as in panel a; 
additional Y-symbols mark locations of dwarf irregulars from 
\citet{vanzee97} and \citet{vzh06}, and 
a cross marks the location of the gas-rich dwarf irregular 
galaxy DDO~154 \citep{vanzee97,ks01}.
Error bars have not been plotted to aid visibility.
Curves of different line type mark loci of closed-box chemical
evolution models with different values of effective oxygen yields 
(by mass) as indicated (see also \citealp{lee03field}).
}
\label{fig_diags}
\end{figure*}

The luminosity-metallicity relation is shown in Fig.~\ref{fig_diags}a.
The solid line is a fit to the sample of nearby dwarfs as described
above.
Cen~A and Scl group dwarfs with \othreea\ measurements are in best
agreement with the fit and with the locus of nearby dwarfs that 
have \othreea\ measurements.
Although the spectra for NGC~5264 are of lower signal-to-noise, 
this galaxy exhibits an unusually high oxygen abundance for its
luminosity, which suggests a more spiral-like nature to the
galaxy and warrants future measurements.
Three Cen~A dwarfs with \othreea\ detections have oxygen abundances
consistent for their optical luminosities.
For the other Cen~A dI's without \othreea\ measurements,
their oxygen abundances generally appear higher than abundances
for nearby dI's at similar luminosities. 
We have found that \othreea\ oxygen abundances exhibit the least
dispersion in the luminosity-metallicity relationship, and there is no
apparent difference between dI's in Cen~A and Scl groups\footnote{
From observations of Virgo Cluster dI's, 
\citet{lee03virgo} showed that the cluster environment appears
to have little effect on the optical luminosity-metallicity relation.
}. 
It would be interesting to examine the relative dispersion
in the luminosity-metallicity relation at near-infrared
wavelengths; see \citet{irlz} and references within.

\subsection{Gas Fraction-Metallicity Relation}

The gas content in galaxies may be sensitive to environmental effects,
and we examine here how environment may affect chemical evolution
directly.
The simplest ``closed-box'' model \citep{schmidt63,ss72}
can be written as $Z_{\rm O} = y_{\rm O} \,\ln (1/\mu)$,
where $Z_{\rm O}$ is the oxygen mass fraction, 
$y_{\rm O}$ is the yield by mass,
and $\mu$ is the baryonic gas fraction equal to the ratio of
the gas mass to the total mass in gas and stars.
%

In Fig.~\ref{fig_diags}b--d, we have plotted the \hi-gas-to-$B$-light
ratio and parameters related to gas fraction against oxygen
abundance.
In the absence of measured optical colours for Cen~A and Scl dI's
to derive color-based stellar mass-to-light ratios described by
\cite{belldejong01}, we compute stellar masses by simply assuming a
constant stellar mass-to-$B$-light ratio equal to one.
Our adopted value is within the range of stellar mass-to-$B$-light
ratios derived by, for example, \cite{vz01} and \cite{lee03field}. 
We compute total gas mass as $M_{\rm gas}$ = 1.36 $M_{\rm HI}$,
which includes helium, but ignores molecular gas.
Under the assumption of closed-box evolution, we have also plotted
various curves with yields varying by a factor of five.
We find that nearby dI's \citep{lee03field} and UGC dI's
\citep{vanzee97} span the range of yields shown.
By inspection, $y_O$ $\approx$ $(3 \pm 2) \times 10^{-3}$ is a good
description of the data shown, in general agreement with the results
of \citet{lee03field} and \citet{vzh06}.
The two Cen~A dI's with direct abundances and low gas fractions are
ESO~272$-$G025 and ESO~383$-$G087.
Generally, Cen~A dI's span the range of yields shown, whereas the
small number of Scl dI's tend to cluster around the lower end of the
yield range. 
However, these results are not definitive because of small-number
statistics, and measurements of additional galaxies could strengthen
possible differences in gas fractions between the two sets of dI's.

\subsection{Relations with Tidal Index and Projected Distance}

\begin{figure*}
\centering
\includegraphics[width=81mm]{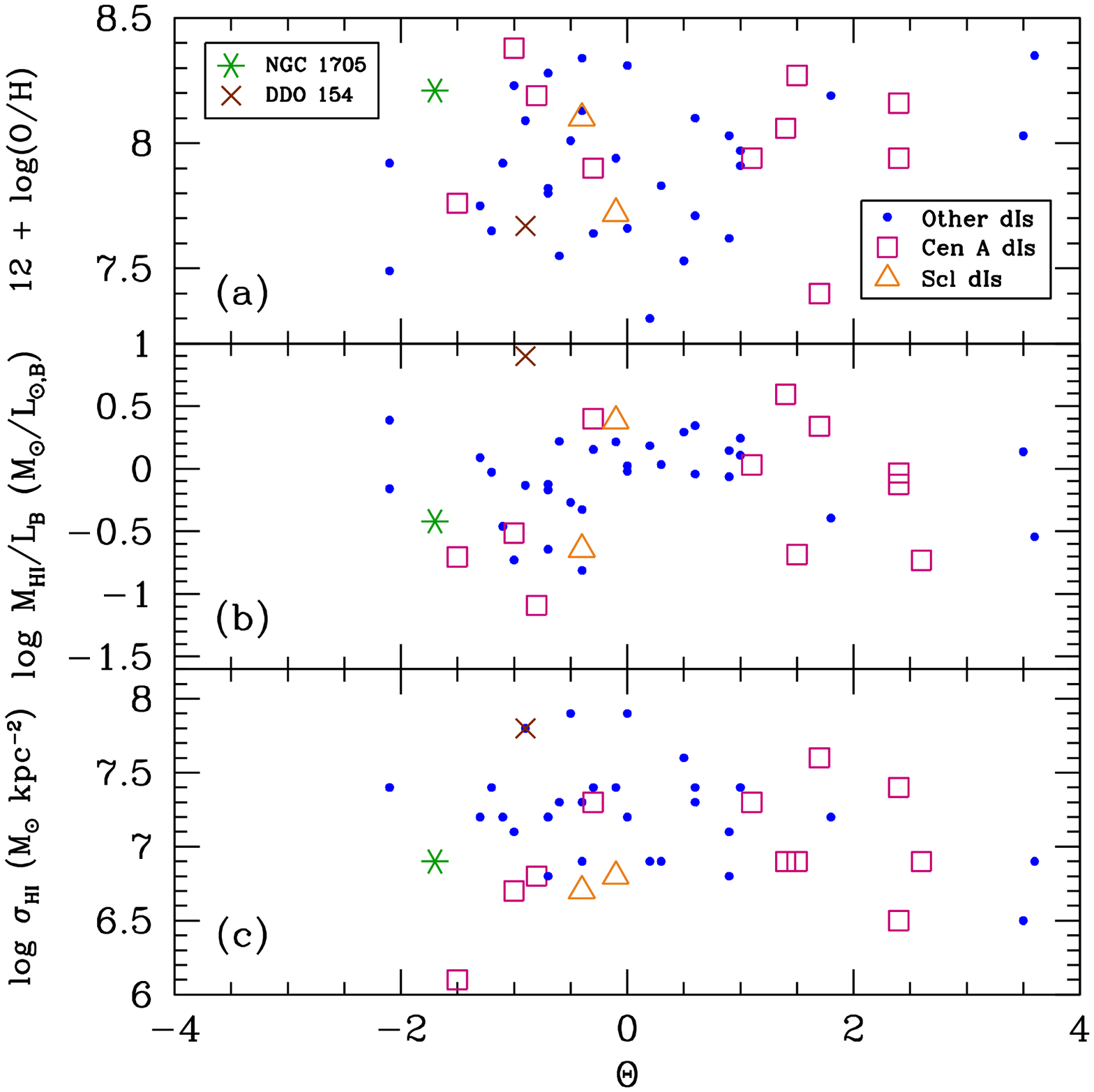} 
\includegraphics[width=81mm]{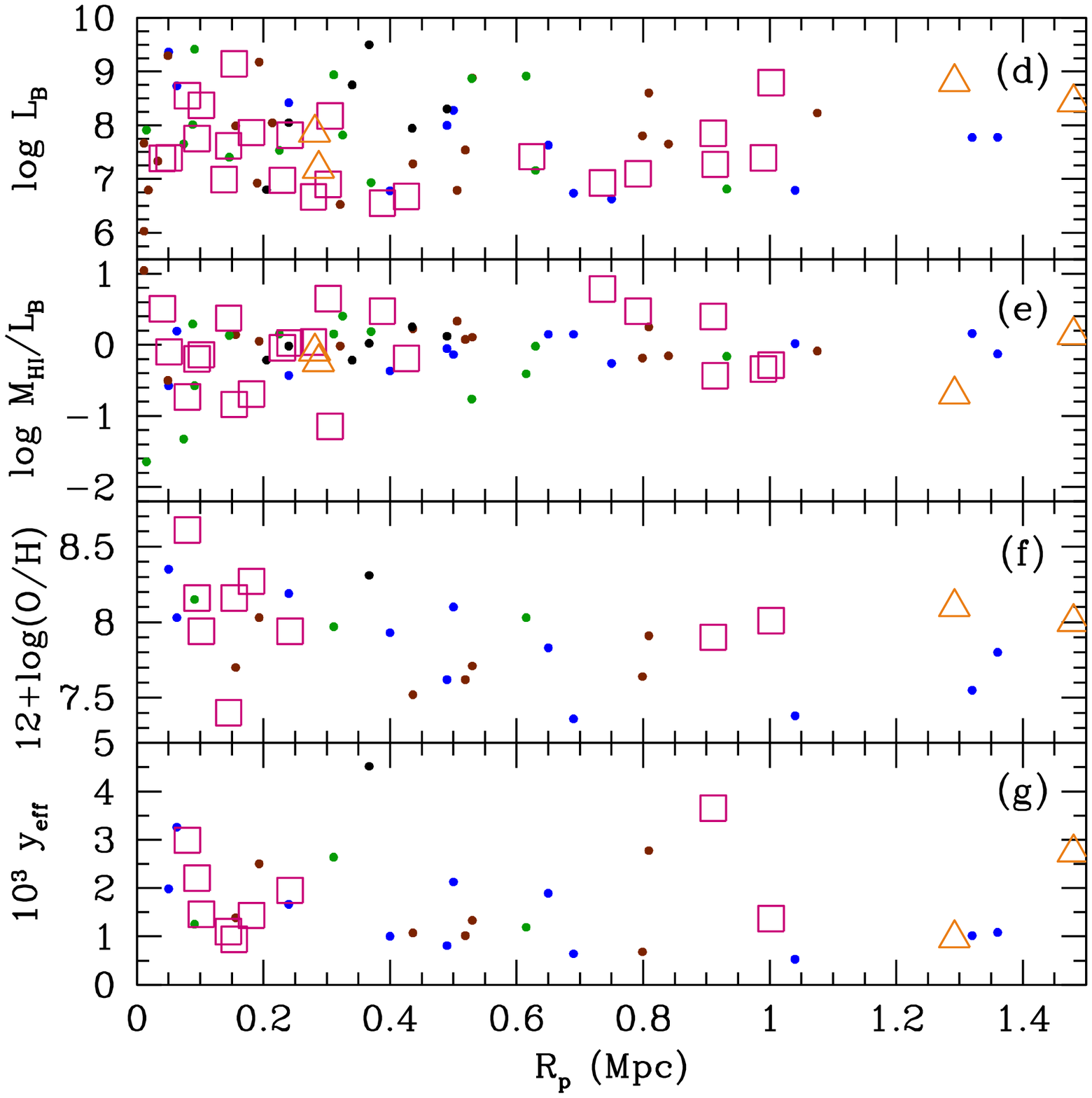}
\caption{
{\em Left panels:}
Oxygen abundance (a),  
\hi\ gas-to-$B$ light fraction (b), and
\hi\ surface mass density (c) 
against tidal index.
Symbols are the same as in Fig.~\ref{fig_diags}.
{\em Right panels:}
Optical $B$ luminosity in solar units (d), 
\hi\ gas-to-$B$ light fraction in solar units (e), 
oxygen abundance (f), and
effective chemical yield (g)
as a function of projected distance of
dwarf galaxies to the primary member in their respective group.
Data are culled from the assigned membership of galaxies
into groups described by \citet{kara05}.
Cen~A group and Scl group dwarfs are marked as open squares in magenta
and open triangles in orange, respectively.
Filled circles denote:
Local Group (Milky Way $+$ M31) dwarfs in blue,
M81 group dwarfs in brown,
IC342 $+$ Maffei group dwarfs in dark green,
and 
CVnI group dwarfs in black.
}
\label{fig_ztidal}
\end{figure*}

To explore further the effects of group environments on
the properties of gas-rich dwarf galaxies, we examine
various properties against tidal index and projected 
distance; the latter two parameters are defined and
compiled in \citet{kara04} and \citet{kara05}.
Briefly, the tidal index of a galaxy is the maximum value 
of an enhancement in mass density caused by all neighbouring galaxies.
Objects with negative tidal indices are isolated in the field, and
objects with positive tidal indices are found within group environments,
and are likely undergoing a tidal interaction.
Within each group, the projected distance of a galaxy, $R_p$, is
given by $R_p = D_m\,\sin (\theta)$, where
$D_m$ is the distance of the main (or primary) group member to the
Milky Way, and $\theta$ is the angular distance (in degrees) of the
galaxy to the main member. 

In Fig.~\ref{fig_ztidal}, we have plotted various parameters
against the tidal index and the projected distance of
galaxies within each group.
On the left side, we have plotted oxygen abundance, \hi\ gas-to-light
fraction, and the \hi\ surface mass density as a function of tidal
index in panels a, b, and c, respectively.
Values for the \hi\ surface mass density were obtained from
\cite{kara04}.
We have compared Cen~A and Scl group dwarfs against other dwarfs (as
in Fig.~\ref{fig_diags}), NGC~1705, and DDO~154. 
There are no obvious trends in the first two panels.
We noted above that ESO~272$-$G025 and ESO~383$-$G087 appeared
to have low gas fractions for their oxygen abundances, compared
to other dwarfs.
Both have negative tidal indices, which indicate only that there
is no strong present-day tidal interactions.
In panel c, the \hi\ surface mass density appears to decrease with
increasing positive tidal index, which is expected as strong tidal
interactions remove gas from galaxies.
We note that ESO~272$-$G025 (Cen~A dI) has an unusually low \hi\ 
surface mass density for its tidal index 
($\Theta = -1.5$, log $\sigma_{\rm HI}$ $\approx$ 6).
This suggests that this dwarf may have encountered a strong
interaction, which has reduced its \hi\ content and has induced
present-day star formation (i.e., presence of \hii\ regions). 
Detailed spatially-resolved observations would be very timely to
confirm this scenario.

In Figs.~\ref{fig_ztidal}d--g, we have plotted
luminosity, \hi-gas-to-light fraction, oxygen abundance, and the
effective yield as a function of the projected distance, respectively.
We have separated dwarfs in different groups by color:
Local Group (Milky Way $+$ M31) in blue,
M81 group dwarfs in brown,
IC342 $+$ Maffei group dwarfs in dark green,
and CVnI group dwarfs in black.
Once more, there are no obvious trends; there is large scatter present
in the plots of oxygen abundance and yield versus distance.
Naturally, more data are required for examination and
for improved statistics in an intergroup comparison.
We note that ESO~381$-$G020 exhibits an unusually high effective yield 
for its projected distance ($\approx$ 0.9~Mpc), although
the dI does not appear to have any other unusual properties.

Although a number of Cen~A dwarfs in our sample have positive tidal
indices, there is little to separate these galaxies from
those with negative tidal indices in the luminosity-metallicity,
metallicity-gas fraction, and metallicity-tidal index diagrams.
There is also no clear trend between the current chemical
enrichment (as represented by either oxygen abundance or yield)
with the projected distance of galaxies within each group.
Because of low-number statistics, it is difficult to say at the
present time whether any separation of Cen~A dwarfs and Scl dwarfs
in these diagnostics is meaningful.

\section{Conclusions}	      
\label{sec_concl}

Results of optical spectroscopy of \hii\ regions from eight dwarf
galaxies in the Centaurus~A group are presented.
For ESO272$-$G025 and ESO324$-$G024, 
direct oxygen abundances of 
12$+$log(O/H) = $7.76 \pm 0.09$ and $7.94 \pm 0.11$ 
are derived, respectively.
Bright-line abundances for the remaining galaxies are derived with  
the McGaugh and the Pilyugin calibrations.
We have also considered data for additional Cen~A dI's
and Sculptor group dI's from the literature.
For their galaxy luminosities, we have found that direct or \othreea\
oxygen abundances agree well with the luminosity-metallicity
relationship for dwarf irregular galaxies.
Despite large tidal indices for a number of Cen~A dwarf galaxies,
there is no difference between galaxies with positive tidal indices
galaxies with negative tidal indices in the luminosity-metallicity,
metallicity-gas fraction, and metallicity-tidal index diagrams.
As expected in strong tidal interactions, the \hi\ surface mass
density appears to decrease with increasing positive tidal index. 
We have also examined global properties of dwarf galaxies based on
their intragroup properties. 
There are no obvious trends in plots of luminosity, HI-to-$B$-light
ratio, oxygen abundance, and yield against projected distance of
galaxies within nearby groups.  
We also report spectra and abundances for three nearby dwarf 
irregular galaxies: DDO~47, NGC~3109, and Sextans~B.
Our direct oxygen abundance ($7.92 \pm 0.06$) for DDO~47 agrees
with the measurement previously reported by \cite{skh89}.
For Sextans~B, our direct oxygen abundance ($7.80 \pm 0.13$) 
is consistent with the low value reported by \cite{skh89}, and
agrees with abundances reported by \cite{kniazev05} and 
\cite{magrini05}.

\section*{acknowledgements}        

We thank the anonymous referee for comments which improved 
the presentation of the manuscript. 
H.~L. is supported by the Gemini Observatory, which is operated by the
Association of Universities for Research in Astronomy, Inc., on
behalf of the international Gemini partnership of Argentina,
Australia, Brazil, Canada, Chile, the United Kingdom, and the United
States of America.
We are grateful to ESO for awarded telescope time, and
we thank Lisa Germany, George Hau, and the staff at ESO La Silla for
help with the observations.
Partial support for this work was provided by NASA through grant 
GO-08192.97A from the Space Telescope Science Institute, which is 
operated by the Association of Universities for Research in 
Astronomy, Inc., under NASA contract NAS5-26555.
D.~B.~Z. acknowledges support from the National Science Foundation
Postdoctoral Fellowship.
E.~K.~G. appreciates support by the Swiss National Science Foundation
through grants 200021-101924 and 200020-105260.
H.~L. is grateful for support from the Max-Planck-Institute for
Astronomy where this project was begun, for partial support
from a NASA LTSARP grant NAG~5-9221, and from the University
of Minnesota. 
H.~L. also thanks Stephanie C\^ot\'e and Evan Skillman for providing
their \halpha\ images, 
and Liese van Zee for discussions regarding bright-line calibrations.
Some data were accessed as Guest User, Canadian Astronomy Data Center,
which is operated by the Dominion Astrophysical Observatory for the 
National Research Council of Canada's Herzberg Institute of Astrophysics. 
This research has made use of the NASA/IPAC Extragalactic Database,
which is operated by the Jet Propulsion Laboratory,
California Institute of Technology, under contract with the
National Aeronautics and Space Administration. 


\label{lastpage}

\bsp 


\begin{thebibliography}{}  


\bibitem[Aller(1984)]{aller84}
        Aller, L. H. 1984, Physics of Thermal Gaseous Nebulae
        (Dordrecht: Reidel)

\bibitem[Asplund et al.(2004)]{asplund04}
        Asplund, M., Grevesse, N., Sauval, A. J., Allende Prieto, C.,
        \& Kiselman, D. 2004, \aap, 417, 751 

\bibitem[Banks et al.(1999)]{banks99}
        Banks, G. D., et al. 1999, \apj, 524, 612

\bibitem[Barnes \& de Blok(2001)]{bdb01}
        Barnes, D. G., \& de Blok, W. J. G. 2001, \aj, 122, 825

\bibitem[Bell \& de Jong(2001)]{belldejong01}
        Bell, E. F., \& de Jong, R. S. 2001, \apj, 550, 212

\bibitem[Bothun et al.(1986)]{bmcm86}
        Bothun, G. D., Mould, J. R., Caldwell, N., \&
        MacGillivray, H. T. 1986, \aj, 92, 1007

\bibitem[Bouchard et al.(2007)]{bouchard07}
        Bouchard, A., Jerjen, H., Da Costa, G. S., \& Ott, J.
        2007, \aj, 133, 261

\bibitem[Bresolin et al.(1993)]{bresolin93}
	Bresolin, F., Capaccioli, M., \& Piotto, G.
	1993, \aj, 105, 1779


\bibitem[Campbell et al.(1986)]{ctm86}
        Campbell, A., Terlevich, R. J., \& Melnick, J. 
        1986, \mnras, 223, 811

\bibitem[Cardelli et al.(1989)]{cardelli89}
        Cardelli, J. A., Clayton, G. C., \& Mathis, J. S.
        1989, \apj, 345, 245

\bibitem[Carignan(1985)]{carignan85}
        Carignan, C. 1985, \apj, 299, 59
 
\bibitem[C\^ot\'e et al.(1997)]{cote97}
        C\^ot\'e, S., Freeman, K. C., Carignan, C., \&
        Quinn, P. J. 1997, \aj, 114, 1313

\bibitem[C\^ot\'e et al.(2000)]{cote00}
	C\^ot\'e, S., Carignan, C., \& Freeman, K. C.
	2000, \aj, 120, 3027


\bibitem[de Vaucouleurs(1979)]{devau79}
	de Vaucouleurs, G. 1979, \aj, 84, 1270

\bibitem[de Vaucouleurs et al.(1991)]{rc3}
	de Vaucouleurs, G., de Vaucouleurs, A., Corwin, H., Jr., 
	Buta, R. J., Paturel, G., \& Fouqu\'e, P.
	1991, Third Reference Catalogue of Bright Galaxies
	(New York: Springer)

\bibitem[Dinerstein(1990)]{dinerstein90}
        Dinerstein, H. L.  1990, in The Interstellar Medium in
        Galaxies, ed. H. A. Thronson \& J. M. Shull 
        (Dordrecht: Kluwer), 257

\bibitem[Dopita et al.(1984)]{dopita84}
        Dopita, M. A., Binette, L., D'Odorico, S., \& Benvenuti, P.
        1984, \apj, 276, 653




\bibitem[Garnett(1992)]{garnett92}
        Garnett, D. R. 1992, \aj, 103, 1330

\bibitem[Graham(1979)]{graham79}
	Graham, J. A. 1979, \apj, 232, 60

\bibitem[Grebel(1997)]{grebel97}
        Grebel, E.~K.\ 1997, Reviews of Modern Astronomy, 10, 29 

\bibitem[Grebel(1999)]{grebel99}
        Grebel, E. K. 1999, in The Stellar Content of the Local
        Group, IAU Symp. 192, ed. P. Whitelock \& R. Cannon 
	(San Francisco: Astron. Soc. of the Pacific), 17 


\bibitem[Grebel et al.(2000)]{grebel00}
        Grebel, E. K., et al. 2000, ASP Conf.~Ser.~221: Stars, Gas and
        Dust in Galaxies: Exploring the Links, ed. D. Alloin, \& K.
        Olsen, \& G. Galaz (San Francisco: ASP), 147 

\bibitem[Grebel(2001a)]{grebel01a}
	Grebel, E. K. 2001a, Ap\&SSS, 277, 231
 
\bibitem[Grebel(2001b)]{grebel01b}
 	Grebel, E. K. 2001b, in Dwarf Galaxies and the Environment,
 	ed. K. S. De Boer, R.-J. Dettmar, \& U. Klein
 	(Aachen: Shaker Verlag), 45

\bibitem[Grebel et al.(2003)]{ggh03}
        Grebel, E. K., Gallagher, J. S., \& Harbeck, D. R.
        2003, \aj, 125, 1926 	

\bibitem[Grossi et al.(2007)]{grossi07}
        Grossi, M., Disney, M. J., Pritzl, B. J., Knezek, P. M., 
        Gallagher, J. S., Minchin, R. F., \& Freeman, K. C.
        2007, \mnras, 374, 107



\bibitem[Hoffman et al.(1996)]{hoffman96}
        Hoffman, G. L., Salpeter, E. E., Farhat, B., Roos, T.,
	Williams, H., \& Helou, G. 1996, \apjs, 105, 269

\bibitem[Jenkins(1987)]{jenkins87} 
        Jenkins, C.~R. 1987, \mnras, 226, 341 


\bibitem[Jerjen et al.(2000)]{jbf00}
	Jerjen, H., Binggeli, B., \& Freeman, K. C.
	2000, \aj, 119, 593

\bibitem[Jobin \& Carignan(1990)]{jc90}
        Jobin, M., \& Carignan, C. 1990, \aj, 100, 648

\bibitem[Karachentsev(2005)]{kara05}
	Karachentsev, I. D. 2005, \aj, 129, 178

\bibitem[Karachentsev et al.(2002)]{kara02a}
 	Karachentsev, I. D., Dolphin, A. E., Geisler, D., et al.
 	2002, \aap, 383, 125

\bibitem[Karachentsev et al.(2002b)]{kara02b}
        Karachentsev, I. D., Sharina, M. E., Dolphin, A. E., et al.
	2002b \aap, 385, 21

\bibitem[Karachentsev et al.(2002c)]{kara02c}
        Karachentsev, I. D., Sharina, M. E., Makarov, D. I., et al.
	2002c, \aap, 389, 812

\bibitem[Karachentsev et al.(2003a)]{kara03a}
 	Karachentsev, I. D., Makarov, D. I., Sharina, M. E., et al. 
 	2003a, \aap, 398, 467

\bibitem[Karachentsev et al.(2003b)]{kara03b}
        Karachentsev, I. D., Makarov, D. I., Sharina, M. E., et al.
	2003b, \aap, 398, 479


\bibitem[Karachentsev et al.(2004)]{kara04}
	Karachentsev, I. D., Karachentseva, V. E., Huchtmeier, W. K.,
	\& Makarov, D. I. 2004, \aj, 127, 2031

\bibitem[Karachentsev et al.(2006)]{kara06}
	Karachentsev, I. D. et al. 2007, \apj, in press
	(astro-ph/0603091)

\bibitem[Kennicutt \& Skillman(1993)]{ks93}
        Kennicutt, R. C., \& Skillman, E. D. 1993, \apj, 411, 655

\bibitem[Kennicutt \& Skillman(2001)]{ks01}
        Kennicutt, R. C., \& Skillman, E. D. 2001, \aj, 121, 1461

\bibitem[Kniazev et al.(2004)]{kniazev04}
	Kniazev, A. Y., Pustilnik, S. A., Grebel, E. K., Lee, H., \&
	Pramskij, A. G. 2004, \apjs, 153, 429

\bibitem[Kniazev et al.(2005)]{kniazev05}
	Kniazev, A. Y., Grebel, E. K., Pustilnik, S. A., Pramskij, A. G.,
 	\& Zucker, D. B. 2005, \aj, 130, 1558 

\bibitem[Kobulnicky et al.(1999)]{chip99}
        Kobulnicky, H. A., Kennicutt, R. C. Jr., \& Pizagno, J. L.
        1999, \apj, 514, 544

\bibitem[H. Lee et al.(2003a)]{lee03south}
        Lee, H., Grebel, E. K., \& Hodge, P. W. 2003a, \aap, 401, 141

\bibitem[H. Lee et al.(2003b)]{lee03field}
        Lee, H., McCall, M. L., Kingsburgh, R., Ross, R., \&
        Stevenson, C. C. 2003b, \aj, 125, 146

\bibitem[H. Lee et al.(2003c)]{lee03virgo}
	Lee, H., McCall, M. L., \& Richer, M. G.
	2003c, \aj, 125, 2975

\bibitem[H. Lee \& Skillman(2004)]{ls04}
	Lee, H., \& Skillman, E. D. 2004, \apj, 614, 698

\bibitem[H. Lee et al.(2005)]{lsv05}
	Lee, H., Skillman, E. D., \& Venn, K. A. 2005, \apj, 620, 223

\bibitem[H. Lee et al.(2006a)]{lsv06}
 	Lee, H., Skillman, E. D., \& Venn, K. A. 2006a, \apj, 642, 813

\bibitem[H. Lee et al.(2006b)]{irlz}
        Lee, H., Skillman, E. D., Cannon, J. M., Jackson, D. C., 
	Gehrz, R. D., Polomski, E. F., \& Woodward, C. E.
	2006b, \apj, 647, 970


\bibitem[Magrini et al.(2005)]{magrini05}
        Magrini, L., Leisy, P., Corradi, R. L. M., Perinotto, M.,
        Mampaso, A., \& V\'{\i}lchez, J. M.
        2005, \aap, 443, 115

\bibitem[Martin(1997)]{martin97}
	Martin, C. 1997,\apj, 491, 561



\bibitem[Mel\'endez(2004)]{melendez04}
        Mel\'endez, J. 2004, \apj, 615, 1042

\bibitem[McCall et al.(1985)]{mrs85}
        McCall, M. L., Rybski, P. M., \& Shields, G. A.
        1985, \apjs, 57, 1

\bibitem[McGaugh(1991)]{mcgaugh91}
        McGaugh, S. S.  1991, \apj, 380, 140

\bibitem[McGaugh(1994)]{mcgaugh94}
        McGaugh, S. S.  1994, \apj, 426, 135

\bibitem[Meier et al.(1989)]{meier89}
	Meier, D. L. et al. 1989, \aj, 98, 27

\bibitem[Mirabel et al.(1999)]{mirabel99} 
        Mirabel, I. F., Laurent, O., Sanders, D. B., et al. 
	1999, \aap, 341, 667

\bibitem[Moles et al.(1990)]{mam90}
	Moles, M., Aparicio, A., \& Masegosa, J.
	1990, \aap, 228, 310

\bibitem[Mulchaey(2000)]{mulchaey00}
        Mulchaey, J. S. 2000, \araa, 38, 289
 
\bibitem[Olive \& Skillman(2001)]{os01}
        Olive, K. A., \& Skillman, E. D. 2001, New Astronomy, 6, 119

\bibitem[Osterbrock(1989)]{osterbrock}
        Osterbrock, D. E. 1989, Astrophysics of Gaseous Nebulae and
        Active Galactic Nuclei (Mill Valley: University Science Books)

\bibitem[Pagel et al.(1979)]{pagel79}
        Pagel, B. E. J., Edmunds, M. G., Blackwell, D. E., et al. 
        1979, \mnras, 189, 95

\bibitem[Pe\~na et al.(2006)]{pena06}
        Pe\~na, M., Richer, M. G., \& Stasi\'nska, G.
	2006, \aap, accepted
	(astro-ph/0612553)

\bibitem[Peng et al.(2002)]{peng02}
	Peng, E. W., Ford, H. W., Freeman, K. C., \& White, R. L.
	2002, \aj, 124, 3144

\bibitem[Pilyugin(2000)]{pilyugin00}
        Pilyugin, L. S. 2000, \aap, 362, 325

\bibitem[Pilyugin(2001a)]{pilyugin01a}
        Pilyugin, L. S. 2001a, \aap, 369, 594




\bibitem[Rejkuba et al.(2006)]{rejkuba06}
        Rejkuba, M., Da Costa, G. S., Jerjen, H., Zoccali, M., \&
        Binggeli, B. 2006, \aap, 448, 983

\bibitem[Richer \& McCall(1992)]{rm92}
        Richer, M. G., \& McCall, M. L. 1992, \aj, 103, 54

\bibitem[Richer \& McCall(1995)]{rm95}
        Richer, M. G., \& McCall, M. L. 1995, \apj, 445, 642

\bibitem[R\"onnback \& Bergvall(1995)]{rb95}
	R\"onnback, J. \& Bergvall, N. 1995, \aap, 302, 353
 

\bibitem[Sakai et al.(1997)]{sakai97}
	Sakai, S., Madore, B. F., Freedman, W. L.
	1997, \apj, 480, 589
        
\bibitem[Schaerer \& Vacca(1998)]{sv98}
          Schaerer D., \& Vacca W.D., 1998, \apj, 497, 168
 
\bibitem[Schmidt(1963)]{schmidt63}
        Schmidt, M. 1963, \apj, 137, 758

\bibitem[Searle \& Sargent(1972)]{ss72}
        Searle, L., \& Sargent, W. L. W. 1972, \apj, 173, 25
 
\bibitem[Shaw \& Dufour(1995)]{sd95}
        Shaw, R. A., \& Dufour, R. J. 1995, \pasp, 107, 896



\bibitem[Skillman(1985)]{skillman85}
        Skillman, E. D. 1985, \apj, 290, 449

\bibitem[Skillman(1989)]{skillman89}
        Skillman, E. D. 1989, \apj, 347, 883

\bibitem[Skillman(1998)]{skillman98}
        Skillman, E. D. 1998, in Stellar Astrophysics of the Local
        Group: VIII Canary Islands Winter School of Astrophysics, 
        ed. A. Aparicio, A. Herrero, \& F. S\'anchez
        (Cambridge: Cambridge Univ. Press), 457

\bibitem[Skillman et al.(1989)]{skh89}
        Skillman, E. D., Kennicutt, R. C., Jr., \& Hodge, P.
        1989, \apj, 347, 875

\bibitem[Skillman et al.(2003a)]{scm03a}
	 Skillman, E. D., C\^ot\'e, S., \& Miller, B. W.
	 2003a, \aj, 125, 593

\bibitem[Skillman et al.(2003b)]{scm03b}
        Skillman, E. D., C\^ot\'e, S., \& Miller, B. W.
        2003b, \aj, 125, 610


\bibitem[Stasi\'{n}ska et al.(1986)]{scv86}
	Stasi\'nska, G., Comte, G., \& Vigroux, L.
	1986, \aap, 154, 352

\bibitem[Storey \& Hummer(1995)]{sh95}
        Storey, P. J., \& Hummer, P. J. 1995, \mnras, 272, 41

\bibitem[Strobel et al.(1991)]{strobel91}
        Strobel, N. V., Hodge, P., \& Kennicutt, R. C., Jr.,
        1991, \apj, 383, 148


\bibitem[Tully et al.(2002)]{tully02}
        Tully, R. B., Somerville, R. S., Trentham, N., \& 
        Verheijen, M. A. W. 2002, \apj, 569, 573 

\bibitem[Tully(2005)]{tully05}
        Tully, R. B. 2005, \apj, 618, 214

\bibitem[van den Bergh(1999)]{vdb99}
        van den Bergh, S. 1999, \apj, 517, L97

\bibitem[van Gorkom et al.(1990)]{vangorkom90}
	van Gorkom, J. H., van der Hulst, J., Haschick, A.,
	\& Tubbs, A. 1990, \aj, 99, 1781

\bibitem[van Zee(2001)]{vz01}
        van Zee, L. 2001, \aj, 121, 2003

\bibitem[van Zee \& Haynes(2006)]{vzh06}
	van Zee, L., \& Haynes, M. P.
	2006, \apj, 636, 214

\bibitem[van Zee et al.(1997)]{vanzee97}
	van Zee, L., Haynes, M. P., \& Salzer, J. J.
	1997, \aj, 114, 2479

\bibitem[van Zee et al.(1998)]{vanzee98}
	van Zee, L., Salzer, J. J., Haynes, M. P.,
	O'Donoghue, A. A., \& Balonek, T. J.
	1998, \aj, 116, 2805
 
\bibitem[van Zee et al.(2004a)]{vanzee04a}
        van Zee, L., Skillman, E. D., \& Haynes, M. P.
        2004, \aj, 128, 121

\bibitem[van Zee et al.(2004b)]{vanzee04b}
        van Zee, L., Barton, E. J., \& Skillman, E. D.
        2004, \aj, 128, 2797

\bibitem[van Zee et al.(2006)]{vzsh06}
	van Zee, L., Skillman, E. D., \& Haynes, M. P.
	2006, \apj, 637, 269

\bibitem[Webster \& Smith(1983)]{ws83}
        Webster, B. L., \& Smith, M. G.  1983, \mnras, 204, 743

\bibitem[Webster et al.(1983)]{webster83}
        Webster, B. L., Longmore, A. J., Hawarden, T. G., \&
        Mebold, U.  1983, \mnras, 205, 643






\end{thebibliography}
\end{document}